\documentclass[12 pt]{article}
\usepackage[normalem]{ulem}
\usepackage{soul}
\usepackage{amsmath,amsfonts,amssymb,mathtools,bbm}
\usepackage[doublespacing]{setspace}
\usepackage{pifont}
\usepackage[bottom]{footmisc}
\usepackage{indentfirst}
\usepackage{endnotes}
\usepackage{rotating}
\usepackage{multirow}
\usepackage{multicol}
\usepackage{bbm}
\usepackage{adjustbox}
\usepackage{changepage}
\usepackage{tabularx,lipsum}
\usepackage[para,online,flushleft]{threeparttable}
\usepackage{verbatim}
\usepackage{graphicx}
\graphicspath{{figures/}}
\usepackage{pdfpages}
\usepackage{fancyhdr}
\usepackage{lscape}
\usepackage[section]{placeins}
\usepackage{hyperref}
\usepackage{natbib}
\usepackage{lineno}
\usepackage{subcaption}
\usepackage{tikz}
\usetikzlibrary{decorations.pathreplacing,positioning, arrows.meta}
\usepackage{caption}
\usepackage[left=1in,right=1in,top=1in,bottom=1in]{geometry}
\usepackage{booktabs}
\usepackage{comment}
\usepackage{breqn}
\usepackage{float}
\usepackage{diagbox}
\usepackage{titlesec}
\usepackage[title]{appendix}
\usepackage{enumitem}
\usepackage{longtable}
\titleformat{\paragraph}
{\normalfont\normalsize\bfseries}{\theparagraph}{1em}{}
\titlespacing*{\paragraph}
{0pt}{3.25ex plus 1ex minus .2ex}{1.5ex plus .2ex}
\usepackage{makecell}
\newtheorem{proposition}{Proposition}
\newtheorem{assumption}{Assumption}

\newcommand{\E}{\mathbb{E}}
\newcommand{\Var}{\operatorname{Var}}
\newcommand{\Cov}{\operatorname{Cov}}
\newcommand{\Corr}{\operatorname{Corr}}
\newcommand{\Gap}{\mathit{Gap}}
\makeatletter
\def\@biblabel#1{\hspace*{-\labelsep}}
\makeatother
\definecolor{yifeiblue}{RGB}{50,30,190}
\hypersetup{colorlinks=true,linkcolor=yifeiblue, urlcolor=yifeiblue, citecolor=yifeiblue}
\makeatletter
\begin{document}
\title{\textbf{Anonymization and Information Loss}\thanks{We are grateful for comments and feedback from Yi Cao, Wenxi Jiang, Yu-Jane Liu, Markus Pelger, Suproteem Sarkar, Olivier Scaillet, Xudong Wen, Zhengyang Xu, Bernard Yeung, Anastasia Zakolyukina, Dake Zhang, Bohui Zhang, and Shunming Zhang, as well as from seminar and conference participants at the Wolfe Research QES Conference, China Fintech Research Conference, Jingshi Scholars Forum, Hong Kong Conference on FinTech and AI in Finance, CICF, Future Scholars in Finance Forum, CES Annual Meeting, and ABFER-JFDS Conference on AI for Finance. }} 
\author{%
Ke Wu\thanks{School of Finance, Renmin University of China.
Email: \href{mailto:ke.wu@ruc.edu.cn}{ke.wu@ruc.edu.cn}.}
\and
Baozhong Yang\thanks{J. Mack Robinson College
of Business, Georgia State University.
Email: \href{mailto:bzyang@gsu.edu}{bzyang@gsu.edu}.}
\and
Zhenkun Ying\thanks{School of Finance, Renmin University of China.
Email: \href{mailto:yingzhenkun0205@ruc.edu.cn}
{yingzhenkun0205@ruc.edu.cn}.}
\and
Dexin Zhou\thanks{
Zicklin School of Business, Baruch College.
Email: \href{mailto:dexin.zhou@baruch.cuny.edu}
{dexin.zhou@baruch.cuny.edu}.}
}
\date{}
\maketitle
\singlespacing
\begin{abstract}
\noindent
Anonymizing financial texts prevents large language models (LLMs) from exploiting look-ahead bias, but inadvertently weakens the extracted signal. We propose a framework disentangling this information loss from bias removal. Predicting S\&P credit downgrades, raw texts significantly outperform anonymized texts. Look-ahead bias does not explain this difference; rather, anonymization degrades predictive performance by masking informative numerical and agent entities, fundamentally altering how LLMs interpret the remaining context. This degradation spans various LLMs, tasks, and text types. To quantify it, we introduce the ``anonymization gap,'' a metric requiring no outcome data.
\end{abstract}

\clearpage
\doublespacing

\section{Introduction}
Large language models (LLMs) have transformed how finance and accounting researchers extract information from textual data. \citet{lopez2023can}, for example, show that LLM-generated sentiment predicts returns more effectively than traditional bag-of-words measures. Researchers use LLMs to extract economic signals from earnings calls, interpret firm policies, and measure geopolitical exposure \citep[e.g.,][]{jha2025chatgptcorporatepolicies, jha2025harnessinggenerativeaieconomic, Clayton2025}. Other applications include mapping global production networks using financial texts \citep{GlobalBusinessNet}, detecting investor sentiment on social media \citep{Chen2025Wisdom}, and generating structured economic representations from news \citep{Sarkar2025EconomicRepresentations}.

Despite their power in analyzing financial texts, several studies caution that LLMs may introduce look-ahead bias in the data extraction process. This bias is particularly relevant for finance and economics, where extracted signals are used to predict future economic performance or stock returns. It arises during training, when LLMs learn from vast corpora spanning long periods of time, potentially causing their outputs to incorporate information that would not have been accessible to an agent prior to the model's knowledge-cutoff
 \citep[e.g.,][]{Levy2024,sarkar2025lookahead,ludwig2025largelanguagemodelsapplied,lopezlira2025memorizationproblemtrustllms}.

Several strategies have been proposed to mitigate this bias. Among them, anonymization has gained significant popularity since it is relatively easy to implement while still allowing researchers to use state-of-the-art models. Anonymization strategies vary in scope. Some replace identifiers such as firm names or tickers with placeholders (e.g., ``FIRM\_1'') \citep{glasserman2023assessinglookaheadbiasstock}, whereas others combine identifier masking with paraphrasing of the entire text to remove firm-specific contextual cues \citep{engelberg2025entity}. The premise is that obscuring explicit identifiers and firm-specific contextual cues forces LLMs to base their analysis exclusively on the document's inherent linguistic and semantic content rather than on firm-specific knowledge.\footnote{Other approaches include training chronologically consistent models that only use data before the date of the textual content of interest \citep[e.g.,][]{sarkar2024storieslm, he2025chronologicallyconsistentlargelanguage} or conducting robustness tests using texts beyond existing models' knowledge-cutoff \citep[e.g.,][]{jha2025chatgptcorporatepolicies, Siano2025}.} Appendix Table~\ref{appendix-tab:prior_anonymization_applications} provides a sample of published and working papers in finance and accounting that explicitly use anonymization to address look-ahead bias.

In this study, we investigate a potentially important limitation of the anonymization methodology: it can lead to information loss while replacing identifiers or paraphrasing firm-specific context. The motivating intuition is that entity identity and numerical detail can sometimes serve as context for interpretation rather than solely as labels. Consider a firm's statement that it will lay off 10,000 employees. For a company undergoing financial distress, such as a struggling retailer facing bankruptcy, this announcement may reinforce a negative outlook by signaling deepening trouble. Conversely, for a highly profitable technology company undertaking strategic restructuring, the same announcement may be interpreted positively as a sign of disciplined cost management. Once the firm name and numerical magnitude are anonymized, the corresponding passages from these two settings may collapse into the same input: ``FIRM\_1 announced it will lay off NUMBER\_1 employees.'' This procedure strips the context necessary to infer what a firm's action signals about its underlying fundamental performance, potentially leading to noisier signals.

To translate this intuition into testable predictions, we develop a parsimonious framework. We express the anonymized score as a component shared with the raw assessment plus an orthogonal, document-specific residual. Under the information-loss interpretation, the shared component captures average signal retention, and the residual captures potential interpretive distortion. The framework yields four propositions. First, under the raw-score encompassing benchmark, the anonymized score has a smaller standalone coefficient and lower explanatory power than the raw score. Second, the anonymized score contributes no incremental predictive content in a horse-race regression that includes both scores. Third, under cross-cutoff stability, positive look-ahead bias generates a positive interaction between the observed score and a pre-cutoff indicator. Fourth, the attenuation-adjusted raw--anonymized gap negatively moderates the predictive slope of the anonymized score. Together, these propositions organize our main empirical tests.

Empirically, we test these predictions using conference call transcripts dated after the models' knowledge-cutoff. We compare the informativeness of signals extracted from the raw and anonymized versions of the same transcripts. Because both the transcripts and their subsequent outcomes postdate the documented cutoff, this sample allows us to examine whether anonymization attenuates genuine predictive content in a setting where such attenuation is unlikely to reflect the removal of memorized future outcomes.

We consider two widely used anonymization strategies. The first relies on the spaCy library's most capable Transformer-based named-entity-recognition model, which we denote TRF, to detect entities and replace them with placeholders such as ``FIRM\_1''. The second is the iterative procedure of \citet{engelberg2025entity}, which uses an LLM to mask entities and paraphrase the surrounding text.

Our main analysis examines an economically important prediction task: forecasting future S\&P credit-rating downgrades \citep[e.g.,][]{donovan2021measuring}. We prompt GPT-4o-mini to extract a downgrade-probability score from each transcript and from its anonymized versions, and we relate these scores to actual rating downgrades over the subsequent twelve months. The ROC curves first show that all three scores predict downgrades, but the raw score has the largest area under the curve. The anonymized scores predict downgrades on their own, but their standalone coefficients and explanatory power are lower than those of the raw score. In a horse-race regression, the raw score remains large and highly significant while the anonymized coefficients become statistically insignificant, the pattern the framework predicts under information loss.

Our tests so far use only the post-cutoff sample. To probe look-ahead bias directly, we extend the sample to the pre-cutoff period (November 2022--October 2023). If look-ahead bias inflated an observed score, our framework predicts a positive interaction between that score and the pre-cutoff indicator. Instead, the interactions are small and statistically insignificant for the raw, TRF-anonymized, and LLM-anonymized scores, providing little evidence that look-ahead bias is a first-order driver of their predictive content.

We also propose a document-level severity diagnostic: the anonymization gap, defined as the attenuation-adjusted absolute difference between the raw and anonymized scores. We find that the anonymized score predicts downgrades less effectively when this gap is larger. The interaction is negative in both windows and significant for both methods post-cutoff, supporting the gap as a proxy for anonymization's document-specific residual.

Having established that anonymization attenuates predictive content and gauged its severity, we next ask where that content resides, beginning with the entity categories that the TRF anonymization algorithm removes. We decompose detected entities into four categories: NUMBERS, AGENTS, PLACES, and OTHERS. Specifically, we apply TRF-based anonymization to each transcript but selectively preserve only one entity category at a time while replacing the other three with placeholders, then extract the downgrade signal and compare its predictive content to that of the raw text. No single preserved category fully recovers the raw text's predictive content. However, preserving NUMBERS (such as financial figures and dates) or AGENTS (such as company and executive names) recovers substantially more of the raw score than preserving PLACES or OTHERS. These TRF-based results indicate that numerical and agent entities serve as particularly important contextual anchors for interpreting financial text.

The entity analysis shows which parts of the source text carry the signal, but not what the model does differently once they are removed. To see what changes, we conduct an audit organized around ten credit-risk factors drawn from S\&P Global Ratings' Corporate Methodology, spanning business risk, financial risk, and rating adjustments. For each transcript version and factor, the model returns a downgrade-risk score and an exact supporting quote, which we match across the raw and anonymized texts to track changes in both the evidence selected and the assessment of shared evidence. The audit reveals two distinct channels. Evidence displacement occurs when anonymization removes a credit-risk reason or redirects the model toward a different supporting passage; interpretive distortion occurs when the model continues to cite the same passage but assigns it a different risk assessment. A composite signal formed from the ten-factor assessments reproduces the predictive loss in the one-shot score, and channel-specific horse-races show that both evidence displacement and the altered reading of shared evidence matter. A Shapley decomposition assigns 40.6\% of the loss under TRF anonymization and 41.3\% under LLM anonymization to evidence displacement; interpretive distortion accounts for the remaining 59.4\% and 58.7\%, respectively.

We next show that these results generalize along three dimensions: the extraction model, the prediction task, and the type of text. First, repeating the extraction with a larger model (GPT-4o) and a reasoning-focused model (o3-mini) yields the same qualitative pattern. In standalone regressions, the raw score consistently outperforms the anonymized score for both models. In horse-race regressions that include both, the raw score remains significant for all three models, whereas the anonymized score is insignificant for GPT-4o-mini and o3-mini and only marginally significant for GPT-4o.

Second, the loss extends well beyond credit risk. Because predicting stock returns is central to asset-pricing research and concerns about look-ahead bias are especially salient in this setting, we first examine whether anonymization attenuates the relation between sentiment extracted by LLMs and stock returns. Sentiment extracted from raw earnings call transcripts strongly predicts announcement-window DGTW-adjusted returns. In standalone regressions, sentiment extracted from TRF- and LLM-anonymized transcripts produces smaller coefficients and lower adjusted $R^2$ than raw sentiment. When the raw and anonymized sentiment measures are both included, the coefficients on both anonymized measures fall by roughly two-thirds, while raw sentiment remains large and highly significant in both specifications. Thus, although the anonymized sentiment measures retain some incremental predictive content, much of their information is subsumed by raw sentiment.

Beyond stock returns, we examine four additional prediction tasks using earnings call transcripts. Following the designs of \citet{WhenIsaLiability}, \citet{jha2025chatgptcorporatepolicies}, and \citet{jha2025harnessinggenerativeaieconomic}, we use textual measures of managerial uncertainty, investment expectations, and macroeconomic outlook to predict future return volatility, capital expenditures, sales growth, and value-added growth. Across all four outcomes, the coefficient on the raw measure exceeds that on its TRF-anonymized counterpart. The anonymized uncertainty measure becomes statistically insignificant once the raw measure is included, while the anonymized measures retain significant but attenuated predictive content in the other three tasks.

Third, to assess whether our conclusions are specific to earnings calls, we replicate the return analysis on short, entity-dense news headlines. The results are consistent: raw sentiment dominates anonymized sentiment in horse-races, showing that the information-loss pattern is not confined to a particular type of disclosure.

Our study makes three contributions to the literature on LLM-based textual analysis in finance and accounting. First, we develop a parsimonious framework that separates two competing explanations for why anonymized and raw scores diverge: the removal of look-ahead bias and the loss of genuine, value-relevant information. While prior work cautions against look-ahead bias \citep{glasserman2023assessinglookaheadbiasstock, sarkar2025lookahead} and proposes anonymization as a remedy \citep{GlobalBusinessNet, engelberg2025entity}, our framework yields testable predictions that let researchers distinguish between these explanations rather than treating anonymization's effect as a black box.

Second, using a post-cutoff sample in which memorized future outcomes are unlikely to account for the raw score's advantage, we show that anonymization carries a significant hidden cost: it removes genuine, value-relevant information along with firm identity, and this cost persists across extraction models, prediction tasks, and types of financial disclosure. Because these tasks follow established designs \citep{donovan2021measuring, WhenIsaLiability, jha2025chatgptcorporatepolicies, jha2025harnessinggenerativeaieconomic}, the loss is not an artifact of settings constructed for this purpose. This evidence speaks to how the raw--anonymized difference should be read. \citet{engelberg2025entity} interpret that difference as an upper bound on look-ahead bias; our results indicate that it can also reflect information loss, since the difference persists in the post-cutoff window. This distinction matters for diagnostics built on that difference \citep{GaoJiangYan2026}.

Third, we develop a set of instruments for diagnosing information loss. The TRF-based entity decomposition and the audit of ten credit-risk factors from S\&P Global Ratings' Corporate Methodology identify its sources and its channels. The entity results highlight numerical and agent context, connecting the anonymization question to a longer tradition of identifying which parts of financial text carry economic signal \citep{WhenIsaLiability, garcia2023colour}, while the audit distinguishes evidence displacement, where anonymization removes or redirects supporting evidence, from interpretive distortion, where the same evidence is read differently once anonymized. The third instrument, the anonymization gap, measures severity document by document and is designed to travel: because it is computed from the raw and anonymized scores alone, it requires no outcome data and can be applied to any corpus a researcher has already anonymized. Documents with larger gaps yield systematically less informative anonymized signals in both the credit-risk and the return settings.

Because we document these costs in settings where look-ahead bias appears to play only a limited role, our results also carry a practical warning: absent a post-cutoff sample for comparison, researchers may mistakenly attribute an attenuation of signal informativeness to the successful mitigation of bias when it is in fact a byproduct of information loss. This concern is compounded by \citet{lopezlira2025memorizationproblemtrustllms}, who find that masked text often remains identifiable because models reconstruct entities and dates from surrounding context: anonymization may therefore impose the informational cost we document while delivering less protection than intended.

\section{A Framework to Separate Information Loss from Look-Ahead Bias}\label{sec:framework}
Intuitively, anonymizing a document strips away identifying and contextual detail, and some of that detail may carry information relevant to predicting future outcomes. The resulting information loss can operate through two channels: it can attenuate the component of the extracted score that tracks the full-context assessment, and it can introduce a document-specific residual. This section formalizes the information-loss hypothesis in a simple linear framework and distinguishes it from the potential removal of look-ahead bias, an alternative mechanism that could drive raw--anonymized score differences. It derives several testable propositions and develops a document-level diagnostic for heterogeneity in this residual. Supporting derivations are reported in Appendix~\ref{app:proofs}.

\subsection{Modeling Information Loss}
Let $Y$ denote a future outcome. Let $R$ and $A$ denote the scores associated with the raw and anonymized versions of a document, both free of look-ahead bias. We center and standardize these scores. Their
correlation,

\begin{equation}
 \rho\equiv\Corr(R,A)\in(0,1),
 \label{eq:rho}
\end{equation}
measures the population attenuation of the anonymized score relative to the raw score. The linear projection of $A$ on $R$ can be written as

\begin{equation}
 A=\rho R+u.
 \label{eq:decomposition}
\end{equation}
Because both scores are standardized, the projection residual satisfies
$\E[u]=0$, $\Cov(R,u)=0$, and $\Var(u)=1-\rho^2$.
Under the motivating information-loss hypothesis, $\rho R$ is interpreted as
the component of the full-context assessment retained after anonymization,
whereas $u$ captures the potential interpretive distortion induced by the
transformation.

To obtain sharp empirical predictions, we add the following benchmark assumption.

\begin{assumption}[Raw-score encompassing]\label{assump:information_loss}
For the prediction task under study, the raw score encompasses the anonymized score:

\begin{equation}
 Y=\mu+\beta R+\varepsilon,\qquad
 \E[\varepsilon\mid R,A]=0.
 \label{eq:outcome}
\end{equation}
\end{assumption}
Assumption~\ref{assump:information_loss} states that the anonymized score contains no outcome-relevant information beyond what is captured by the raw score. Together
with the projection in equation~\eqref{eq:decomposition}, it yields Propositions~\ref{prop:attenuation} and~\ref{prop:horse_race} below.

\begin{proposition}[Attenuation]\label{prop:attenuation}
Under equation~\eqref{eq:decomposition} and
Assumption~\ref{assump:information_loss}, the standalone
population slopes satisfy $b_R=\beta$ and $b_A=\rho\beta$, and the corresponding
fits satisfy $R_A^2=\rho^2R_R^2$. Because $0<\rho<1$, the anonymized-score
coefficient is smaller in absolute value than the full-context coefficient.
For a positively oriented signal ($\beta>0$), $0<b_A<b_R$, so $b_A$ is a lower
bound on the full-context predictive association.
\end{proposition}
Proposition~\ref{prop:attenuation} has a direct applied implication. If Assumption~\ref{assump:information_loss} holds, a statistically significant coefficient estimated from an anonymized score is conservative: the corresponding raw-score coefficient is larger in absolute value, and the raw score explains more of the outcome. Anonymization thus provides a lower bound on the estimated predictive association.

\begin{proposition}[No incremental information]\label{prop:horse_race}
Under equation~\eqref{eq:decomposition} and
Assumption~\ref{assump:information_loss}, a population
horse-race regression that includes both $R$ and $A$ assigns coefficient
$\beta$ to $R$ and coefficient zero to $A$.
\end{proposition}
Together, Propositions~\ref{prop:attenuation} and~\ref{prop:horse_race} provide
two testable implications of the benchmark information-loss hypothesis: the anonymized score should have an attenuated standalone association with the outcome and, under the raw-score encompassing condition, should contribute no incremental predictive content relative to the raw score.

\subsection{Modeling Look-Ahead Bias}
We now introduce look-ahead bias into the framework. For documents dated before the model's knowledge-cutoff, a score may partly reflect outcomes that entered the model's training data rather than information contained in the document. Let $P$ equal one for pre-cutoff documents and zero otherwise. For either score $S\in\{R,A\}$, let $\widetilde S$ denote the score we observe. Following \citet{GaoJiangYan2026}, we represent look-ahead bias as

\begin{equation}
 \widetilde S=S+\gamma_S P\varepsilon,\qquad
 0\leq\gamma_A\leq\gamma_R\leq 1.
 \label{eq:lookaheadbias}
\end{equation}
Here $\gamma_S$ measures, in standardized units, the share of the future innovation absorbed by the observed score. The upper bound rules out look-ahead bias larger than the full innovation, while the ordering $\gamma_A\leq\gamma_R$ reflects the intended ability of anonymization to weaken the link between the document and information the model may have learned about the firm.

To diagnose look-ahead bias empirically, consider the cutoff-interaction regression

\begin{equation}
 Y=d_0+d_1\widetilde S+d_2P+d_3(\widetilde S\times P)+e.
 \label{eq:cutoffregression}
\end{equation}
In this specification, $d_1$ is the post-cutoff predictive slope and $d_3$ captures the additional pre-cutoff slope, that is, the extra predictive power associated with look-ahead bias. If the positive interaction appears for the raw score but not for its anonymized counterpart, the pattern is consistent with anonymization removing look-ahead bias. If neither score exhibits a positive interaction, the pattern provides little evidence that look-ahead bias accounts for the raw score's advantage.

The interpretation of $d_3$ as look-ahead bias requires that the underlying score--outcome relation is stable across the cutoff. We formalize this as follows.

\begin{assumption}[Cross-cutoff stability]\label{assump:stability}
$\Cov(S,\varepsilon\mid P=p)=0$ in both periods; the score--outcome slope and the variances of the score and the future innovation are stable across the cutoff; the future innovation has strictly positive variance; and the same positive score scaling is used in both periods.
\end{assumption}

Under this stability assumption, a finding of $d_3=0$ implies $\gamma_S=0$: the score carries no look-ahead bias, so any attenuation after anonymization reflects information loss rather than bias removal. Appendix~\ref{app:proofs} derives the exact interaction coefficient.

\begin{proposition}[Look-ahead bias diagnostic]\label{prop:cutoff}
Under Assumptions~\ref{assump:information_loss} and~\ref{assump:stability}, consider a positively oriented score. If $\gamma_S=0$, then $d_3=0$ in equation~\eqref{eq:cutoffregression}. If $0<\gamma_S\leq 1$, then $d_3>0$. Thus, under the maintained assumptions, the model yields only two cases: no look-ahead bias produces $d_3=0$, whereas positive look-ahead bias produces $d_3>0$.
\end{proposition}

\subsection{The Anonymization Gap}
Beyond the average patterns characterized above, raw--anonymized disagreement may vary substantially across documents. From
equation~\eqref{eq:decomposition}, define the rescaled projection residual $v$ and its
absolute magnitude as

\begin{equation}
 v\equiv\frac{A}{\rho}-R=\frac{u}{\rho},
 \qquad
 \Gap\equiv\lvert v\rvert
 =\left\lvert\frac{A}{\rho}-R\right\rvert.
 \label{eq:gap}
\end{equation}
Rescaling $A$ by $\rho$ and subtracting $R$ removes the common projection
component, leaving the signed document-specific residual $v$; $\Gap$ records its magnitude. $\Gap$ is simply a scaled measure of disagreement between
the two scores, and under the information-loss benchmark, it can be interpreted as
a proxy for heterogeneity in this document-specific residual.

\begin{assumption}[Symmetry and independence]\label{assump:symmetry}
The rescaled residual $v$ is symmetric about zero and independent of the raw score $R$.
\end{assumption}

\noindent Under Assumptions~\ref{assump:information_loss} and~\ref{assump:symmetry}, the predictive slope of the anonymized score conditional on $\Gap=g$ is

\begin{equation}
 b(g)\equiv
 \frac{\Cov(Y,A\mid\Gap=g)}{\Var(A\mid\Gap=g)}
 =\frac{\beta}{\rho(1+g^2)}.
 \label{eq:conditionalslope}
\end{equation}
Appendix~\ref{app:proofs} gives the full derivation. For the positively oriented scores used in our tests, this slope decreases as the gap rises: a larger gap adds variation to the anonymized assessment without adding covariance with the outcome. This prediction motivates the interaction regression

\begin{equation}
 Y=c_0+c_1A+c_2\Gap+c_3(A\times\Gap)+e.
 \label{eq:gapregression}
\end{equation}
The benchmark predicts $c_3<0$. An empirically negative interaction would therefore support interpreting $\Gap$ as a proxy for the document-specific residual.

\begin{proposition}[The gap and conditional attenuation]\label{prop:gap}
Under Assumptions~\ref{assump:information_loss} and~\ref{assump:symmetry}, the conditional slope $b(g)=\beta/[\rho(1+g^2)]$ decreases with $g>0$ for a positively oriented signal. Thus, the linear interaction specification in equation~\eqref{eq:gapregression} has $c_3<0$.
\end{proposition}

\section{Methodology and Data}
\subsection{Anonymization Algorithms}\label{subsec:anonymization_algorithms}
We employ two approaches for text anonymization. The first uses a specialized named-entity-recognition (NER) model from the Python spaCy package, \texttt{en\_core\_web\_trf} (TRF), which is based on a Transformer architecture. The second approach follows \citet{engelberg2025entity} and uses a commercial LLM, specifically GPT-4o-mini, in conjunction with custom prompts to perform anonymization.

\begin{figure}[htbp]
  \centering
  \includegraphics[width=1\textwidth]{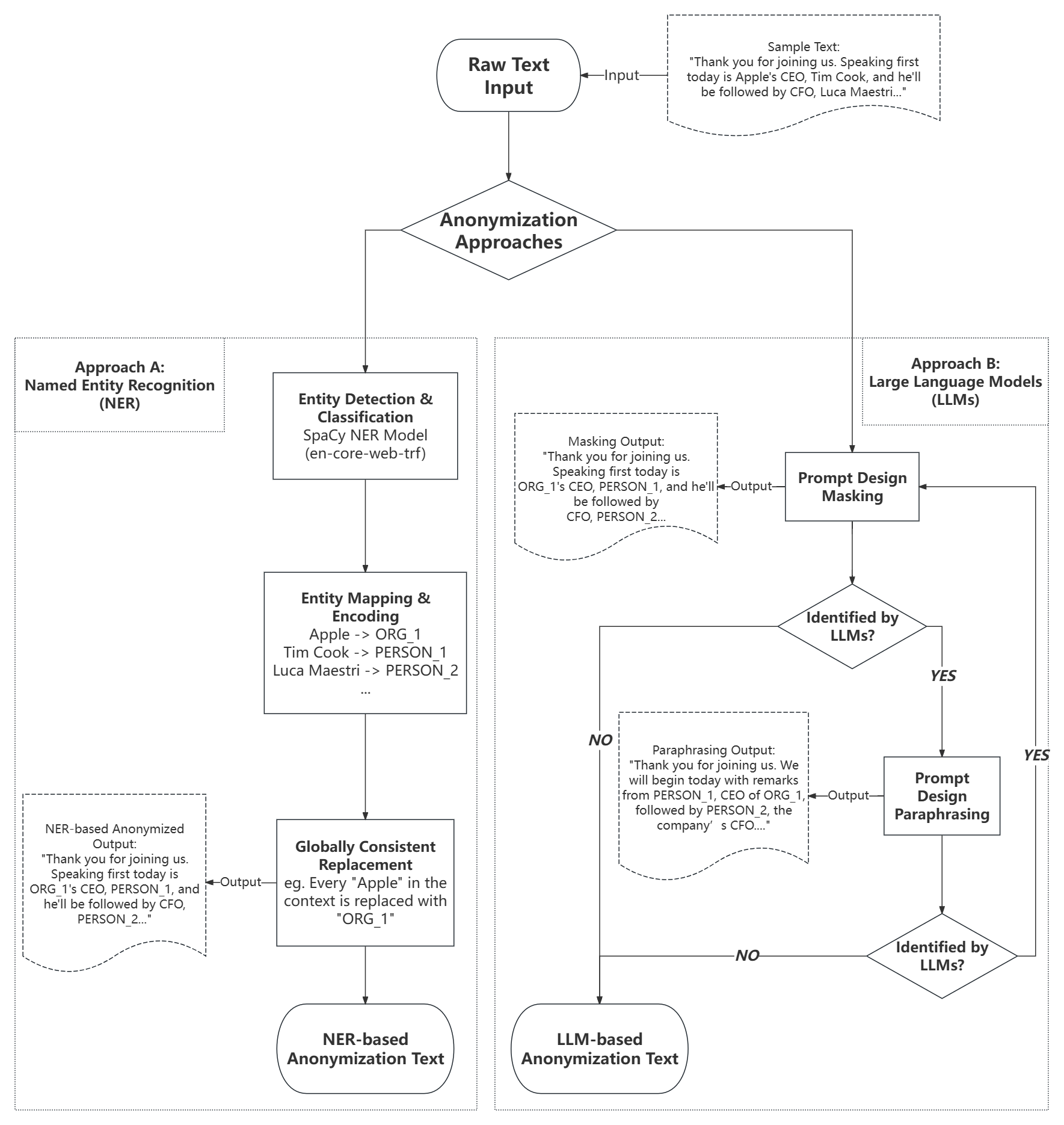}
  \caption{{\bf Anonymization Pipeline}\\\\
  \footnotesize{This figure shows two text anonymization approaches applied to a sample from an Apple Inc. earnings call transcript. Approach A shows a Named Entity Recognition (NER) approach using the spaCy TRF model. This method employs a global mapping strategy, under which each unique entity is consistently replaced by a numbered placeholder (e.g., PERSON\_1, ORG\_2) according to its order of appearance. Approach B uses GPT-4o-mini in the fixed sequence masking $\rightarrow$ re-identification test $\rightarrow$ conditional paraphrasing $\rightarrow$ re-identification test. A masked output that passes the first test is accepted without paraphrasing; an output that fails proceeds to paraphrasing and is tested again. Failure of the second test starts another round, for at most three rounds.}}
  \label{fig:anonymization_flowchart}
\end{figure}

These two models differ substantially in size. The \texttt{en\_core\_web\_trf} model contains approximately 0.1 billion parameters, reflecting its RoBERTa-base backbone.\footnote{\url{https://huggingface.co/roberta-base}} By contrast, GPT-4o-mini is larger by roughly two orders of magnitude; although its exact parameter count is undisclosed, industry estimates place it on the order of 8 billion parameters.\footnote{\url{https://techcrunch.com/2024/07/18/openai-unveils-gpt-4o-mini-a-small-ai-model-powering-chatgpt}}
Approach A in Figure~\ref{fig:anonymization_flowchart} illustrates the workflow for the spaCy NER model (\texttt{en\_core\_web\_trf}). The model first scans the raw text to recognize specific entities, such as ``Apple,'' ``Tim Cook,'' and ``Luca Maestri'' in the example shown. It then generates a consistent global map based on the entity type and order of appearance, for instance mapping ``Apple'' to ORG\_1 as the first organization identified, and ``Tim Cook'' to PERSON\_1 as the first person.\footnote{spaCy recognizes 18 entity types. As reported in Appendix Table~\ref{appendix-tab:Entity Category}, we group them into four main categories to structure our anonymization: NUMBERS (which includes DATE, CARDINAL, MONEY, PERCENT, ORDINAL, TIME, and QUANTITY), PLACES (GPE, LOC, FAC), AGENTS (PERSON, ORG, NORP), and OTHERS (PRODUCT, EVENT, LAW, WORK\_OF\_ART, and LANGUAGE).} Finally, the system applies these mappings globally to the entire text unit, producing an anonymized output that preserves textual structure while obscuring identities.

Approach B in Figure~\ref{fig:anonymization_flowchart} describes the strategy using a large language model, specifically GPT-4o-mini. Unlike the single-pass NER approach, this method follows a fixed four-step sequence: \emph{masking, re-identification testing, paraphrasing, and re-identification testing again}. First, the model masks entities by replacing them with placeholders. Second, an independent LLM attempts to re-identify the firm and date from the masked output.\footnote{We consider the text to be ``identified'' if the LLM's guessed firm name has a \texttt{fuzz.token\_sort\_ratio} score of 75 or higher when matched against the actual firm name, or if the estimated conference call date falls within 7 days of the actual date, following \citet{engelberg2025entity}.} If neither item is identified under these criteria, the masked output passes and becomes the final text. If either item is identified, the masked text proceeds to the third step, paraphrasing, in which sentence structures and recognizable expressions are rewritten to remove stylistic identifiers (e.g., changing ``Speaking first today...'' to ``We will begin today with remarks from...''). Fourth, the same re-identification test is applied to the paraphrased output. If the paraphrased text passes, it becomes the final output; if it fails, a new round begins with more aggressive masking instructions. The masking--test--paraphrasing--test cycle is repeated for at most three rounds, after which the last output is retained if the round limit is reached. Our prompt design follows \citet{engelberg2025entity}, and Appendix~\ref{subsec:anonymization_prompt} reports the prompts in this operational order.

Table~\ref{tab:anonymization_comparison} presents a text excerpt from Apple Inc.'s 2024 Q1 earnings call transcript, along with the anonymization results from the TRF and LLM approaches. The TRF model exhibits a rigid, category-based replacement strategy; for instance, it anonymizes temporal expressions like ``today'' and ``annual'' into ``DATE\_1'' and ``DATE\_2,'' and correctly identifies ``SEC'' as an organization (``ORG\_2''). In contrast, the LLM approach follows a distinct progression.
After the masking output fails its initial re-identification test, the conditional paraphrasing step offers additional flexibility by altering sentence structure, for example, by changing ``Please note'' to ``Please be aware,'' and by replacing specific identifiers with semantic descriptions, such as converting ``SEC'' into ``the regulatory body.''

\subsection{Data}
\subsubsection{Firm Data}
The sample for this research consists of constituent stocks from the iShares Russell 3000 ETF. Our primary post-cutoff window runs from November 2023 through December 2024; as described below, we extend the sample back to November 2022 for the tests that span the knowledge-cutoff. We obtain return data from CRSP, corporate accounting data from Compustat, credit-downgrade dates from S\&P, and EPS forecast data from the I/B/E/S Detail History database. Detailed variable definitions can be found in Appendix Table~\ref{appendix-tab:variable_definitions_full}.

\subsubsection{Earnings Call Transcripts}
The primary textual source we examine in this study is corporate conference calls, which
have been shown to carry rich value-relevant information about firms
\citep[e.g.,][]{frankel1999empirical,brown2004conference}. Many recent studies
leverage LLMs to extract information from these texts
\citep[e.g.,][]{jha2025chatgptcorporatepolicies,Clayton2025}. To isolate the effect of anonymization unconfounded by look-ahead bias, we use data generated after the
knowledge-cutoff date of the LLMs employed. The models
used in this study---GPT-4o-mini-2024-07-18, GPT-4o-2024-08-06, and
o3-mini-2025-01-31---all share a knowledge-cutoff date of October
2023.\footnote{\url{https://github.com/HaoooWang/llm-knowledge-cutoff-dates}}
Accordingly, we download earnings call transcripts dated from November 2023
onward from Seeking Alpha.

Because the earnings-call-related tasks in this paper rely on different matched data and matching procedures, we employ two distinct sample-filtering schemes tailored to the data requirements of our prediction tasks. The first applies to the credit-downgrade prediction; the full procedure is detailed in Panel A of Appendix Table~\ref{appendix-tab:Sample Selection Criteria}. The second applies to the earnings-call return analysis and the alternative transcript-based prediction tasks in Section~\ref{sec:beyond_credit_risk}, detailed in Panel B of Appendix Table~\ref{appendix-tab:Sample Selection Criteria}.

The two schemes share the same initial steps. Both are strictly restricted to the
November 2023--December 2024 window. We remove duplicate entries and those with
incorrect timestamps, and we exclude transcripts exceeding 15,000 tokens. This
threshold is established because the GPT-4o-mini model used for the anonymization
procedure has a maximum output limit of 16,000 tokens; longer input transcripts
would therefore result in truncated anonymized text.

The schemes diverge thereafter. For the downgrade prediction, following
\citet{donovan2021measuring}, we exclude utilities and financial firms (SIC
4000--4999 and 6000--6299), require the full set of control variables, and
require an S\&P credit rating observed before the earnings call. These sequential
filters leave 2,488 post-cutoff transcripts.\footnote{The sequential sample
selection in Appendix Table~\ref{appendix-tab:Sample Selection Criteria} leaves
2,488 transcripts. The main regressions with SIC3 $\times$ Quarter fixed effects
report 2,243 observations because the fixed-effect estimator drops 245 singleton
observations. The same singleton-removal procedure applies to the
other fixed-effects regressions in the paper, although the number of
observations excluded may vary across specifications.} For the return-prediction and other transcript-based tasks, we
match the transcript data with the relevant returns, analyst EPS forecast errors,
and accounting outcomes using date and PERMNO, and drop observations with
missing controls for the corresponding specification.

To facilitate our analysis of the pre-cutoff period, we extend our dataset to
include earnings call transcripts from November 2022 to October 2023. We apply the
same filtering criteria to this earlier sample for both the downgrade-prediction
and return-prediction tasks. To ensure a consistent comparison across the pre- and
post-cutoff periods and to mitigate potential selection bias, our combined analysis
is restricted to a balanced panel of firms that have valid observations in both the
pre-cutoff (November 2022--October 2023) and post-cutoff (November
2023--December 2024) periods.

\subsubsection{News Headlines}
In addition to conference calls, news articles have also been shown to deliver rich information about firms \citep[e.g.,][]{TETLOCK2008,tetlock2010does}. Moreover, many studies attempt to extract signals from news headlines and articles \citep[e.g.,][]{Bybee2023,lopez2023can,engelberg2025entity}.

To ensure the generalizability of our results, we also collect daily news headlines from the Finnhub API. Consistent with our post-cutoff sample design, our data collection begins on November 1, 2023. Following the methodology of \citet{NBER2019}, we first exclude headlines associated with multiple firm tickers, retaining only those that feature a single firm. This restriction is necessary because when we prompt the GPT model to assess the price implications for ``the corresponding firm,'' headlines mentioning multiple or no specific firms would create ambiguity and confuse the model. To handle duplicate articles that arise from re-syndication across different news outlets, we retain headlines only from the most frequent source in our dataset, which is Yahoo Finance. Furthermore, we remove headlines containing the word ``stock'' or ``stocks,'' as these titles may directly allude to stock price movements or trading volume, thus confounding the signal.

Guided by \citet{Bybee2023}, we also exclude articles published on weekends. Finally, the filtered headlines are matched with other relevant stock and corporate financial data. The complete sample filtering procedure is detailed in Panel C of Appendix Table~\ref{appendix-tab:Sample Selection Criteria}.

\subsection{LLM-Generated Measurements}\label{subsec:llm_generated}
We design a suite of prompts to extract several measures from the aforementioned textual data, including the probability of a credit-rating downgrade, uncertainty, investment (the firm's outlook for future capital expenditures), economy (the firm's outlook for the future macroeconomy), and sentiment. For each measure, we generate a quantitative score. The specific definitions of these scores and the complete prompts are reported in Appendix~\ref{app:prompts}, Sections~\ref{subsec:downgrade_prompt}--\ref{subsec:sentiment_news_prompt}.

For the credit-risk mechanism analysis, we additionally conduct a structured score-and-quote extraction based on ten factors of S\&P Global Ratings' Corporate Methodology using GPT-4o-mini-2024-07-18, and
Appendix~\ref{subsec:sp10_prompt} gives the complete prompt.

\subsection{Anonymization Output Characteristics and Re-identification}
We first provide descriptive statistics for the earnings-call transcripts. We tokenize the raw transcripts using the Xenova/GPT-4o tokenizer,\footnote{\url{https://huggingface.co/Xenova/gpt-4o}} which yields the length of each transcript as well as the length of the entities identified by the TRF method; dividing the latter by the former gives the entity share recognized by TRF. For the LLM-based method, because it generates plain text rather than structurally tagged tokens, we compute the entity share as the proportion of anonymized placeholders in the anonymized output rather than in the raw text. This construction is intended to approximate the TRF measure, though the two are not strictly comparable: placeholders need not occupy the same number of tokens as the text they replace, and paraphrasing alters the denominator. Panel A of Table~\ref{tab:summary_stats_anonymization} reports transcript characteristics and the share of tokens replaced by each method. The average transcript length is 9,539 tokens, with the 25th and 75th percentiles at 7,977 and 11,237 tokens, respectively. The TRF method identifies \(18.92\%\) of the text as entities, while the LLM-based method identifies \(11.51\%\).

We next examine the effectiveness of the two anonymization algorithms. Following prior literature \citep[e.g.,][]{engelberg2025entity}, we ask GPT-4o-mini to identify the masked date, firm name, ticker, and industry associated with each anonymized transcript. The re-identification prompt is detailed in Appendix~\ref{subsec:anonymization_test_prompt}, and Panel B reports the resulting identification rates. We find that the TRF algorithm exhibits weaker anonymization performance: GPT-4o-mini successfully identifies \(66.88\%\) of firm names and \(53.90\%\) of tickers, although date identification remains low at \(6.03\%\). In contrast, the LLM-based approach is substantially more thorough. Consistent with \citet{engelberg2025entity}, it drastically reduces re-identification rates, leaving only \(9.04\%\) of firm names and \(2.81\%\) of dates identifiable. This comparison shows that the iterative LLM approach obscures firm-specific information more effectively than standard NER-based substitution. The two methods therefore bracket a wide range of anonymization strength, which is useful for the tests that follow: an information loss common to both cannot be attributed to one method masking too little.

\section{Testing the Framework: Credit-Rating Downgrades}\label{sec:main_results}
Using credit-rating downgrades as our primary outcome, this section evaluates the four propositions developed in Section~\ref{sec:framework}. We first document post-cutoff predictive performance using ROC curves and then test standalone attenuation and raw-score encompassing in regressions (Propositions~\ref{prop:attenuation} and~\ref{prop:horse_race}). We next examine the knowledge-cutoff interaction in Proposition~\ref{prop:cutoff} and the score--gap interaction in Proposition~\ref{prop:gap}.

\subsection{Post-Cutoff Predictive Performance}
\begin{figure}[htbp]
  \centering
  \includegraphics[width=0.7\textwidth]{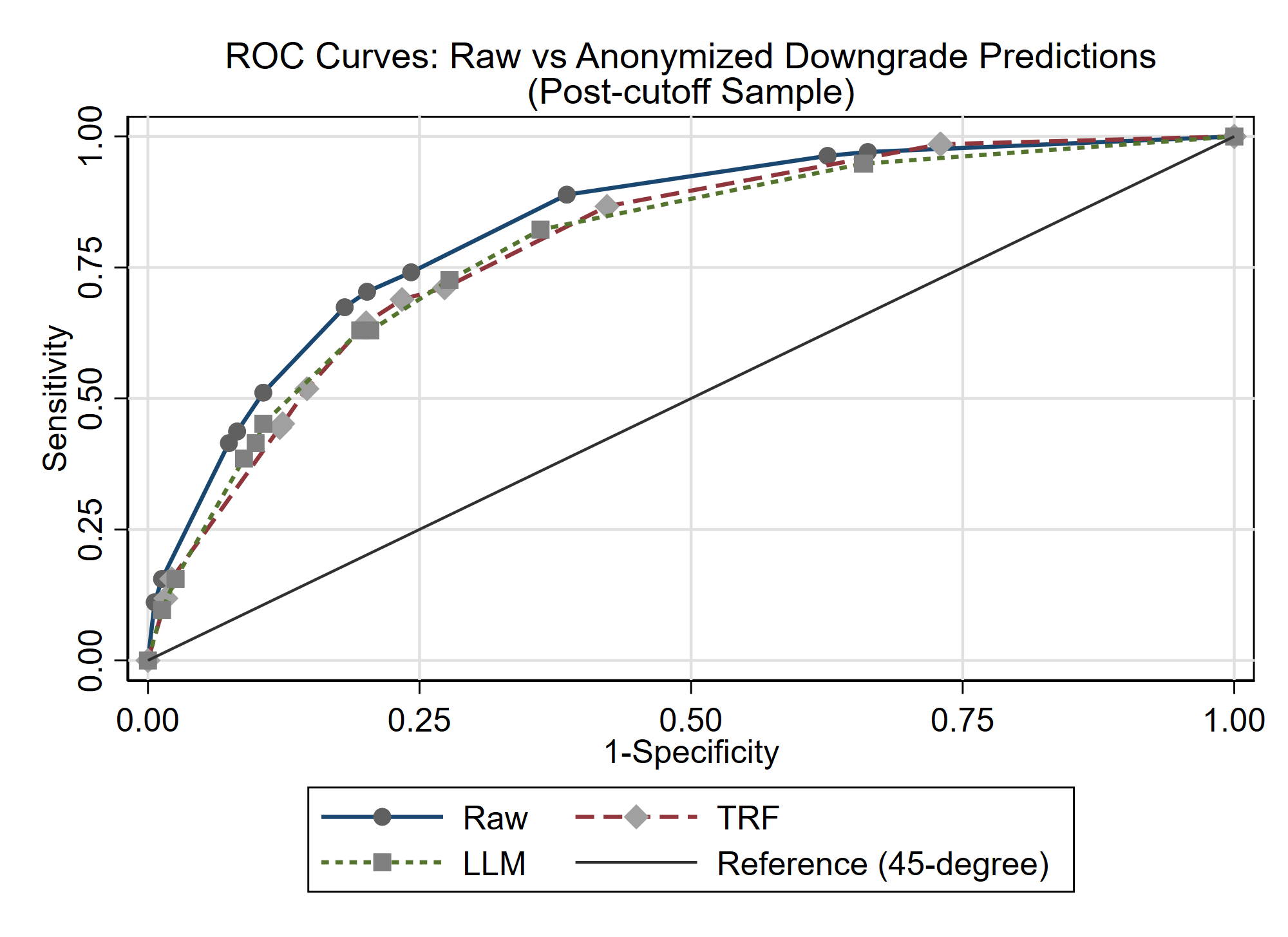}
  \caption{{\bf ROC Curves of Downgrade Predictions Across Anonymization Methods}\\
  \footnotesize{This figure plots the receiver operating characteristic (ROC) curves for predicting the realized 12-month S\&P credit rating downgrade using the GPT-4o-mini-extracted downgrade probability from RAW, TRF-anonymized, and LLM-anonymized transcripts. The sample contains 2,488 post-cutoff transcripts from November 2023 through December 2024.}}
  \label{fig:roc_all_scores}
\end{figure}

Following \citet{donovan2021measuring}, who show that machine-learning methods can extract qualitative credit-risk signals from conference calls, we use GPT-4o-mini to construct a downgrade-probability score from the raw transcript and from its TRF- and LLM-anonymized versions; the extraction prompt is detailed in Appendix~\ref{subsec:downgrade_prompt}. Before turning to formal regressions, we present the first outcome-based evidence consistent with information loss in Figure~\ref{fig:roc_all_scores}, which plots the receiver operating characteristic (ROC) curves for predicting the realized 12-month downgrade using the RAW, TRF-, and LLM-anonymized scores. All three curves lie well above the 45-degree reference line, indicating that each score carries predictive content for downgrades. The ordering among them, however, is unambiguous: the RAW curve lies above both anonymized
curves across essentially the entire range of the false-positive axis and correspondingly has a larger estimated area under the curve (AUC). The TRF and LLM curves, by contrast, are nearly indistinguishable from
each other and lie consistently below RAW.

Crucially, the transcripts underlying this figure are drawn entirely from the
post-cutoff window (November 2023--December 2024), which postdates the October 2023
knowledge-cutoff of the extraction model. Under the maintained knowledge-cutoff design, the ordering is therefore unlikely to reflect the model having learned the subsequently realized downgrade outcomes during training.

We now quantify this attenuation by regressing the realized downgrade indicator $Downgrade_{i,t+1}$---equal to one if S\&P downgrades the firm's credit rating within the twelve months following the earnings call, and zero otherwise---on the extracted downgrade-probability scores:

\begin{equation}
Downgrade_{i,t+1} = \beta\,\widehat{Downgrade}_{i,t} + \gamma X_{i,t} + \delta_t + \epsilon_{i,t},
\label{eq:downgrade_main}
\end{equation}
where $\widehat{Downgrade}_{i,t}$ is the score extracted from the raw, TRF-, or LLM-anonymized text, and $\delta_t$ denotes SIC3 $\times$ Quarter fixed effects. The control vector $X_{i,t}$ contains $\ln(FirmValue)$, $Leverage_{Asset}$, $MTB$, $ROA$, $RevGrowth$, $Std(Income)$, $Std(Revenue)$, $Zscore$, and $Oscore$.

All independent variables are standardized to facilitate comparison across text treatments. We report the results in Table~\ref{tab:downgrade_main}, without firm-level controls in Panel A and with the full set of controls in Panel B.

The standalone specifications in Columns 1--3 of Panel B display the directional ordering predicted by Proposition~\ref{prop:attenuation}. The raw score is a strong and highly significant predictor of downgrades (coefficient $0.061$, $t=5.52$), and both the TRF- and LLM-anonymized scores are attenuated: their coefficients remain significant ($0.047$, $t=4.80$; and $0.049$, $t=4.92$) but are smaller, with lower adjusted $R^2$. Because the scores are standardized and the outcome is an indicator, these coefficients are directly interpretable: a one-standard-deviation increase in the raw score is associated with a 6.1-percentage-point higher probability of a downgrade, compared with 4.7 and 4.9 percentage points for the TRF- and LLM-anonymized scores. Under Assumption~\ref{assump:information_loss}, this ordering means the anonymized coefficients are conservative: the full-context association is larger still.

The horse-race specifications in Columns 4 and 5 are consistent with the zero-incremental-information benchmark. When the raw score and an anonymized score enter the same regression, the coefficient on $\widehat{Downgrade}_{RAW}$ remains large and significant ($0.073$, $t=3.58$ in Column 4), while the coefficient on $\widehat{Downgrade}_{TRF}$ falls to $-0.012$ ($t=-0.72$), statistically indistinguishable from zero; the same pattern holds for the LLM-anonymized score ($-0.007$, $t=-0.52$ in Column 5). Thus, the anonymized scores remain predictive on their own but add no statistically detectable incremental content once the raw score is included, which is the joint pattern the framework implies. As shown in Panel A, the results are robust to excluding control variables.

\subsection{Knowledge-Cutoff Diagnostic for Look-Ahead Bias}
We next evaluate Proposition~\ref{prop:cutoff}'s diagnostic for look-ahead bias. An LLM trained on data through its knowledge-cutoff may absorb a firm's subsequently realized downgrade, in which case the observed score should have a stronger predictive slope in the pre-cutoff period under the proposition's cross-period stability condition. We test this implication by extending the downgrade sample to include a pre-knowledge-cutoff window from November 2022 through October 2023, retaining only firms that appear in both windows.

\begin{figure}[htbp]
  \centering
  \begin{subfigure}[b]{0.6\textwidth}
    \centering
    \includegraphics[width=\textwidth]{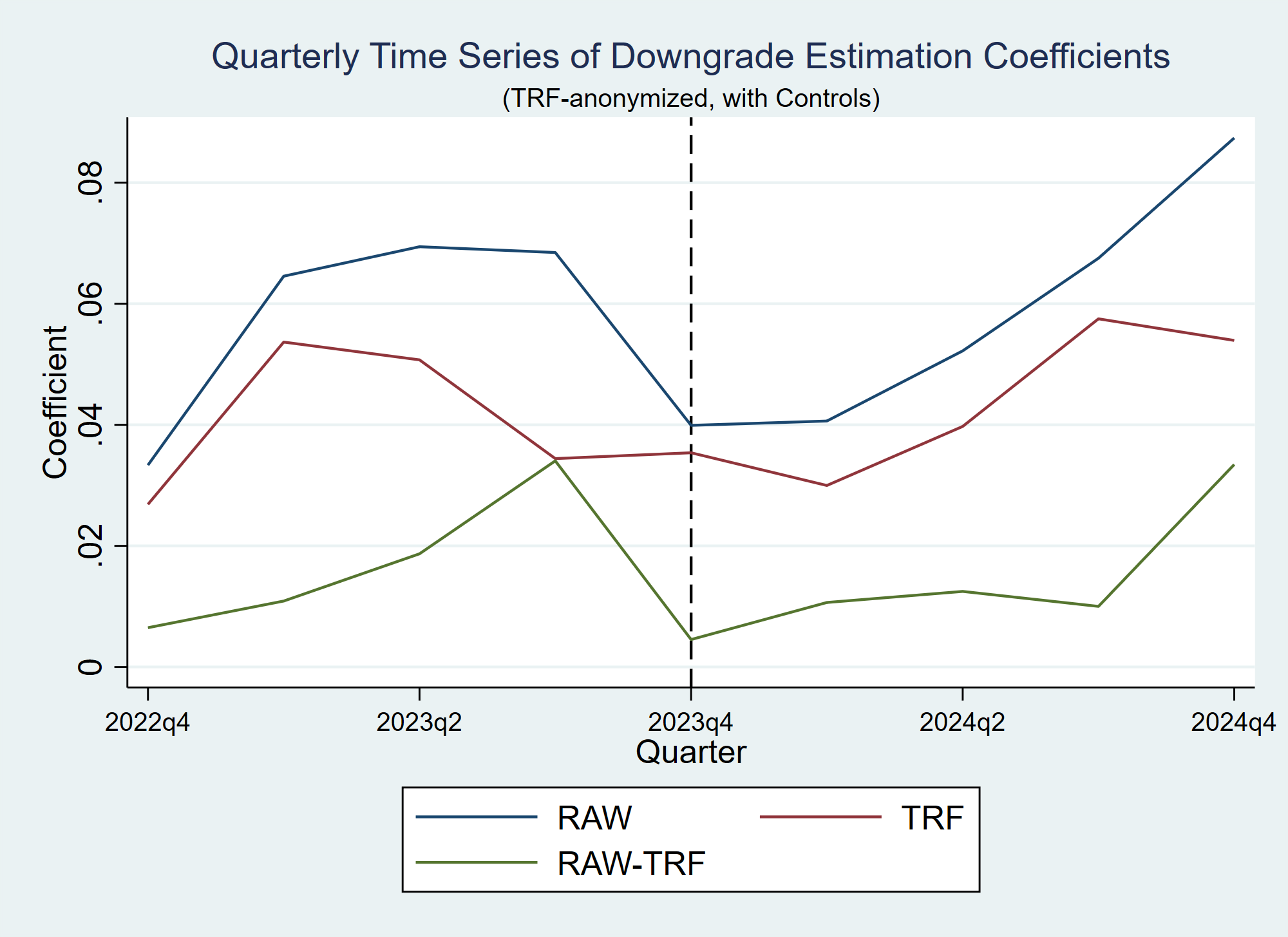}
    \caption{\centering Quarterly Downgrade Predictive Coefficients \\ (RAW and TRF-anonymized)}
    \label{fig:downgrade_qbeta_trf}
  \end{subfigure}
  \vspace{1em}
  \begin{subfigure}[b]{0.6\textwidth}
    \centering
    \includegraphics[width=\textwidth]{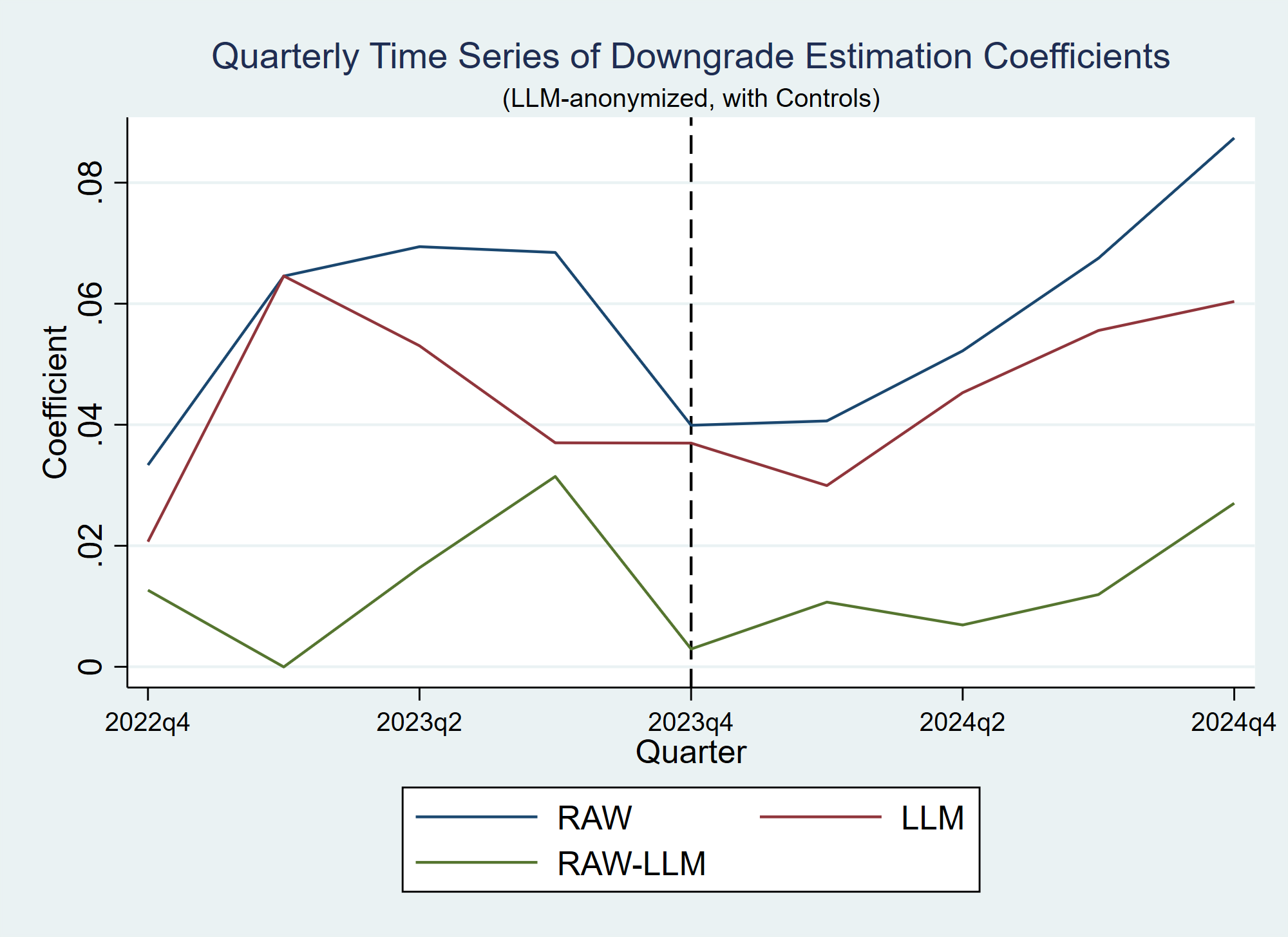}
    \caption{\centering Quarterly Downgrade Predictive Coefficients \\ (RAW and LLM-anonymized)}
    \label{fig:downgrade_qbeta_llm}
  \end{subfigure}
  \caption{{\bf Quarterly Time Series of Downgrade Predictive Coefficients}\\\\
  \footnotesize{Each subfigure plots three quarterly time series on the balanced panel of firms appearing in both the pre- and post-cutoff samples: the quarterly regression coefficient of $Downgrade_{i,t+1}$ on $\widehat{Downgrade}_{RAW,i,t}$ (dark blue), on the anonymized counterpart (red), and the difference between the two coefficients. Each quarterly regression includes the same firm-level controls as Panel B of Table~\ref{tab:downgrade_main} and 3-digit SIC industry fixed effects, with standard errors clustered at the 3-digit SIC level. The vertical dashed line marks the LLM knowledge-cutoff (2023 Q4). Subfigure (a) compares the coefficients derived from the RAW transcripts against those from the TRF-anonymized samples. Subfigure (b) compares the coefficients from the RAW transcripts against those from the LLM-anonymized samples.}}
  \label{fig:downgrade_quarterly_beta}
\end{figure}

We first present the evidence visually. Figure~\ref{fig:downgrade_quarterly_beta} plots quarter-by-quarter coefficients on the downgrade score. The coefficient on the raw score consistently exceeds those on the anonymized scores on both sides of the cutoff, marked by the dashed vertical line, with no visible increase as the sample moves into the pre-cutoff period. We then estimate

\begin{equation}
Downgrade_{i,t+1}
=
\beta_1 \widehat{Downgrade}_{i,t}
+
\beta_2
\bigl(
\widehat{Downgrade}_{i,t}\times Z_{i,t}
\bigr)
+
\beta_3 Z_{i,t}
+
\gamma X_{i,t}
+
\delta_t
+
\epsilon_{i,t},
\label{eq:downgrade_interaction}
\end{equation}
where $Z_{i,t}$ is a moderating variable. Setting $Z_{i,t}=Pre$, an indicator equal to one for transcripts in the pre-cutoff window, makes $\beta_2$ the empirical counterpart of $d_3$ in Proposition~\ref{prop:cutoff}. Table~\ref{tab:downgrade_pre_interaction} therefore evaluates the positive-interaction diagnostic under its cross-period stability conditions. The downgrade score is positive and significant in all three specifications: the coefficients are $0.061$ ($t=5.35$), $0.047$ ($t=4.55$), and $0.050$ ($t=4.82$) for the RAW, TRF, and LLM scores, respectively. By contrast, the corresponding interactions with $Pre$ are small and statistically insignificant: $-0.005$ ($t=-0.31$), $-0.009$ ($t=-0.59$), and $-0.010$ ($t=-0.63$). None has the positive sign the look-ahead bias model predicts.

To further tighten the diagnostic, we also shorten the prediction horizon from twelve months to three months. We redefine the dependent variable to indicate whether S\&P downgrades the firm within three months after the earnings call and re-extract the downgrade-probability scores using an otherwise identical prompt that replaces ``twelve months'' with ``three months.'' The shorter horizon sharply limits the overlap between the pre-cutoff prediction windows and post-cutoff dates, making realized pre-cutoff downgrades substantially more likely to have been available in the model's training data.

Appendix Table~\ref{appendix-tab:downgrade_pre_interaction_3m} repeats the diagnostic with the shorter horizon. The three-month downgrade scores remain positive and significant in the post-cutoff period, with coefficients of $0.026$ ($t=4.52$), $0.023$ ($t=3.86$), and $0.020$ ($t=3.71$) for the RAW, TRF, and LLM scores, respectively. More importantly, none of the interactions with $Pre$ is positive or statistically significant. The interaction coefficients are $-0.006$ ($t=-0.68$) for RAW, $-0.010$ ($t=-1.30$) for TRF, and $-0.002$ ($t=-0.30$) for LLM. Across both horizons, the interaction is never positive or statistically significant, contrary to what the diagnostic predicts when look-ahead bias inflates pre-cutoff scores.

Together with the raw score's post-cutoff advantage, these estimates provide little evidence that look-ahead bias is a first-order explanation for the observed attenuation and are consistent with anonymization removing or altering contemporaneously available credit-risk information.

\subsection{The Anonymization Gap and Conditional Attenuation}
We next implement the gap in equation~\eqref{eq:gap}. Let $R$ and $A$ denote the winsorized and standardized raw and anonymized downgrade scores. For each anonymization method, we estimate $\rho$ over the pooled balanced sample and construct $\Gap=\lvert A/\rho-R\rvert$; we then winsorize and standardize $\Gap$ and interact it with the standardized anonymized score. The estimated values of $\rho$ are 0.8993 for TRF and 0.8997 for LLM. Setting $Z_{i,t}=\Gap$ in equation~\eqref{eq:downgrade_interaction} makes the score--gap interaction the empirical analogue of $c_3$ in Proposition~\ref{prop:gap}. Table~\ref{tab:downgrade_gap_pre_post} evaluates that proposition's negative-interaction prediction in the full sample and in the pre- and post-cutoff subsamples.

The interaction is negative throughout, as Proposition~\ref{prop:gap} predicts. In the full sample, the coefficients are $-0.019$ ($t=-3.62$) for TRF and $-0.012$ ($t=-2.37$) for LLM. Because both the score and $\Gap$ are standardized, a one-standard-deviation increase in $\Gap$ reduces the marginal predictive effect of the anonymized score, relative to its main-effect coefficient at the mean gap, by approximately 37\% for TRF ($0.019/0.051$) and 24\% for LLM ($0.012/0.049$). The pre-cutoff interactions are $-0.015$ ($t=-1.88$) and $-0.013$ ($t=-1.46$), while the post-cutoff interactions are $-0.023$ ($t=-3.02$) and $-0.013$ ($t=-2.19$), respectively. Thus, the anonymized score predicts downgrades least effectively where its attenuation-adjusted deviation from the raw score is greatest. The statistically significant post-cutoff estimates provide the cleanest support for that directional prediction: because those documents' outcomes postdate the maintained cutoff, removal of memorized outcomes is unlikely to explain the negative interaction. This pattern supports using the gap as a proxy for the document-specific residual.

\section{Mechanisms of Information Loss}\label{sec:mechanisms}

Having evaluated the four proposition-level predictions in the main downgrade analysis, we next examine how anonymization produces information loss. We proceed in two stages. First, a TRF-based decomposition examines which of four entity categories---NUMBERS, PLACES, AGENTS, and OTHERS---are most important for retaining predictive content. Second, a quotation-backed audit organized around ten factors from S\&P Global Ratings' Corporate Methodology traces whether anonymization changes the evidence selected by the model or changes how preserved evidence is interpreted.

\subsection{A TRF-Based Entity Decomposition}\label{subsec:entity_decomposition}
To localize the textual sources of information loss, we re-extract the downgrade score from transcripts in which the TRF model preserves only one entity category---NUMBERS, PLACES, AGENTS, or OTHERS---while anonymizing the other three. As shown in Appendix Table~\ref{appendix-tab:Entity Category}, NUMBERS and AGENTS are considerably more frequent in our corpus than PLACES and OTHERS.

Table~\ref{tab:downgrade_correlation} provides an initial comparison. Panel A shows that the raw score correlates 0.907 with both fully anonymized scores, while Panel B shows that the correlation with the raw score is 0.946 when only NUMBERS are preserved and 0.932 when only AGENTS are preserved, compared with 0.916 for PLACES and 0.914 for OTHERS. These high but imperfect correlations show that anonymization changes the model-extracted assessment and that the four entity categories retain different shares of the raw signal.

Table~\ref{tab:downgrade_keep_one} then compares the predictive content of the single-category-preserved variants. In the standalone specifications, the coefficients are 0.058 for both the NUMBERS- and AGENTS-preserved scores, compared with 0.049 for PLACES and 0.051 for OTHERS; their adjusted $R^2$ values show the same ordering. In percentage-point terms, preserving NUMBERS or AGENTS recovers a 5.8-point association per standard deviation, against 6.1 points for the raw score and 4.7 points for the fully TRF-anonymized score. No single category, however, recovers the full-context signal. In Columns 6--9, the raw score remains statistically significant when entered alongside every single-category-preserved variant, whereas none of the four variants has statistically detectable incremental predictive content once the raw score is included. Taken together, the correlation, standalone, and horse-race evidence shows that preserving NUMBERS or AGENTS retains more of the raw text's predictive content than preserving PLACES or OTHERS, but no individual category fully reproduces the signal extracted from the original transcript.

\subsection{A Ten-Factor Credit-Risk Audit}\label{sec:sp10_audit}

The entity decomposition shows which context matters, but not what the model does differently once that context is gone, leaving information loss a label rather than an observable process. We therefore audit each transcript against the ten factors of S\&P Global Ratings' Corporate Methodology; because these are the criteria the agency itself applies, they span the credit-relevant content a transcript can carry. For each factor we require an exact supporting quote alongside the score. The quote ties an otherwise opaque number to a specific passage, letting us track evidence across versions and separate changes in which evidence is used from changes in how it is read.

\subsubsection{Audit Design and the Two Channels}\label{subsec:sp10_design}

The ten factors occupy four positions in S\&P's criteria sequence. Industry Risk, Country
Risk, and Competitive Position describe the business risk profile;
Cash Flow/Leverage describes the financial risk profile;
Diversification/Portfolio Effect, Capital Structure, Financial Policy,
Liquidity, and Management and Governance operate as post-anchor modifiers; and
Comparable Ratings Analysis provides a final relative-positioning assessment.
Appendix~\ref{appendixC1} reports the S\&P credit rating framework and
definitions of the ten factors.

We use the same GPT-4o-mini-2024-07-18 snapshot as in the preceding analysis.
For each factor, we design a prompt asking the model to assess that factor's
implication for future downgrade risk, and apply the same prompt and output
schema separately to the raw earnings-call transcript and its TRF- and
LLM-anonymized versions. For each factor, the model returns a score in
$[0,1]$ and one exact supporting quote; a higher score represents a higher
probability of being downgraded in the future based on that factor, and the
quote is the passage the model identifies as best supporting the score.
If the transcript contains no qualifying
issuer-specific evidence for a factor, both fields are null. The full prompt
appears in Appendix~\ref{subsec:sp10_prompt}. The audit covers the
2,488 transcripts with valid extractions. As in the preceding analysis, we use
only the post-cutoff sample, so that the comparisons reported below are
unlikely to reflect outcomes the model absorbed during training.

We next compare both the quoted passages and their associated scores. We remove
entities, placeholders, and numerical tokens that would mechanically prevent a
match, compute a similarity score between the raw and anonymized quotes, and
divide the extracted evidence into three intuitive groups: \emph{shared
reasons}, for which both versions cite the same passage; \emph{RAW-only
reasons}, which disappear or switch passages after anonymization; and
\emph{anonymized-only reasons}, which replace them.\footnote{We clean each
quote by removing named entities, placeholders, and numerical tokens, then
compute a fuzzy token-sort ratio; two quotes are treated as the same passage
when the ratio is at least 75. Appendix~\ref{appendixC2} reports the full
procedure.}

\begin{figure}[htbp]
  \centering
  \includegraphics[width=0.7\textwidth]{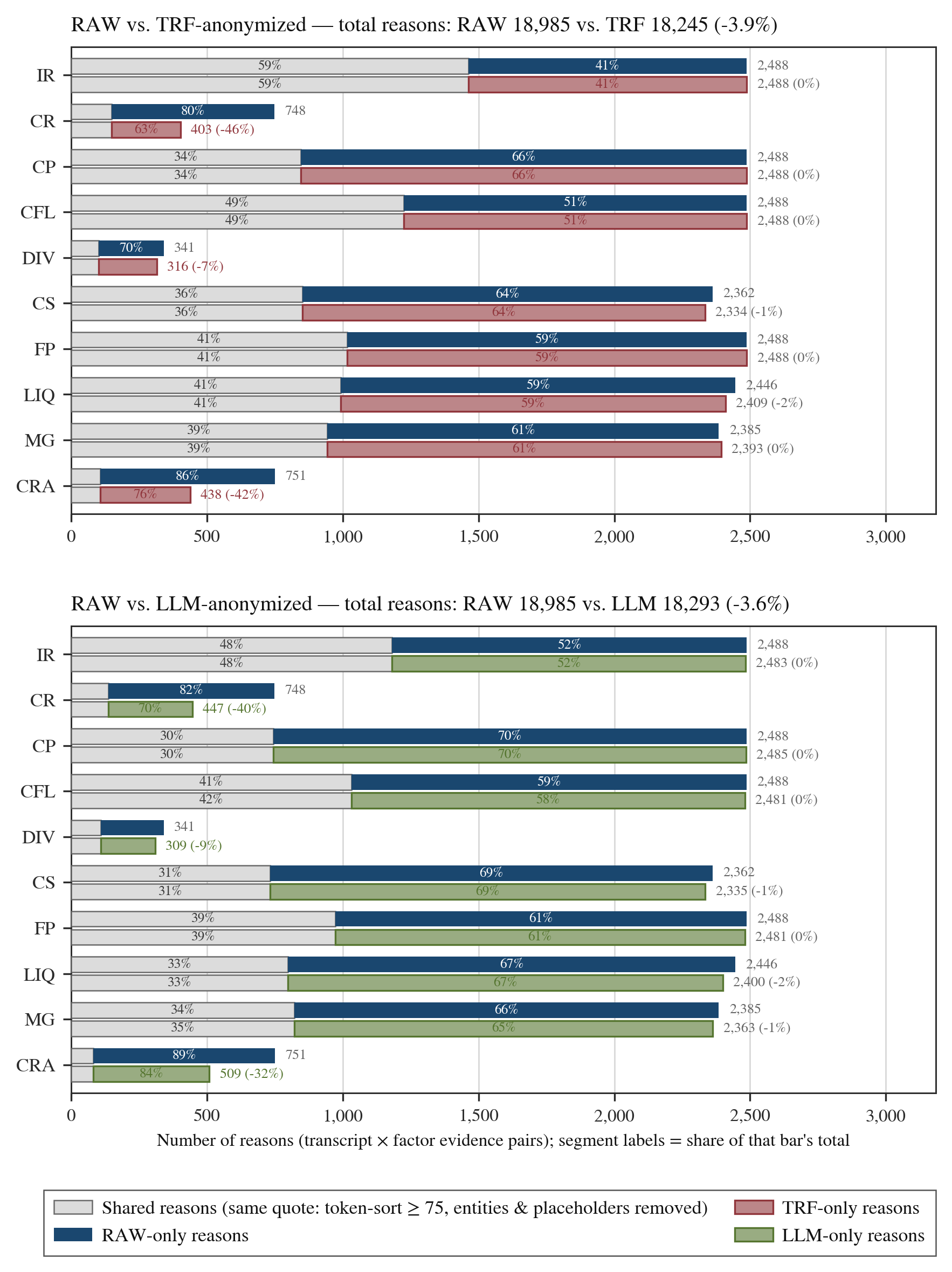}
  \caption{{\bf S\&P Ten-Factor Reason Coverage and Turnover After Anonymization}\\
  \footnotesize{This figure compares transcript-grounded credit-risk reasons extracted from the raw and anonymized versions of 2,488 post-cutoff earnings call transcripts. A reason is an evidence-backed transcript--factor assessment for which the model returns both a non-null score and a supporting quote under the ten-factor taxonomy described in Appendix~\ref{app:sp10_method}.
  For each anonymization method, the upper bar for a factor reports the raw-text reasons and the lower bar reports the anonymized-text reasons. Evidence is classified as matched when the raw and anonymized quotes have a token-sort similarity of at least 75 after spaCy-recognized entities are removed from the raw quote and placeholders and tokens containing numbers are removed from the anonymized quote. The remaining segments constitute evidence displacement: the raw segment combines evidence that disappears after anonymization and the raw side of a pair that cites a nonmatched anchor, while the anonymized segment is defined symmetrically. Segment labels report shares within each bar; numbers at the right report factor totals and the anonymized percentage change.
  IR, CR, CP, CFL, DIV, CS, FP, LIQ, MG, and CRA denote Industry Risk, Country Risk, Competitive Position, Cash Flow/Leverage, Diversification/Portfolio Effect, Capital Structure, Financial Policy, Liquidity, Management and Governance, and Comparable Ratings Analysis, respectively.
  }}
  \label{fig:sp_reason_flow}
\end{figure}

Figure~\ref{fig:sp_reason_flow} reports evidence coverage by factor. The total
number of reasons falls from 18,985 in the raw extractions to 18,245 under
TRF and 18,293 under LLM anonymization, declines of 3.9\% and 3.6\%. This
aggregate decline is the first sign that anonymization narrows evidence
coverage. Some dimensions lose evidence outright: Country Risk reasons fall by
46\% under TRF and 40\% under LLM anonymization, while Comparable Ratings
Analysis falls by 42\% and 32\%. More importantly, replacement of the cited
evidence is far more pervasive than the aggregate counts suggest. Competitive
Position, for example, is populated almost as often after anonymization as
before, yet only 34\% of raw reasons retain the same TRF quote anchor and only
30\% retain the same LLM anchor. The same broad pattern appears across the
factor bars: many dimensions preserve their reason counts while replacing a
large share of the passages used to support those reasons. Anonymization
therefore changes what the model treats as evidence even when it does not
reduce how much evidence the model reports.

These patterns point to two distinct channels, which we define as follows. The first is
\emph{evidence displacement}: anonymization causes a credit-risk reason to
disappear or be replaced by a different supporting passage. The second is
\emph{interpretive distortion}: the model continues to cite the same passage
but changes the risk assessment attached to it after identifying information
and quantitative context are removed.

\begin{figure}[htbp]
  \centering
  \includegraphics[width=0.88\textwidth]{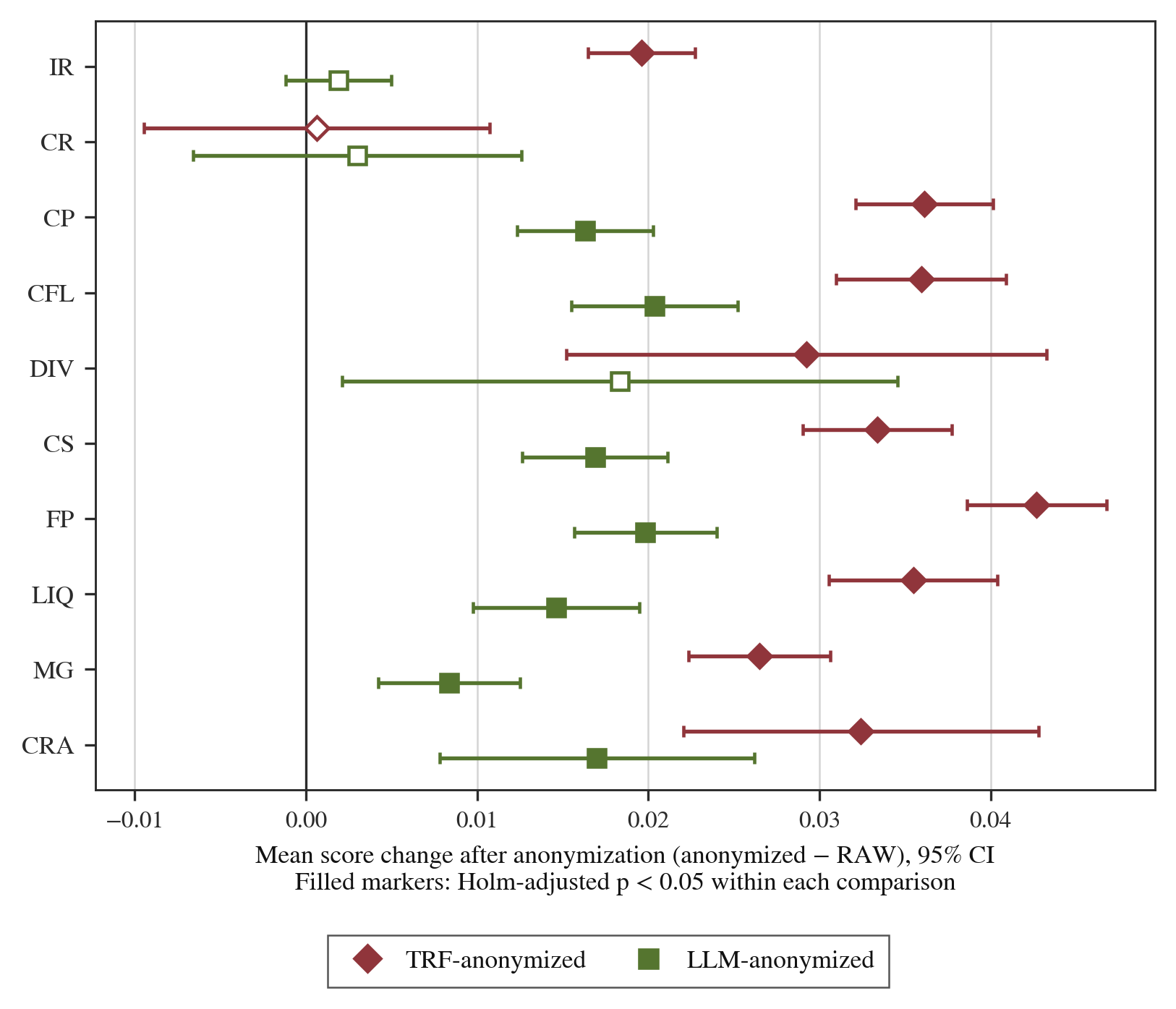}
  \caption{{\bf Conditional Drift in S\&P-Aligned Factor Scores After Anonymization}\\
  \footnotesize{This figure plots paired mean changes in the researcher-constructed factor-level downgrade-risk score after anonymization, computed as the anonymized score minus the raw score. The sample for each estimate is restricted to transcript--factor pairs for which both text versions return a non-null score and supporting quote. Scores range from zero to one, with larger values indicating a more adverse transcript-grounded credit-risk assessment within the factor's scope. Horizontal bars show 95\% confidence intervals. Filled markers indicate that the paired mean difference remains significant at the 5\% level after Holm adjustment across the ten factor-level tests within each raw--anonymized comparison. The factors and sample are defined as in Figure~\ref{fig:sp_reason_flow}.}}
  \label{fig:sp_score_drift}
\end{figure}

Besides changing the evidence selected, anonymization also changes how the
selected evidence is read. Figure~\ref{fig:sp_score_drift} compares factor
scores when both versions contain evidence. The changes are predominantly
positive, and nine of the ten TRF differences and seven of the ten LLM differences
remain significant at the 5\% family-wise level after the Holm correction.
Under TRF anonymization, the average Cash Flow/Leverage, Capital Structure,
Financial Policy, and Liquidity scores rise by roughly three to four
percentage points; many LLM-anonymized scores rise by one to two percentage
points. Removing identity thus shifts the model's factor-level assessments
systematically upward; that is, the model generally forms more pessimistic
expectations for anonymized text.

\subsubsection{Predictive Content of the Composite}\label{subsec:sp10_composite}

To test whether these descriptive changes carry predictive consequences, we
first aggregate the factor-level scores into a transcript-level composite.
Let $j$ index transcripts, $k$ index credit-risk factors, and
$v\in\{RAW,TRF,LLM\}$ denote the transcript version. Let
$\mathcal K_j^v$ be the set of factors for which version $v$ returns both a
non-null score and a non-null supporting quote. Let $s_{jk}^v$ denote the score
assigned to factor $k$ in version $v$. The notation $|\mathcal K_j^v|$ denotes
the number of factors with valid score--quote pairs. The composite is the
average score across these factors:
\begin{equation}
 C_j^v=\frac{1}{|\mathcal K_j^v|}
       \sum_{k\in\mathcal K_j^v}s_{jk}^v.
\label{eq:sp10_composite}
\end{equation}
A factor enters the average only when both its score and quote are non-null.
The composite therefore measures the average assessed probability of downgrade
across the factors for which the transcript supplies qualifying evidence.

The first regression asks whether the ten-factor audit reproduces the
information loss in the one-shot downgrade score. Let $D_j$ indicate whether
its firm is downgraded over the next twelve months. $C_j^{RAW}$ and
$C_j^{\mathrm{Anon}}$ represent the average score across ten factors from raw
and anonymized text separately, and \textit{Anon} denotes TRF or LLM. For each
anonymization method, we estimate standalone specifications and the horse-race
\begin{equation}
 D_j
 =\alpha+\beta_R C_j^{RAW}
 +\beta_A C_j^{\mathrm{Anon}}
 +\gamma X_j+\delta_t+\varepsilon_j.
\label{eq:sp10_composite_regression}
\end{equation}
The control vector $X_j$, SIC3 $\times$ Quarter fixed effects
$\delta_t$, standardization, and SIC3-clustered standard errors are identical
to those in equation~\eqref{eq:downgrade_main}. Standalone specifications
include one composite at a time.

Before turning to the regressions in Table~\ref{tab:sp10_composite}, we
validate the composite against the one-shot downgrade score. The ten-factor
composites correlate 0.819, 0.798, and 0.809 with the corresponding one-shot
downgrade scores for raw, TRF-anonymized, and LLM-anonymized transcripts,
respectively. These high correlations indicate that the factor-based
decomposition largely recovers the information captured by the one-shot scores,
suggesting that the ten factors provide a meaningful representation of the
underlying credit-risk dimensions. In the standalone regressions, a
one-standard-deviation increase in the raw composite predicts a
4.5-percentage-point increase in downgrade probability ($t=4.41$), compared
with 3.5 percentage points for both TRF ($t=4.22$) and LLM anonymization
($t=4.10$).

The standalone estimates display the ordering in
Proposition~\ref{prop:attenuation}: both anonymized composites carry smaller
coefficients than the raw composite, and their adjusted $R^2$
values are also lower. The
horse-races then test Proposition~\ref{prop:horse_race}'s
zero-incremental-coefficient benchmark. Against TRF, the raw coefficient is
0.045 and remains significant, while the anonymized coefficient falls to
$-0.000$ ($t=-0.04$). Against LLM, the raw coefficient is 0.046,
while the anonymized coefficient falls to $-0.001$ ($t=-0.09$).
The ten-factor composite therefore reproduces the empirical attenuation and
horse-race patterns found with the one-shot score and is consistent with the
observable implications of Propositions~\ref{prop:attenuation}
and~\ref{prop:horse_race}. This result gives economic content to the
descriptive audit: the disappearance and replacement of credit-risk reasons,
together with the systematic upward drift in factor assessments, are not merely
changes in model output. Collectively, they accompany a loss of predictive
information about future downgrades.

\subsubsection{Isolating the Two Channels}\label{subsec:sp10_channel}

The second regression asks whether the predictive loss operates through evidence
displacement, interpretive distortion, or both. To separate the two channels,
we calculate four variables whose names describe their construction:
\textit{SharedRaw} and \textit{SharedAnon} are the average raw and anonymized
scores on shared reasons; \textit{RawOnly} is the average score on RAW-only
reasons; and \textit{AnonOnly} is the average score on anonymized-only reasons.
Comparing \textit{SharedRaw} with \textit{SharedAnon} holds the cited passage
fixed, so the difference between them measures interpretive distortion. Comparing \textit{RawOnly}
with \textit{AnonOnly} instead measures evidence displacement. The overall
composite is a weighted average of the relevant shared and version-specific
reasons. Appendix~\ref{appendixC2} gives the detailed procedure and the exact
set-based formulas.

The second regression replaces the two composites with the four conditional
means:
\begin{equation}
\begin{aligned}
 D_j={}&\alpha
 +\beta_1 \mathrm{SharedRaw}_j
 +\beta_2 \mathrm{SharedAnon}_j\\
 &+\beta_3 \mathrm{RawOnly}_j
 +\beta_4 \mathrm{AnonOnly}_j
 +\gamma X_j+\delta_t+\varepsilon_j.
\end{aligned}
\label{eq:sp10_reason_regression}
\end{equation}
The shared pair compares how the same evidence is interpreted; the
version-specific pair compares the average predictive content of the evidence
displaced by and introduced through anonymization. The controls, fixed effects,
and error clustering are the same as in
equation~\eqref{eq:sp10_composite_regression}; the four conditional means are
winsorized but retain their original $[0,1]$ units.

Table~\ref{tab:sp10_reason_decomposition} is a channel-level analogue of
Proposition~\ref{prop:horse_race}: within a common reason set, or across the
reasons displaced by anonymization, it asks whether the anonymized assessment
adds predictive information once the corresponding raw assessment is included.
The table first examines whether changes in the assessment assigned to a common
evidence anchor are associated with predictive attenuation. For reasons shared
between the raw and TRF readings, the two conditional means have positive
standalone coefficients of 0.293 and 0.234. When both enter together, the
raw-reading coefficient is 0.287 ($t=3.74$), while the anonymized-reading
coefficient falls to 0.007 ($t=0.12$). The LLM comparison is similar:
standalone coefficients of 0.364 and 0.272 become 0.385 ($t=4.29$) and
$-0.026$ ($t=-0.35$) in the horse-race. The model continues to point to the
same broad passage, but the assessment formed without the original context
contains no statistically detectable positive incremental information. Holding the evidence anchor fixed, then, the anonymized reading is statistically
subsumed by the full-context reading in this sample.

A parallel predictive pattern appears in the evidence-displacement comparison.
In the raw--TRF comparison, the RAW-only and TRF-only means have standalone
coefficients of 0.331 and 0.270. In their horse-race, the RAW-only coefficient
remains 0.287 ($t=2.80$), whereas the TRF-only coefficient falls to 0.056
($t=0.94$). Under LLM anonymization, standalone coefficients of 0.314 and 0.258
become 0.247 ($t=2.08$) and 0.074 ($t=0.90$). The reasons displaced by
anonymization retain statistically detectable incremental predictive content in
these pairwise specifications, whereas the replacement reasons do not. This
comparison is again consistent with the zero-incremental-information benchmark,
now through evidence displacement rather
than reinterpretation of a shared passage.

The full specifications bring both channels together in the last column. For
TRF, \textit{SharedRaw} remains positive and significant at 0.216 ($t=2.68$), while
\textit{RawOnly} remains positive at 0.151 but is estimated less precisely
($t=1.37$); both anonymized measures are insignificant. For LLM, the shared RAW mean remains
positive and significant at 0.352 ($t=4.17$); the RAW-only mean falls to 0.035
($t=0.27$), and both anonymized means are again insignificant. The significance of
\textit{SharedRaw} but not \textit{RawOnly} suggests that the predictive signal
is concentrated in evidence that remains recognizable after anonymization,
whereas RAW-only evidence provides limited incremental information once the
shared component is accounted for. Taken together, the estimates are consistent
with both changes in the evidence set and changes in the reading of preserved
evidence contributing to predictive attenuation.

\subsubsection{Relative Contributions of Each Channel}\label{subsec:sp10_shapley}

The component regressions provide evidence that both channels are associated
with the predictive loss, but their coefficients do not provide an additive
attribution of the total loss. The conditional means are defined on different
reason sets and are highly correlated, and the measured contribution of one
channel may depend on whether it is introduced before or after the other. We
address this path dependence with a Shapley decomposition of the decline in
adjusted $R^2$.

For each anonymization method, we construct four states with descriptive names.
\textit{Raw} is the original composite score based on the unaltered transcript.
\textit{Replace} substitutes anonymized-only reasons for RAW-only reasons while
retaining the raw interpretation of shared reasons; it therefore introduces only
evidence displacement. \textit{Reread} retains the raw evidence set but applies
the anonymized interpretation to shared reasons; it therefore introduces only
interpretive distortion. \textit{Anon} is the fully anonymized composite score,
incorporating both evidence displacement and interpretive distortion. For each
state, we aggregate the underlying factor-level scores into a transcript-level
composite score. We then estimate the same predictive regression separately using
each composite score as the primary explanatory variable and collect the
corresponding $R^2$. Within each anonymization method, all four regressions use
the common sample for which the four composites, the outcome, and the controls
are observed; the TRF and LLM common samples are constructed separately. These
four predictive performances provide the basis for the subsequent Shapley
decomposition of information loss. The detailed procedure is described in
Appendix~\ref{appendixC3}.

Let $R^2_{\mathrm{Raw}}$, $R^2_{\mathrm{Replace}}$,
$R^2_{\mathrm{Reread}}$, and $R^2_{\mathrm{Anon}}$ denote the
adjusted $R^2$ from entering each state separately into the same downgrade
regression. Total information loss is
$L=R^2_{\mathrm{Raw}}-R^2_{\mathrm{Anon}}$. The Shapley
contributions of evidence displacement ($D$) and interpretive distortion ($I$)
are
\[
\begin{aligned}
 \phi_D
 &=\frac{1}{2}\left[
   \left(R^2_{\mathrm{Raw}}-R^2_{\mathrm{Replace}}\right)
  +\left(R^2_{\mathrm{Reread}}-R^2_{\mathrm{Anon}}\right)\right],\\
 \phi_I
 &=\frac{1}{2}\left[
   \left(R^2_{\mathrm{Raw}}-R^2_{\mathrm{Reread}}\right)
  +\left(R^2_{\mathrm{Replace}}-R^2_{\mathrm{Anon}}\right)\right].
\end{aligned}
\]
Each contribution averages the channel's marginal loss across the two possible
orderings, and $\phi_D+\phi_I=L$ by construction.

\begin{figure}[htbp]
  \centering
  \includegraphics[width=0.82\textwidth]{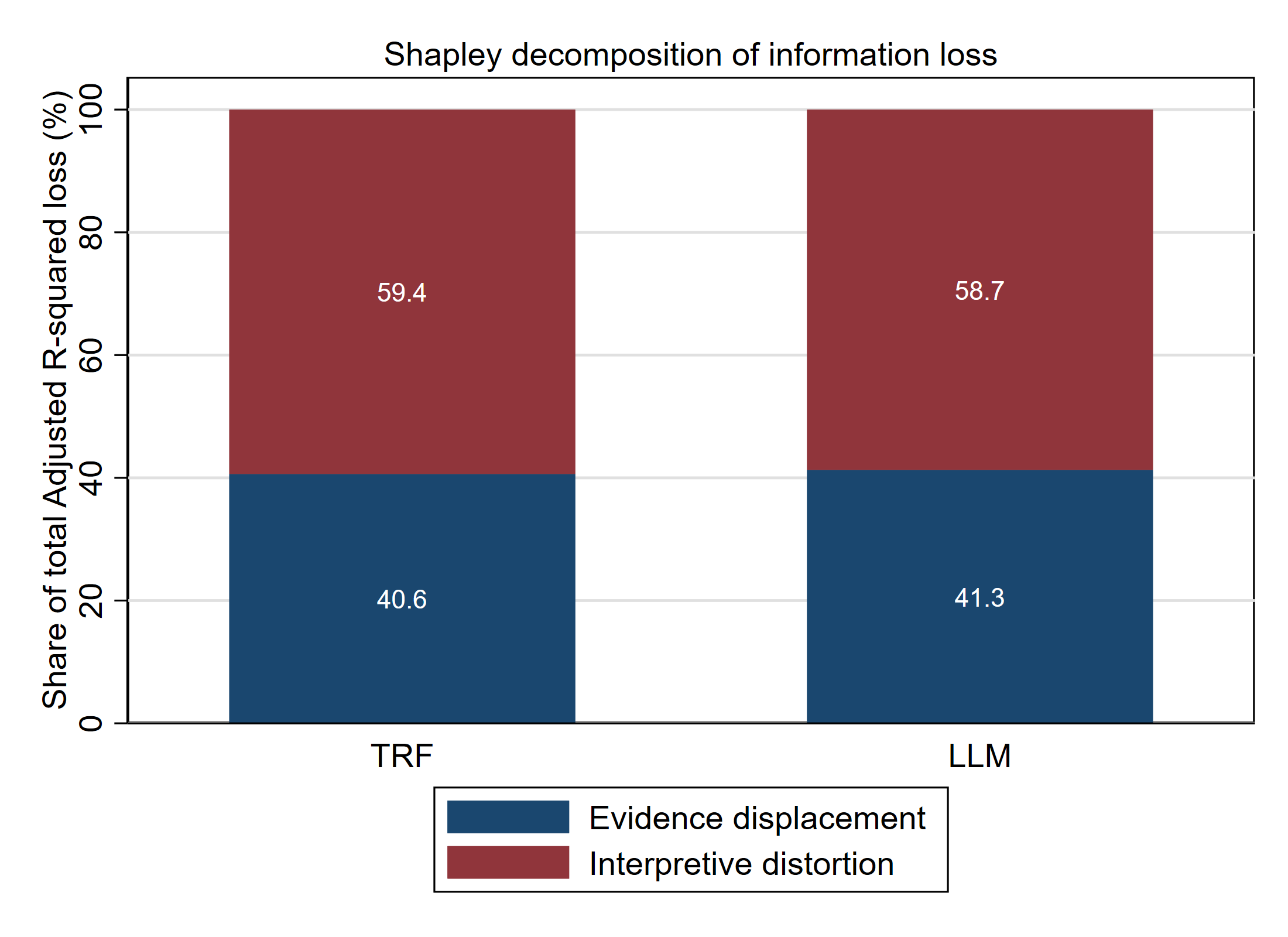}
  \caption{{\bf Shapley Decomposition of Credit-Risk Information Loss}\\
  \footnotesize{This figure decomposes the decline in adjusted $R^2$ from the raw ten-factor composite to the fully anonymized composite into two channels: evidence displacement and interpretive distortion. Each stacked bar reports the Shapley share of the total adjusted $R^2$
 decline for the TRF- and LLM-anonymized transcripts, respectively; the two components sum to 100 percent within each anonymization method. The values shown inside the bars are percentage shares. The decomposition procedure is described in detail in Appendix~\ref{appendixC3}.
  }}
  \label{fig:sp10_shapley}
\end{figure}

Figure~\ref{fig:sp10_shapley} reports the accounting shares assigned to the two
channels within the constructed four-state decomposition. Evidence displacement
accounts for 40.6\% of the adjusted $R^2$ decline under TRF anonymization and
41.3\% under LLM anonymization, while interpretive distortion accounts for the
remaining 59.4\% and 58.7\%, respectively. Within this accounting exercise,
approximately three-fifths of the decline is associated with changing the
assessment of evidence whose anchor remains recognizable, consistent with entity
and numerical context affecting how the model reads the surviving passage.

These shares are not in tension with the channel regressions above, because the
two exercises answer different questions. The regressions ask whether one
channel's assessment adds predictive content \emph{conditional on} the other,
and the conditional means are defined on different reason sets and are highly
correlated with one another. The Shapley shares instead average each channel's
marginal contribution to explanatory power across the two possible orderings. A
channel can therefore carry little incremental coefficient in the joint
specification while still accounting for a substantial share of the total
decline.

\section{Robustness Across Models, Tasks, and Texts}\label{sec:robustness}
We assess the robustness of the information-loss pattern along three dimensions. First, we vary the LLM used to extract the textual signals. Second, holding earnings calls fixed as the textual source, we examine stock-return prediction and four additional prediction tasks. Third, we vary the source text by applying the return-prediction analysis to news headlines.

\subsection{Alternative Extraction Models}\label{subsec:alternative_llms}
To verify that the main findings are not specific to GPT-4o-mini, we repeat the downgrade extraction using GPT-4o and o3-mini. Table~\ref{tab:downgrade_3llms} repeats the comparisons associated with Propositions~\ref{prop:attenuation} and~\ref{prop:horse_race}. Across all three models, the raw score predicts downgrades (Columns 1--3), while the TRF-anonymized score has a smaller coefficient and lower adjusted $R^2$ in every standalone specification (Columns 4--6), displaying the predicted ordering. In the horse-races (Columns 7--9), the anonymized coefficient is $-0.012$ ($t=-0.72$) for GPT-4o-mini, $0.025$ ($t=1.73$) for GPT-4o, and approximately zero ($t=-0.04$) for o3-mini; the raw score remains significant in every case. GPT-4o-mini and o3-mini are consistent with the zero-incremental-information benchmark, while GPT-4o leaves a small, marginally significant incremental coefficient. Overall, the directional attenuation and horse-race patterns recur across extraction LLMs.

The same model-robustness pattern appears in the stock-return application examined below. Appendix Table~\ref{appendix-tab:sentiment_3llms} repeats the return analysis using GPT-4o-mini, GPT-4o, and o3-mini. Across all three models, the anonymized standalone coefficient is attenuated, while the raw sentiment coefficient remains larger in the horse-race. The information-loss pattern is therefore not specific to either the workhorse extraction model or the credit-downgrade task.

\subsection{Alternative Prediction Tasks Using Earnings Calls}\label{sec:beyond_credit_risk}
Holding the textual source fixed, we next examine whether the information-loss pattern extends beyond credit-rating prediction. We begin with earnings-call sentiment and announcement returns, an application in which concerns about look-ahead bias are particularly salient \citep[e.g.,][]{TETLOCK2008, engelberg2025entity}, and then examine four additional forward-looking outcomes.

\subsubsection{Earnings Call Sentiment and Stock Returns}
We extract sentiment from the original and anonymized transcripts using GPT-4o-mini, with a prompt that closely follows \citet{engelberg2025entity} (detailed in Appendix~\ref{subsec:sentiment_prompt}). We tabulate the pairwise correlation between the extracted sentiment scores in Appendix Table~\ref{appendix-tab:sentiment_correlation}. The correlation between the sentiment score extracted from the raw and the TRF-anonymized transcripts is 0.864, while the correlation with the LLM-anonymized transcripts is 0.858. These high but imperfect correlations show that anonymization changes the extracted sentiment measures without displacing them entirely.

We therefore evaluate the predictive consequences of those changes using the DGTW-adjusted announcement return $DGTW_{i,t+1}$, measured in percentage points and defined as in \citet{DGTW1997}.

For the return-prediction analysis, we first assign each conference call to an announcement trading date based on when its information becomes available to the market. A call held before the U.S. market closes at 16:00 ET is assigned to the same trading day, whereas a call held after the market closes is assigned to the next trading day. Thus, for example, a call held at 15:00 ET on Monday is matched to Monday's return, while a call held at 17:00 ET on Monday is matched to Tuesday's return. This convention ensures that the closing price used to calculate the announcement return is observed after the transcript becomes public, so that $DGTW_{i,t+1}$ captures the market's response to it.

Empirically, we adopt the following regression model:

\begin{equation}
DGTW_{i,t+1} = \beta\,\text{Sentiment}_{i,t} + \gamma X_{i,t}+\delta_t + \epsilon_{i,t}
\label{eq:main_regression1}
\end{equation}
The key independent variables are the sentiment scores extracted from the raw text and from the texts anonymized using the TRF and LLM methods, respectively. Following \citet{TETLOCK2008} and \citet{engelberg2025entity}, we include a standard set of control variables, $X_{i,t}$: analyst forecast error ($FE$), shown by prior work to explain abnormal returns \citep{HIRSHLEIFER2009}; DGTW-adjusted returns over the run-up window ($DGTW_{i,t}$, $DGTW_{i,t-1}$, $DGTW_{i,t-2}$, $DGTW_{i,t-21,t-3}$, $DGTW_{i,t-252,t-22}$); and firm characteristics including the logarithm of firm size ($\ln(Size)$), the book-to-market ratio ($\ln(BM)$), and turnover ($\ln(Turnover)$). All regressions include date fixed effects, and all independent variables are standardized to allow a direct comparison of coefficients across text treatments.

In the specification with controls (reported in Panel B of Table~\ref{tab:earnings_call_sentiment}), Columns 1--3 evaluate the ordering in Proposition~\ref{prop:attenuation}. The sentiment score from the raw text exhibits a strong and highly significant relationship with the return: as reported in Column 1, the coefficient on the raw score is $2.487$ ($t=18.83$), confirming that LLMs effectively capture sentiment signals. As shown in Columns 2 and 3, sentiment scores from anonymized texts (TRF and LLM) remain significantly associated with returns, but their coefficients are attenuated to 2.331 and 2.329, respectively, accompanied by a decline in adjusted $R^2$ from 0.132 to 0.124.

To compare the signals more formally, Columns 4 and 5 evaluate the zero-incremental-information benchmark by including both the raw and an anonymized sentiment score. The coefficient on $Sentiment_{RAW}$ remains large and highly significant, whereas the coefficients on the anonymized scores contract sharply: the $Sentiment_{TRF}$ coefficient falls from 2.331 to 0.775 (Column 4), a 67\% reduction, and the $Sentiment_{LLM}$ coefficient drops to 0.798 (Column 5). The sentiment-relevant information in the anonymized text is thus largely subsumed by that of the original text, though both anonymized coefficients remain statistically significant. As shown in Panel A of Table~\ref{tab:earnings_call_sentiment}, the results are robust to excluding control variables.

\subsubsection{Four Additional Earnings Call Prediction Tasks}
We next ask whether anonymization attenuates signals constructed for outcomes other than credit ratings and stock returns. If identifying context helps an LLM interpret forward-looking statements, removing that context should attenuate the text's ability to explain subsequent risk, investment, and operating performance. Table~\ref{tab:multitask} extends the horse-race test by including both the raw and TRF-anonymized score in each task.

First, prior research establishes a positive relation between textual uncertainty in corporate disclosures and subsequent return volatility \citep{WhenIsaLiability}. In Column 1, the raw uncertainty score is positive and significant ($0.026$, $t=2.03$), whereas the TRF-anonymized score is smaller and statistically insignificant ($0.012$, $t=0.78$). Second, following \citet{jha2025chatgptcorporatepolicies}, Column 2 uses management's investment outlook to predict capital expenditures two quarters ahead. Both scores are significant, but the raw coefficient is larger ($0.043$ versus $0.033$).

Third, drawing on \citet{jha2025harnessinggenerativeaieconomic}, Columns 3 and 4 use the firm's macroeconomic outlook to predict sales growth and value-added growth two fiscal quarters ahead. Both raw and anonymized scores remain significant, but the raw coefficient is larger in each task: $1.803$ versus $1.550$ for sales growth and $2.460$ versus $1.787$ for value-added growth. Across all four tasks, the raw coefficient dominates the anonymized one.

\subsubsection{Additional Evidence for the Return Task}
The return setting also permits three additional checks. First, when only one entity category is preserved at a time, raw sentiment remains significant when entered with every single-category-preserved variant, so none of those variants statistically subsumes the full-context signal in this sample (Appendix Table~\ref{appendix-tab:sentiment_keep_one}). The NUMBERS- and AGENTS-preserved variants retain somewhat more incremental predictive content than the PLACES- and OTHERS-preserved variants, consistent with the mechanism evidence in Section~\ref{sec:mechanisms}.

Second, Appendix Table~\ref{appendix-tab:sentiment_pre_interaction} and Appendix Figure~\ref{appendix-fig:sentiment_quarterly_beta} extend the knowledge-cutoff analysis to returns. Quarter-by-quarter coefficients show raw sentiment exceeding its anonymized counterparts in every quarter except the earliest, 2022 Q4, where the ordering reverses for both methods. The $Sentiment\times Pre$ interaction is significantly negative for all three sentiment measures, which falls outside the two cases the cutoff diagnostic admits: under its maintained assumptions the interaction is zero without look-ahead bias and positive with it. A significantly negative estimate therefore signals that one of those assumptions fails, most plausibly the stability of the sentiment--return relation across the cutoff. We accordingly treat the diagnostic as uninformative for returns and rely on the downgrade setting, where the interaction is small and insignificant, for the look-ahead assessment.

Third, Appendix Table~\ref{appendix-tab:sentiment_gap_pre_post} provides an out-of-task evaluation of Proposition~\ref{prop:gap}. Using $\Gap=\lvert A/\rho-R\rvert$, with $\rho=0.8529$ for TRF and 0.8510 for LLM, the $Sentiment\times\Gap$ coefficients are $-0.543$ and $-0.541$ in the full sample, $-0.431$ and $-0.401$ in the pre-cutoff sample, and $-0.635$ and $-0.666$ in the post-cutoff sample; all are significant at the 1\% level. The uniformly negative interactions match the proposition's directional prediction: anonymized sentiment is least informative where its attenuation-adjusted deviation from raw sentiment is largest. The post-cutoff estimates provide the strongest evidence for the document-specific-residual interpretation.

\subsection{Alternative Text: News Headlines}\label{subsec:news_robustness}
Finally, we vary the textual source by turning to news headlines, another widely used form of financial text \citep[e.g.,][]{Bybee2023,lopez2023can,engelberg2025entity}. Compared with the verbose nature of earnings call transcripts, news headlines are extremely short but far denser in information, particularly about entities.

Because a single stock may have multiple headlines on a given day, we average all headline scores for each PERMNO-date to create a daily score. Our final sample comprises 52,716 news headlines from Yahoo Finance, corresponding to 30,950 firm-day observations, with the detailed sample-selection criteria reported in Panel C of Appendix Table~\ref{appendix-tab:Sample Selection Criteria}. The average headline length is 15.55 tokens, and entities recognized by the TRF model account for 38.30\% of headline tokens on average, substantially higher than the 18.92\% observed in earnings call transcripts under the same approach.

Following the methodology used for transcripts, we design a prompt to extract sentiment from headlines (Appendix~\ref{subsec:sentiment_news_prompt}) and average the scores across all headlines for a given firm-day.

Table~\ref{tab:news_prediction} presents regressions explaining the DGTW-adjusted announcement return using news sentiment, where the return follows the same timing convention as for earnings calls. In both the single-score specifications (Columns 1 and 2) and the specifications with standard controls (Columns 4 and 5), sentiment derived from the raw ($Sentiment_{RAW}^{News}$) and TRF-anonymized ($Sentiment_{TRF}^{News}$) headlines is a significant positive predictor of returns, and the anonymized coefficient is the smaller of the two, matching the predicted ordering. The standalone attenuation is modest, however, and the adjusted $R^2$ values are indistinguishable once controls are included. Columns 3 and 6 turn to the horse-race. Without controls (Column 3), the coefficient on $Sentiment_{RAW}^{News}$ (0.315) remains significant while that on $Sentiment_{TRF}^{News}$ (0.135) is attenuated to less than a third of its standalone magnitude; the pattern persists with controls (Column 6): 0.314 versus 0.131. The sharp conditional attenuation shows that named entities contribute materially to explanatory value in both transcripts and headlines.

\section{Conclusion}
This study provides empirical evidence that anonymization, a widely used technique for mitigating look-ahead bias in LLM-based financial text analysis, carries a substantial and underappreciated cost: information loss. By removing or transforming identifying and contextual information, anonymization can attenuate the ability of LLMs to extract economically meaningful signals from financial texts.

Our main analysis examines the prediction of future S\&P credit-rating downgrades, evaluating in sequence the four predictions the framework delivers. The ROC curves show that all three scores predict downgrades, but the raw score has the largest area under the curve. In the regressions, anonymized scores remain informative on their own but have smaller coefficients and lower explanatory power than the raw score. Once the raw score is included, the anonymized coefficients become statistically indistinguishable from zero. Extending the analysis across the knowledge-cutoff produces small and insignificant score--pre-cutoff interactions, providing little evidence of the positive pre-cutoff increment that look-ahead bias would generate. Finally, the negative score--gap interactions, especially in the post-cutoff sample, support interpreting the gap as a proxy for the document-specific residual.

We next examine the mechanisms underlying this predictive attenuation. The TRF-based decomposition shows that no single preserved entity category fully recovers the full-context signal, although preserving numerical entities, such as financial figures and dates, or agent entities, such as firm and executive names, recovers substantially more than preserving place or other entities. The ten-factor credit-risk audit then shows that anonymization changes both the evidence selected and the assessments assigned to shared evidence, the channels we call evidence displacement and interpretive distortion. A composite constructed from the ten-factor assessments reproduces the predictive loss in the one-shot downgrade score. Within the audit's four-state accounting exercise, the Shapley decomposition attributes 40.6\% of the adjusted $R^2$ decline under TRF anonymization and 41.3\% under LLM anonymization to evidence displacement; interpretive distortion accounts for the remaining 59.4\% and 58.7\%, respectively.

We then assess robustness across extraction models, prediction tasks, and financial text types. Repeating the downgrade analysis with GPT-4o and o3-mini produces the same qualitative attenuation and horse-race patterns, showing that the main result is not specific to GPT-4o-mini. In earnings-call return prediction, sentiment extracted from raw transcripts has greater explanatory power for announcement-window stock returns and substantially subsumes anonymized sentiment in horse-races. Across four additional transcript-based tasks, the raw measures also have greater predictive power for future return volatility, capital expenditures, sales growth, and value-added growth than their anonymized counterparts. The same ordering appears in the analysis of news headlines, showing that the information-loss pattern is not confined to one model, outcome, or type of financial disclosure.

Our findings highlight a fundamental methodological trade-off. Anonymization may reduce an LLM's exposure to memorized firm-specific information, but it can also remove genuine economic context that is necessary for accurate textual interpretation. Researchers should therefore not interpret a decline in predictive power after anonymization as evidence that look-ahead bias has been successfully eliminated. Without a post-cutoff benchmark, the same decline may instead reflect the loss of economically relevant information. More constructively, the anonymization gap we introduce lets researchers gauge the severity of this loss in their own corpora, without requiring outcome data. Future work should develop methods that mitigate look-ahead bias while preserving the contextual information required for meaningful financial text analysis.

\clearpage

\bibliographystyle{apalike}

\bibliography{new_reference_updated}

\clearpage

\clearpage

\clearpage

\clearpage

\clearpage

\clearpage

\clearpage

\clearpage

\begin{table}[htbp]
\footnotesize
\caption{{\bf Comparison of Text Anonymization Approaches}\\\\
\footnotesize{This table displays an excerpt from Apple Inc.'s 2024 Q1 earnings call transcript alongside the anonymized versions produced by the two approaches described in Section~\ref{subsec:anonymization_algorithms}: the spaCy \texttt{en\_core\_web\_trf} named-entity-recognition model (TRF) and the iterative LLM procedure (GPT-4o-mini). Replaced entities in the anonymized text are highlighted in bold for clarity.}}
\label{tab:anonymization_comparison}
\vspace{0.2cm}
\centering
\newcolumntype{P}[1]{>{\raggedright\arraybackslash}p{#1}}
\begin{tabular}{| P{0.28\textwidth} | P{0.28\textwidth} | P{0.28\textwidth} | }
    \hline
    \textbf{Raw Text} & \textbf{TRF Anonymized} &  \textbf{LLM Anonymized} \\
    \hline
    Please note that some of the information you'll hear during our discussion today will consist of forward-looking statements, including, without limitation, those regarding revenue, gross margin, operating expenses, other income and expense, taxes, capital allocation and future business outlook, including the potential impact of macroeconomic conditions on the firm's business and results of operations. These statements involve risks and uncertainties that may cause actual results or trends to differ materially from our forecast. For more information, please refer to the risk factors discussed in Apple's most recently filed annual report on Form 10-K and the form 8-K filed with the SEC today along with the associated press release. Apple assumes no obligation to update any forward-looking statements, which speak only as of the date they are made.
    &
    Please note that some of the information you'll hear during our discussion \textbf{DATE\_1} will consist of forward-looking statements, including, without limitation, those regarding revenue, gross margin, operating expenses, other income and expense, taxes, capital allocation and future business outlook, including the potential impact of macroeconomic conditions on the firm's business and results of operations. These statements involve risks and uncertainties that may cause actual results or trends to differ materially from our forecast. For more information, please refer to the risk factors discussed in \textbf{ORG\_1}'s most recently filed \textbf{DATE\_2} report on Form 10-K and the form 8-K filed with the \textbf{ORG\_2} \textbf{DATE\_1} along with the associated press release. \textbf{ORG\_1} assumes no obligation to update any forward-looking statements, which speak only as of the date they are made.
    &
    Please be aware that some of the information shared during our conversation today will include forward-looking statements, which may encompass, but are not limited to, projections regarding revenue, gross margin, operating expenses, other income and expenses, taxes, capital allocation, and future business outlook, including potential effects of macroeconomic conditions on the entity's operations and results. These statements carry risks and uncertainties that could lead to actual outcomes differing significantly from our predictions. For further details, please refer to the risk factors outlined in \textbf{Entity\_A}'s most recent annual report on Form 10-K and the Form 8-K submitted to the regulatory body today, along with the related press release. \textbf{Entity\_A} has no obligation to update any forward-looking statements, which are valid only as of the date they are made.
    \\
    \hline
\end{tabular}
\end{table}

\clearpage

\begin{table}[htbp]
\centering
\caption{{\bf Conference Call Entity Count and Anonymization Performance (Downgrade Prediction Subsample)}\\\\
\footnotesize{This table reports the entity count and anonymization performance for the downgrade-prediction post-cutoff subsample of 2,488 transcripts. Panel A reports descriptive statistics on transcript length and the share of tokens that are entities after anonymization, for the TRF and LLM (GPT-4o-mini) approaches. Length is measured in number of tokens using the Xenova/GPT-4o tokenizer. Entity\% is the share of tokens accounted for by entities. For TRF, it is the token length of the entities the model identifies in the raw transcript divided by the raw transcript length; for the LLM approach, which returns plain text rather than tagged spans, it is the share of placeholder tokens in the anonymized output. The two shares are therefore constructed on different bases and are not strictly comparable. Panel B reports the re-identification rates for four key attributes (Date, Firm Name, Ticker, and Industry) under the two anonymization approaches. We define a correct match using the following criteria: the estimated Date must fall within 7 days of the actual conference call date; Firm Name and Industry are evaluated using fuzzy string matching (Levenshtein token sort ratio), where a score of 75 or higher after text normalization is considered a match; and Ticker symbols require an exact case-insensitive match with the ground truth.}}
\label{tab:summary_stats_anonymization}
\vspace{0.2cm}
\begin{tabular}{@{\extracolsep{8pt}}lccccc@{}}
\toprule
\multicolumn{6}{l}{Panel A: Transcript Length and Entity Percentages}\\
\midrule
 & \textbf{Mean} & \textbf{Std. Dev.} & \textbf{25th Pct.} & \textbf{Median} & \textbf{75th Pct.} \\
\midrule
\addlinespace[0.5em]
Length & 9,539 & 2,323 & 7,977 & 9,888 & 11,237 \\
\addlinespace[0.3em]
Entity\% (TRF) & 18.92 & 2.53 & 17.20 & 18.71 & 20.55 \\
\addlinespace[0.3em]
Entity\% (LLM) & 11.51 & 3.42 & 9.47 & 11.26 & 13.19 \\
\addlinespace[0.2em]
\bottomrule
\end{tabular}
\vspace{0.5em}
\begin{tabular}{l c c}
\toprule
\multicolumn{3}{l}{Panel B: Identification Ratio \%}\\
\midrule
Date       & 6.03  & 2.81  \\
Firm Name  & 66.88 & 9.04  \\
Ticker     & 53.90 & 9.12  \\
Industry   & 50.56 & 44.01 \\
\midrule
Anonymization Approach  & TRF & GPT-4o-mini \\
Entity Remove           & ALL & ALL \\
Recognition Approach    & GPT-4o-mini & GPT-4o-mini \\
\bottomrule
\end{tabular}
\end{table}

\clearpage

\begin{table}[htbp]
  \centering
  \footnotesize
  \caption{{\bf Downgrade Predictions: Standalone and Horse-Race Regressions}\\\\
  \footnotesize{This table reports regressions of the realized S\&P credit-rating downgrade indicator $Downgrade_{i,t+1}$ on the model-extracted downgrade probability $\widehat{Downgrade}_{RAW}$, $\widehat{Downgrade}_{TRF}$, and $\widehat{Downgrade}_{LLM}$. Panel A reports regressions without firm-level controls; Panel B includes the full set of firm-level controls: $\ln(FirmValue)$, $Leverage_{Asset}$, $MTB$, $ROA$, $RevGrowth$, $Std(Income)$, $Std(Revenue)$, $Zscore$, and $Oscore$, all defined as in Appendix Table~\ref{appendix-tab:variable_definitions_full}. Within each panel, Columns (1)--(3) report standalone single-score regressions; Columns (4) and (5) report horse-race specifications that include the raw score together with one anonymized score. The sample is the downgrade-task post-cutoff subsample of 2,488 transcripts (Nov 2023--Dec 2024), and the unit of observation is the earnings call transcript. All variables are winsorized at the 1st and 99th percentiles, and all independent variables are standardized within the regression sample. All regressions include SIC3 $\times$ Quarter fixed effects. T-statistics based on SIC3-clustered standard errors are reported in parentheses. * $p < 0.10$, ** $p < 0.05$, *** $p < 0.01$.}}
  \label{tab:downgrade_main}
  \vspace{0.2cm}
  \setlength{\tabcolsep}{6pt}
  \begin{tabular}{@{}lccccc@{}}
    \toprule
    & \multicolumn{5}{c}{Dependent Variable: $Downgrade_{i,t+1}$} \\
    \cmidrule(lr){2-6}
    VARIABLES & (1) & (2) & (3) & (4) & (5) \\
    \midrule
    \multicolumn{6}{l}{\textit{Panel A: Without Controls}}\\
    \addlinespace
    $\widehat{Downgrade}_{RAW}$ & 0.077*** & & & 0.099*** & 0.095*** \\
    & (6.471) & & & (4.520) & (5.483) \\
    \addlinespace
    $\widehat{Downgrade}_{TRF}$ & & 0.062*** & & -0.024 & \\
    & & (5.820) & & (-1.350) & \\
    \addlinespace
    $\widehat{Downgrade}_{LLM}$ & & & 0.064*** & & -0.019 \\
    & & & (5.864) & & (-1.520) \\
    \addlinespace
    \midrule
    Controls & No & No & No & No & No \\
    SIC3 $\times$ Quarter FE & Yes & Yes & Yes & Yes & Yes \\
    Observations & 2,243 & 2,243 & 2,243 & 2,243 & 2,243 \\
    Adj. $R^2$ & 0.223 & 0.195 & 0.198 & 0.225 & 0.224 \\
    \midrule
    \multicolumn{6}{l}{\textit{Panel B: With Controls}}\\
    \addlinespace
    $\widehat{Downgrade}_{RAW}$ & 0.061*** & & & 0.073*** & 0.068*** \\
    & (5.515) & & & (3.576) & (4.215) \\
    \addlinespace
    $\widehat{Downgrade}_{TRF}$ & & 0.047*** & & -0.012 & \\
    & & (4.799) & & (-0.719) & \\
    \addlinespace
    $\widehat{Downgrade}_{LLM}$ & & & 0.049*** & & -0.007 \\
    & & & (4.919) & & (-0.522) \\
    \addlinespace
    \midrule
    Controls & Yes & Yes & Yes & Yes & Yes \\
    SIC3 $\times$ Quarter FE & Yes & Yes & Yes & Yes & Yes \\
    Observations & 2,243 & 2,243 & 2,243 & 2,243 & 2,243 \\
    Adj. $R^2$ & 0.246 & 0.232 & 0.234 & 0.246 & 0.245 \\
    \bottomrule
  \end{tabular}
\end{table}

\clearpage

\begin{table}[htbp]
  \centering
  \small
  \caption{{\bf Sample Period and Predictive Power of Downgrade Estimates}\\\\
  \footnotesize{This table examines whether the predictive power of downgrade-probability scores extracted from raw and anonymized earnings call transcripts differs between pre- and post-knowledge-cutoff subsamples. The dependent variable is $Downgrade_{i,t+1}$, an indicator equal to 1 if S\&P downgrades the firm's credit rating within the following 12 months from the earnings call and 0 otherwise. $\widehat{Downgrade}_{RAW}$, $\widehat{Downgrade}_{TRF}$, and $\widehat{Downgrade}_{LLM}$ denote the predicted downgrade probabilities from raw, TRF-anonymized, and LLM-anonymized transcripts. $Pre$ is an indicator variable equal to 1 if the transcript falls in the pre-cutoff window (Nov 2022--Oct 2023) and 0 in the post-cutoff window (Nov 2023--Dec 2024). Columns (1)--(3) use the full balanced panel of firms appearing in both periods. All variables are winsorized at the 1st and 99th percentiles, and all independent variables are standardized (except dummy variable $Pre$). Control variables are identical to those in Panel B of Table~\ref{tab:downgrade_main}. All regressions include SIC3 $\times$ Quarter fixed effects. T-statistics based on SIC3-clustered standard errors are reported in parentheses. * $p < 0.10$, ** $p < 0.05$, *** $p < 0.01$.}}
  \label{tab:downgrade_pre_interaction}
  \vspace{0.2cm}
  \setlength{\tabcolsep}{8pt}
  \begin{tabular}{@{}lccc@{}}
    \toprule
    & \multicolumn{3}{c}{Dependent Variable: $Downgrade_{i,t+1}$} \\
    \cmidrule(lr){2-4}
    VARIABLES & (1) & (2) & (3) \\
    \midrule
    $\widehat{Downgrade}_{RAW}$ & 0.061*** & & \\
    & (5.351) & & \\
    \addlinespace
    $\widehat{Downgrade}_{RAW} \times Pre$ & -0.005 & & \\
    & (-0.306) & & \\
    \addlinespace
    $\widehat{Downgrade}_{TRF}$ & & 0.047*** & \\
    & & (4.551) & \\
    \addlinespace
    $\widehat{Downgrade}_{TRF} \times Pre$ & & -0.009 & \\
    & & (-0.587) & \\
    \addlinespace
    $\widehat{Downgrade}_{LLM}$ & & & 0.050*** \\
    & & & (4.821) \\
    \addlinespace
    $\widehat{Downgrade}_{LLM} \times Pre$ & & & -0.010 \\
    & & & (-0.631) \\
    \midrule
    Controls & Yes & Yes & Yes \\
    SIC3 $\times$ Quarter FE & Yes & Yes & Yes \\
    Observations & 3,943 & 3,943 & 3,943 \\
    Adj. $R^2$ & 0.218 & 0.202 & 0.205 \\
    \bottomrule
  \end{tabular}
\end{table}

\clearpage

\begin{table}[htbp]
  \centering
  \footnotesize
  \caption{{\bf Anonymization Gap and Predictive Power of Downgrade Estimates}\\\\
  \footnotesize{This table examines whether the attenuation-adjusted gap between raw and anonymized downgrade scores tracks where anonymization attenuates predictive content. The dependent variable is $Downgrade_{i,t+1}$, an indicator equal to one if S\&P downgrades the firm's credit rating within twelve months after the earnings call. For each anonymization method, the raw score $R$ and anonymized score $A$ are winsorized at the 1st and 99th percentiles and standardized over the pooled balanced sample. We estimate $\rho=\Corr(R,A)$ on that sample and define $\Gap=\lvert A/\rho-R\rvert$. The gap is then winsorized at the 1st and 99th percentiles and standardized. Each specification includes the standardized anonymized score, the standardized gap, and their product; the interaction itself is not re-standardized. The estimated values are $\rho_{TRF}=0.8993$ and $\rho_{LLM}=0.8997$, and the same pooled-sample estimate of $\rho$ is used in all six columns. Columns (1)--(2) use the full balanced sample, Columns (3)--(4) the pre-cutoff window (Nov. 2022--Oct. 2023), and Columns (5)--(6) the post-cutoff window (Nov. 2023--Dec. 2024). Controls are identical to Panel B of Table~\ref{tab:downgrade_main}. All regressions include SIC3 $\times$ Quarter fixed effects. T-statistics based on SIC3-clustered standard errors are in parentheses. * $p < 0.10$, ** $p < 0.05$, *** $p < 0.01$.}}
  \label{tab:downgrade_gap_pre_post}
  \vspace{0.2cm}
  \setlength{\tabcolsep}{3pt}
  \begin{tabular}{@{}lcccccc@{}}
    \toprule
    & \multicolumn{6}{c}{Dependent Variable: $Downgrade_{i,t+1}$} \\
    \cmidrule(lr){2-7}
    VARIABLES & (1) & (2) & (3) & (4) & (5) & (6) \\
    \midrule
    $\widehat{Downgrade}_{TRF}$ & 0.051*** & & 0.049*** & & 0.054*** & \\
    & (4.755) & & (3.280) & & (4.338) & \\
    \addlinespace
    $\widehat{Downgrade}_{TRF} \times \Gap_{TRF}$ & -0.019*** & & -0.015* & & -0.023*** & \\
    & (-3.615) & & (-1.881) & & (-3.015) & \\
    \addlinespace
    $\Gap_{TRF}$ & 0.006 & & -0.002 & & 0.010 & \\
    & (0.830) & & (-0.188) & & (1.139) & \\
    \addlinespace
    $\widehat{Downgrade}_{LLM}$ & & 0.049*** & & 0.043*** & & 0.054*** \\
    & & (4.584) & & (2.821) & & (4.307) \\
    \addlinespace
    $\widehat{Downgrade}_{LLM} \times \Gap_{LLM}$ & & -0.012** & & -0.013 & & -0.013** \\
    & & (-2.371) & & (-1.461) & & (-2.194) \\
    \addlinespace
    $\Gap_{LLM}$ & & 0.006 & & 0.013 & & -0.000 \\
    & & (0.877) & & (1.065) & & (-0.032) \\
    \midrule
    Sample & Full & Full & Pre & Pre & Post & Post \\
    Controls & Yes & Yes & Yes & Yes & Yes & Yes \\
    SIC3 $\times$ Quarter FE & Yes & Yes & Yes & Yes & Yes & Yes \\
    Observations & 3,943 & 3,943 & 1,760 & 1,760 & 2,133 & 2,133 \\
    Adj. $R^2$ & 0.211 & 0.208 & 0.181 & 0.179 & 0.250 & 0.245 \\
    \bottomrule
  \end{tabular}
\end{table}

\clearpage

\begin{table}[htbp]
\centering
\footnotesize
\caption{{\bf Correlations of Downgrade Prediction Signals}\\
\footnotesize{This table reports Pearson correlations among downgrade probabilities extracted using GPT-4o-mini. Panel A compares scores extracted from raw transcripts (RAW) with scores extracted from transcripts fully anonymized using TRF or LLM. Panel B compares the RAW score with scores extracted from transcripts in which only one entity category is preserved and the remaining categories are anonymized using TRF. NUM, PLC, AGT, and OTH denote specifications preserving only NUMBERS, PLACES, AGENTS, and OTHERS, respectively. All correlations are computed on the 2,488-transcript post-cutoff sample.}}
\label{tab:downgrade_correlation}
\vspace{0.2cm}
\setlength{\tabcolsep}{6pt}
\begin{tabular}{lccc}
\multicolumn{4}{l}{\textbf{Panel A: Raw and Fully Anonymized Downgrade Scores}}\\
\toprule
& $\widehat{Downgrade}_{RAW}$ & $\widehat{Downgrade}_{TRF}$ & $\widehat{Downgrade}_{LLM}$ \\
\midrule
$\widehat{Downgrade}_{RAW}$ & 1.000 &  &  \\
$\widehat{Downgrade}_{TRF}$ & 0.907 & 1.000 &  \\
$\widehat{Downgrade}_{LLM}$ & 0.907 & 0.931 & 1.000 \\
\bottomrule
\end{tabular}
\vspace{0.6em}
\setlength{\tabcolsep}{3pt}
\begin{tabular}{lccccc}
\multicolumn{6}{l}{\textbf{Panel B: Raw and Single-Category-Preserved Downgrade Scores}}\\
\toprule
& $\widehat{Downgrade}_{RAW}$ & $\widehat{Downgrade}_{NUM}$ & $\widehat{Downgrade}_{PLC}$ & $\widehat{Downgrade}_{AGT}$ & $\widehat{Downgrade}_{OTH}$ \\
\midrule
$\widehat{Downgrade}_{RAW}$ & 1.000 &   &   &   &   \\
$\widehat{Downgrade}_{NUM}$ & 0.946 & 1.000 &   &   &   \\
$\widehat{Downgrade}_{PLC}$ & 0.916 & 0.939 & 1.000 &   &   \\
$\widehat{Downgrade}_{AGT}$ & 0.932 & 0.899 & 0.912 & 1.000 &   \\
$\widehat{Downgrade}_{OTH}$ & 0.914 & 0.935 & 0.976 & 0.916 & 1.000 \\
\bottomrule
\end{tabular}
\end{table}

\clearpage

\begin{table}[htbp]
  \centering
  \footnotesize
  \caption{{\bf Comparing Downgrade Predictions When Preserving Different Categories of Entities}\\\\
  \footnotesize{This table reports regressions of $Downgrade_{i,t+1}$ on the downgrade probability extracted from transcripts where only one of the four entity categories is preserved at a time, while the other three categories are anonymized using the TRF model. The category definitions are provided in Appendix Table~\ref{appendix-tab:Entity Category}. Specifically, $\widehat{Downgrade}_{NUM}$, $\widehat{Downgrade}_{PLC}$, $\widehat{Downgrade}_{AGT}$, and $\widehat{Downgrade}_{OTH}$ are extracted from transcripts where the TRF model preserves only the NUMBERS, PLACES, AGENTS, or OTHERS category, respectively, while anonymizing all entities belonging to the other three categories. Columns (1)--(5) report standalone single-score regressions for the RAW score and the four single-category-preserved variants; Columns (6)--(9) include the RAW score together with each single-category-preserved variant. The sample is the downgrade-task post-cutoff subsample of 2,488 transcripts (Nov 2023--Dec 2024). Control variables are identical to those in Panel B of Table~\ref{tab:downgrade_main}. All variables are winsorized at the 1st and 99th percentiles, and all independent variables are standardized. All regressions include SIC3 $\times$ Quarter fixed effects. T-statistics based on SIC3-clustered standard errors are reported in parentheses. * $p < 0.10$, ** $p < 0.05$, *** $p < 0.01$.}}
  \label{tab:downgrade_keep_one}
  \vspace{0.2cm}
  \setlength{\tabcolsep}{2pt}
  \begin{tabular}{@{}lccccccccc@{}}
    \toprule
    & \multicolumn{9}{c}{Dependent Variable: $Downgrade_{i,t+1}$} \\
    \cmidrule(lr){2-10}
    VARIABLES & (1) & (2) & (3) & (4) & (5) & (6) & (7) & (8) & (9) \\
    \midrule
    $\widehat{Downgrade}_{RAW}$ & 0.061*** & & & & & 0.044** & 0.072*** & 0.045** & 0.059*** \\
    & (5.515) & & & & & (2.380) & (3.535) & (2.272) & (2.976) \\
    \addlinespace
    $\widehat{Downgrade}_{NUM}$ & & 0.058*** & & & & 0.019 & & & \\
    & & (5.323) & & & & (1.060) & & & \\
    \addlinespace
    $\widehat{Downgrade}_{PLC}$ & & & 0.049*** & & & & -0.011 & & \\
    & & & (4.890) & & & & (-0.653) & & \\
    \addlinespace
    $\widehat{Downgrade}_{AGT}$ & & & & 0.058*** & & & & 0.017 & \\
    & & & & (5.227) & & & & (0.895) & \\
    \addlinespace
    $\widehat{Downgrade}_{OTH}$ & & & & & 0.051*** & & & & 0.003 \\
    & & & & & (5.075) & & & & (0.157) \\
    \midrule
    Controls & Yes & Yes & Yes & Yes & Yes & Yes & Yes & Yes & Yes \\
    SIC3 $\times$ Quarter FE & Yes & Yes & Yes & Yes & Yes & Yes & Yes & Yes & Yes \\
    Observations & 2,243 & 2,243 & 2,243 & 2,243 & 2,243 & 2,243 & 2,243 & 2,243 & 2,243 \\
    Adj. $R^2$ & 0.246 & 0.243 & 0.233 & 0.242 & 0.237 & 0.246 & 0.246 & 0.246 & 0.245 \\
    \bottomrule
  \end{tabular}
\end{table}

\clearpage

\begin{table}[htbp]
  \centering
  \footnotesize
  \caption{{\bf S\&P Ten-Factor Composite Scores and Downgrade Prediction}\\\\
  \footnotesize{This table reports regressions of the realized twelve-month downgrade indicator on researcher-constructed composite scores from the S\&P-aligned ten-factor audit. For each text version, the composite is the average score across factors for which the model returns both a score and a supporting quote. Evidence coverage is reported separately in Figure~\ref{fig:sp_reason_flow}. Composite scores are winsorized at the 1st and 99th percentiles and standardized. Columns (1)--(3) report standalone regressions for the raw, TRF-anonymized, and LLM-anonymized composites. Columns (4)--(5) include the raw composite with one anonymized composite. All specifications include the controls in Panel B of Table~\ref{tab:downgrade_main} and SIC3 $\times$ Quarter fixed effects.
  The sample is the downgrade-task post-cutoff subsample of 2,488 transcripts (Nov 2023--Dec 2024). T-statistics based on SIC3-clustered standard errors are in parentheses. * $p < 0.10$, ** $p < 0.05$, *** $p < 0.01$.}}
  \label{tab:sp10_composite}
  \vspace{0.2cm}
  \setlength{\tabcolsep}{5pt}
  \begin{tabular}{@{}lccccc@{}}
    \toprule
    & \multicolumn{5}{c}{Dependent Variable: $Downgrade_{i,t+1}$} \\
    \cmidrule(lr){2-6}
    VARIABLES & (1) & (2) & (3) & (4) & (5) \\
    & Raw & TRF & LLM & Raw + TRF & Raw + LLM \\
    \midrule
    $\widetilde C^{RAW}$ & 0.045*** & & & 0.045*** & 0.046*** \\
    & (4.405) & & & (3.648) & (3.068) \\
    \addlinespace
    $\widetilde C^{TRF}$ & & 0.035*** & & -0.000 & \\
    & & (4.217) & & (-0.040) & \\
    \addlinespace
    $\widetilde C^{LLM}$ & & & 0.035*** & & -0.001 \\
    & & & (4.099) & & (-0.093) \\
    \midrule
    Controls & Yes & Yes & Yes & Yes & Yes \\
    SIC3 $\times$ Quarter FE & Yes & Yes & Yes & Yes & Yes \\
    Observations & 2,243 & 2,243 & 2,241 & 2,243 & 2,241 \\
    Adj. $R^2$ & 0.226 & 0.218 & 0.218 & 0.225 & 0.225 \\
    \bottomrule
  \end{tabular}
\end{table}

\clearpage

\begin{table}[htbp]
  \centering
  \scriptsize
  \caption{{\bf Predictive Content of Shared and Version-Specific Credit-Risk Reasons}\\\\
  \footnotesize{This table reports regressions using the four variables that measure the two channels defined in Section~\ref{sec:sp10_audit} and Appendix~\ref{app:sp10_method}. Shared reasons are factor assessments for which the raw and anonymized versions cite matched normalized quote anchors. \textit{SharedRaw} and \textit{SharedAnon} are the mean raw and anonymized scores across those reasons; the difference between them measures interpretive distortion. RAW-only reasons comprise raw evidence that disappears after anonymization and the raw side of a factor for which the cited passage changes. Anonymized-only reasons are defined symmetrically. \textit{RawOnly} and \textit{AnonOnly} are the corresponding means, and the difference between them measures evidence displacement. A mean is missing when its reason set is empty, so observations vary across columns. Panel A compares raw and TRF-anonymized transcripts, and Panel B compares raw and LLM-anonymized transcripts. Columns (1)--(2) report separate regressions for \textit{SharedRaw} and \textit{SharedAnon}, and Column (3) includes both. Columns (4)--(5) report separate regressions for \textit{RawOnly} and \textit{AnonOnly}, and Column (6) includes both. Column (7) includes all four measures. The variables are winsorized at the 1st and 99th percentiles and retain their original $[0,1]$ units. All specifications include the controls in Panel B of Table~\ref{tab:downgrade_main} and SIC3 $\times$ Quarter fixed effects. The sample is the downgrade-task post-cutoff subsample (Nov 2023--Dec 2024), and the unit of observation is the earnings call transcript. T-statistics based on SIC3-clustered standard errors are in parentheses. * $p < 0.10$, ** $p < 0.05$, *** $p < 0.01$.}}
  \label{tab:sp10_reason_decomposition}
  \vspace{0.15cm}
  \setlength{\tabcolsep}{2pt}
  \begin{adjustbox}{max width=\textwidth}
  \begin{tabular}{@{}lccccccc@{}}
    \toprule
    & (1) & (2) & (3) & (4) & (5) & (6) & (7) \\
    & \makecell{Shared\\RAW}
    & \makecell{Shared\\Anon.}
    & \makecell{Shared\\Joint}
    & \makecell{RAW-\\only}
    & \makecell{Anon.-\\only}
    & \makecell{Only\\Joint}
    & \makecell{All\\Four} \\
    \midrule
    \multicolumn{8}{l}{\textit{Panel A: RAW versus TRF-anonymized transcripts}}\\
    \addlinespace
    \makecell[l]{$\mathrm{SharedRaw}$} & 0.293*** & & 0.287*** & & & & 0.216*** \\
    & (4.240) & & (3.744) & & & & (2.677) \\
    \addlinespace
    \makecell[l]{$\mathrm{SharedAnon}$} & & 0.234*** & 0.007 & & & & -0.005 \\
    & & (3.734) & (0.118) & & & & (-0.085) \\
    \addlinespace
    \makecell[l]{$\mathrm{RawOnly}$} & & & & 0.331*** & & 0.287*** & 0.151 \\
    & & & & (3.984) & & (2.802) & (1.372) \\
    \addlinespace
    \makecell[l]{$\mathrm{AnonOnly}$} & & & & & 0.270*** & 0.056 & 0.004 \\
    & & & & & (4.163) & (0.938) & (0.061) \\
    \midrule
    Controls & Yes & Yes & Yes & Yes & Yes & Yes & Yes \\
    SIC3 $\times$ Quarter FE & Yes & Yes & Yes & Yes & Yes & Yes & Yes \\
    Observations & 2,172 & 2,172 & 2,172 & 2,227 & 2,231 & 2,227 & 2,155 \\
    Adj. $R^2$ & 0.222 & 0.215 & 0.222 & 0.226 & 0.219 & 0.225 & 0.230 \\
    \midrule
    \multicolumn{8}{l}{\textit{Panel B: RAW versus LLM-anonymized transcripts}}\\
    \addlinespace
    \makecell[l]{$\mathrm{SharedRaw}$} & 0.364*** & & 0.385*** & & & & 0.352*** \\
    & (5.139) & & (4.288) & & & & (4.172) \\
    \addlinespace
    \makecell[l]{$\mathrm{SharedAnon}$} & & 0.272*** & -0.026 & & & & -0.062 \\
    & & (4.263) & (-0.347) & & & & (-0.859) \\
    \addlinespace
    \makecell[l]{$\mathrm{RawOnly}$} & & & & 0.314*** & & 0.247** & 0.035 \\
    & & & & (3.601) & & (2.076) & (0.274) \\
    \addlinespace
    \makecell[l]{$\mathrm{AnonOnly}$} & & & & & 0.258*** & 0.074 & 0.079 \\
    & & & & & (3.805) & (0.895) & (0.988) \\
    \midrule
    Controls & Yes & Yes & Yes & Yes & Yes & Yes & Yes \\
    SIC3 $\times$ Quarter FE & Yes & Yes & Yes & Yes & Yes & Yes & Yes \\
    Observations & 2,015 & 2,015 & 2,015 & 2,233 & 2,228 & 2,227 & 2,001 \\
    Adj. $R^2$ & 0.225 & 0.212 & 0.224 & 0.207 & 0.201 & 0.205 & 0.209 \\
    \bottomrule
  \end{tabular}
  \end{adjustbox}
\end{table}

\clearpage

\begin{table}[htbp]
  \centering
  \footnotesize
  \caption{{\bf Information Loss of Alternative LLMs for Downgrade Predictions}\\\\
  \footnotesize{This table compares downgrade probabilities generated by three distinct models: GPT-4o-mini-2024-07-18, GPT-4o-2024-08-06, and o3-mini-2025-01-31. All regressions include the full set of control variables and use the realized S\&P credit-rating downgrade indicator $Downgrade_{i,t+1}$ as the dependent variable. Columns (1) through (3) report the results for $\widehat{Downgrade}_{RAW}$ extracted from the raw transcripts by each LLM. Columns (4) through (6) report the results for $\widehat{Downgrade}_{TRF}$ extracted from the TRF-anonymized text. Columns (7) through (9) conduct horse-race regressions for each LLM-extracted signal. The sample is the downgrade-task post-cutoff subsample of 2,488 transcripts (Nov 2023--Dec 2024). Control variables are identical to those in Panel B of Table~\ref{tab:downgrade_main}. All variables are winsorized at the 1st and 99th percentiles, and all independent variables are standardized. All regressions include SIC3 $\times$ Quarter fixed effects. T-statistics based on SIC3-clustered standard errors are reported in parentheses. * $p < 0.10$, ** $p < 0.05$, *** $p < 0.01$.}}
  \label{tab:downgrade_3llms}
  \vspace{0.2cm}
  \setlength{\tabcolsep}{2pt}
  \begin{tabular}{@{}lccccccccc@{}}
    \toprule
    & \multicolumn{9}{c}{Dependent Variable: $Downgrade_{i,t+1}$} \\
    \cmidrule(lr){2-10}
    VARIABLES & (1) & (2) & (3) & (4) & (5) & (6) & (7) & (8) & (9) \\
    \midrule
    $\widehat{Downgrade}_{RAW}$ & 0.061*** & 0.058*** & 0.070*** & & & & 0.073*** & 0.039** & 0.071*** \\
    & (5.515) & (4.782) & (5.905) & & & & (3.576) & (2.488) & (5.569) \\
    \addlinespace
    $\widehat{Downgrade}_{TRF}$ & & & & 0.047*** & 0.049*** & 0.046*** & -0.012 & 0.025* & -0.000 \\
    & & & & (4.799) & (4.321) & (4.225) & (-0.719) & (1.731) & (-0.040) \\
    \midrule
    LLM & 4o-mini & 4o & o3-mini & 4o-mini & 4o & o3-mini & 4o-mini & 4o & o3-mini \\
    Controls & Yes & Yes & Yes & Yes & Yes & Yes & Yes & Yes & Yes \\
    SIC3 $\times$ Quarter FE & Yes & Yes & Yes & Yes & Yes & Yes & Yes & Yes & Yes \\
    Observations & 2,243 & 2,243 & 2,243 & 2,243 & 2,243 & 2,243 & 2,243 & 2,243 & 2,243 \\
    Adj. $R^2$ & 0.246 & 0.237 & 0.260 & 0.232 & 0.233 & 0.229 & 0.246 & 0.241 & 0.260 \\
    \bottomrule
  \end{tabular}
\end{table}

\clearpage

\begin{table}[htbp]
  \centering
  \footnotesize
  \setlength{\tabcolsep}{4pt}
  \caption{{\bf Predictive Power of Earnings Call Sentiment for DGTW Returns}\\\\
  \footnotesize{This table presents regressions of DGTW-adjusted stock returns ($DGTW_{i,t+1}$) on sentiment measures extracted using GPT-4o-mini from earnings conference call transcripts. Following the matching convention described in the text, the dependent variable captures the market's reaction to the call: for calls held before the U.S. market close (16:00 ET) it is measured on the same trading day, and for calls held after the close on the next trading day, so that it fully incorporates the information disclosed in the transcript; we denote it $DGTW_{i,t+1}$. Panel A reports specifications without control variables, and Panel B adds the full set of controls. The control variables are collapsed into a single ``Controls'' row for brevity; they comprise analyst forecast error ($FE$), DGTW-adjusted returns over the run-up window ($DGTW_{i,t}$, $DGTW_{i,t-1}$, $DGTW_{i,t-2}$, $DGTW_{i,t-21,t-3}$, $DGTW_{i,t-252,t-22}$), and firm characteristics including the logarithm of firm size ($\ln(Size)$), book-to-market ratio ($\ln(BM)$), and turnover ($\ln(Turnover)$). In each panel, Columns (1) through (3) individually add the sentiment scores derived from the raw, TRF-anonymized, and LLM-anonymized transcripts. Columns (4) and (5) include both the sentiment extracted from the raw transcript and from an anonymized transcript. All variables are winsorized at the 1st and 99th percentiles. All independent variables are standardized. All regressions include date fixed effects. The sample covers post-cutoff earnings call transcripts (Nov 2023--Dec 2024), and the unit of observation is the earnings call. T-statistics based on time-clustered standard errors are reported in parentheses. * $p < 0.10$, ** $p < 0.05$, *** $p < 0.01$.}}
  \label{tab:earnings_call_sentiment}
  \vspace{0.2cm}
  \begin{tabular}{@{}lccccc@{}}
    \toprule
    & \multicolumn{5}{c}{Dependent Variable: $DGTW_{i,t+1}$} \\
    \cmidrule(lr){2-6}
    VARIABLES & (1) & (2) & (3) & (4) & (5) \\
    \midrule
    \multicolumn{6}{l}{\textit{Panel A: Without Controls}}\\
    \addlinespace
    $Sentiment_{RAW}$ & 2.549*** & & & 1.842*** & 1.833*** \\
     & (19.602) & & & (8.741) & (7.237) \\
    \addlinespace
    $Sentiment_{TRF}$ & & 2.413*** & & 0.820*** & \\
     & & (17.329) & & (3.766) & \\
    \addlinespace
    $Sentiment_{LLM}$ & & & 2.410*** & & 0.838*** \\
     & & & (17.521) & & (3.212) \\
    \addlinespace
    \midrule
    Controls & No & No & No & No & No \\
    Date FE & Yes & Yes & Yes & Yes & Yes \\
    Observations & 7,352 & 7,352 & 7,352 & 7,352 & 7,352 \\
    Adj. $R^2$ & 0.078 & 0.070 & 0.070 & 0.080 & 0.080 \\
    \midrule
    \multicolumn{6}{l}{\textit{Panel B: With Controls}}\\
    \addlinespace
    $Sentiment_{RAW}$ & 2.487*** & & & 1.822*** & 1.808*** \\
    & (18.832) & & & (8.475) & (7.234) \\
    \addlinespace
    $Sentiment_{TRF}$ & & 2.331*** & & 0.775*** & \\
    & & (17.287) & & (3.662) & \\
    \addlinespace
    $Sentiment_{LLM}$ & & & 2.329*** & & 0.798*** \\
    & & & (17.466) & & (3.247) \\
    \addlinespace
    \midrule
    Controls & Yes & Yes & Yes & Yes & Yes \\
    Date FE & Yes & Yes & Yes & Yes & Yes \\
    Observations & 7,352 & 7,352 & 7,352 & 7,352 & 7,352 \\
    Adj. $R^2$ & 0.132 & 0.124 & 0.124 & 0.133 & 0.133 \\
    \bottomrule
  \end{tabular}
\end{table}

\clearpage

\begin{table}[htbp]
  \centering
  \small
  \caption{{\bf Information Loss in Alternative Tasks}\\
  \footnotesize{This table compares the predictive content of textual scores extracted from RAW and TRF-anonymized earnings-call transcripts across four forward-looking outcomes. Each column reports a horse-race regression that includes both the RAW and TRF-anonymized scores. Column (1) examines realized post-announcement return volatility ($Vol_{post}$) using managerial uncertainty, controlling for pre-announcement volatility ($Vol_{pre}$), pre-announcement alpha ($Alpha_{pre}$), absolute abnormal return ($|Abnormal\ Ret|$), firm size ($\ln(Size)$), and book-to-market ratio ($\ln(BM)$). Column (2) predicts capital expenditures two quarters ahead ($Capx_{t+2}$) using investment expectations, controlling for current capital expenditures ($Capx_t$), market leverage ($Leverage_{Market}$), and firm size ($Total\ Asset$). Columns (3) and (4) predict two-quarter-ahead sales growth ($SaleChange_{t+2}$) and value-added growth ($ValueAddChange_{t+2}$), respectively, using firms' expressed macroeconomic outlook and controlling for current sales growth ($SaleChange_{t}$) or value-added growth ($ValueAddChange_{t}$), as well as book-to-market ratio ($BM$), firm size ($Total\ Asset$), and asset tangibility ($Tangibility$). The post-announcement window underlying $Vol_{post}$ is matched on the calendar (trading-day) window following each call, whereas $Capx_{t+2}$, $SaleChange_{t+2}$, and $ValueAddChange_{t+2}$ are matched on the firm's fiscal quarters following the call, ensuring that each outcome is realized strictly after, and is therefore genuinely predicted by, the earnings call. Detailed variable definitions are reported in Appendix Table~\ref{appendix-tab:variable_definitions_full}. All variables are winsorized at the 1st and 99th percentiles, and all independent variables are standardized. All regressions include date fixed effects. The sample covers post-cutoff earnings call transcripts (Nov 2023--Dec 2024). T-statistics based on time-clustered standard errors are reported in parentheses. * $p<0.10$, ** $p<0.05$, *** $p<0.01$.}}
  \label{tab:multitask}
  \vspace{0.2cm}
  \setlength{\tabcolsep}{6pt}
  \begin{tabular}{@{}lcccc@{}}
    \toprule
    & $Vol_{post}$ & $Capx_{t+2}$ & $SaleChange_{t+2}$ & $ValueAddChange_{t+2}$ \\
    \cmidrule(lr){2-5}
    VARIABLES & (1) & (2) & (3) & (4) \\
    \midrule
    $Uncertainty_{RAW}$ & 0.026** & & & \\
    & (2.028) & & & \\
    \addlinespace
    $Uncertainty_{TRF}$ & 0.012 & & & \\
    & (0.777) & & & \\
    \addlinespace
    $Investment_{RAW}$ & & 0.043*** & & \\
    & & (3.482) & & \\
    \addlinespace
    $Investment_{TRF}$ & & 0.033** & & \\
    & & (2.502) & & \\
    \addlinespace
    $Economy_{RAW}$ & & & 1.803*** & 2.460*** \\
    & & & (5.473) & (5.506) \\
    \addlinespace
    $Economy_{TRF}$ & & & 1.550*** & 1.787*** \\
    & & & (4.376) & (3.491) \\
    \midrule
    Controls & Yes & Yes & Yes & Yes \\
    Date FE & Yes & Yes & Yes & Yes \\
    Observations & 7,255 & 5,692 & 6,650 & 6,398 \\
    Adj. $R^2$ & 0.776 & 0.730 & 0.237 & 0.278 \\
    \bottomrule
  \end{tabular}
\end{table}

\clearpage

\begin{table}[htbp]
  \centering
  \small
    \caption{{\bf Predictive Power of News Headlines Sentiment for DGTW Returns}\\\\
    \footnotesize{This table examines whether information loss also affects news headlines. The dependent variable is the DGTW-adjusted stock return ($DGTW_{i,t+1}$). Following the same matching convention used for earnings calls, this return is measured on the same trading day for news released before the U.S. market close (16:00 ET) and on the next trading day for news released after the close, so that it fully incorporates the information in the headline. The sentiment scores are derived from RAW and TRF-anonymized headlines. Columns (1) and (2) show single-score regressions for the RAW and TRF-anonymized scores, respectively. Column (3) conducts a horse-race regression between the RAW and TRF-anonymized sentiment scores. Columns (4) through (6) replicate this analysis with control variables. The control variables are collapsed into a single ``Controls'' row for brevity; they comprise past returns over the run-up window ($DGTW_{i,t}$, $DGTW_{i,t-1}$, $DGTW_{i,t-2}$, $DGTW_{i,t-21,t-3}$, $DGTW_{i,t-252,t-22}$) and firm characteristics ($\ln(Size)$, $\ln(BM)$, $\ln(Turnover)$). All variables are winsorized at the 1st and 99th percentiles, and all independent variables are standardized. All regressions include date fixed effects. The sample comprises 52,716 Yahoo Finance headlines aggregated to 30,950 firm-day observations over the post-cutoff window (Nov 2023--Dec 2024); the unit of observation is the firm-day. T-statistics based on time-clustered standard errors are reported in parentheses. * $p < 0.10$, ** $p < 0.05$, *** $p < 0.01$.}}
    \label{tab:news_prediction}
    \vspace{0.2cm}
    \setlength{\tabcolsep}{4pt}
    \begin{tabular}{@{}lcccccc@{}}
      \toprule
      & \multicolumn{6}{c}{Dependent Variable: $DGTW_{i,t+1}$} \\
      \cmidrule(lr){2-7}
      VARIABLES & (1) & (2) & (3) & (4) & (5) & (6) \\
      \midrule
      $Sentiment_{RAW}^{News}$ & 0.444*** & & 0.315*** & 0.439*** & & 0.314*** \\
      & (17.477) & & (4.664) & (17.353) & & (4.633) \\
      \addlinespace
      $Sentiment_{TRF}^{News}$ & & 0.437*** & 0.135** & & 0.432*** & 0.131** \\
      & & (17.325) & (2.036) & & (17.212) & (1.974) \\
      \midrule
      Controls & No & No & No & Yes & Yes & Yes \\
      Date FE & Yes & Yes & Yes & Yes & Yes & Yes \\
      Observations & 30,950 & 30,950 & 30,950 & 30,950 & 30,950 & 30,950 \\
      Adj. $R^2$ & 0.016 & 0.015 & 0.016 & 0.016 & 0.016 & 0.016 \\
      \bottomrule
    \end{tabular}
\end{table}

\clearpage

\appendix
\counterwithin{table}{section}
\counterwithin{figure}{section}

\section{Supporting Derivations for the Framework}\label{app:proofs}
This appendix derives the restrictions implied by the maintained benchmarks in
Section~\ref{sec:framework}. Unless stated otherwise, variables are centered
and population-standardized.

\subsection{Modeling Information Loss}
Equation~\eqref{eq:decomposition} gives $u=A-\rho R$. Because $R$ and $A$
have unit variance and correlation $\rho$,
\[
 \Cov(R,u)=\Cov(R,A)-\rho\Var(R)=\rho-\rho=0,
 \qquad
 \Var(u)=1-\rho^2.
\]
Thus $u$ is the part of the anonymized score that does not move with the raw
score; this orthogonality is mechanical because $u$ is the population
projection residual. Under Assumption~\ref{assump:information_loss}, $u$ is
uncorrelated with the outcome and therefore reflects document-specific
variation rather than incremental predictive content.

Now use $Y=\mu+\beta R+\varepsilon$. In a regression with a single
standardized score, the slope is its covariance with $Y$. Therefore,
\[
 b_R=\Cov(Y,R)=\beta,
 \qquad
 b_A=\Cov(Y,A)=\rho\beta.
\]
The anonymized slope is the raw slope multiplied by $\rho$. Let $R_R^2$ and
$R_A^2$ denote the population coefficients of determination from the two
standalone linear projections, each including an intercept. Squaring the slope
attenuation gives the corresponding fit:
\[
 R_A^2=\rho^2R_R^2.
\]
Because $0<\rho<1$, $|b_A|<|b_R|$ whenever $\beta\neq0$; for $\beta>0$,
$0<b_A<b_R$. This proves the attenuation and lower-bound result in
Proposition~\ref{prop:attenuation}.

In a horse-race containing both $R$ and $A$, write $A=\rho R+u$. Once $R$ is
included, only $u$ is left to identify the coefficient on $A$. Under
Assumption~\ref{assump:information_loss}, $u$ is uncorrelated with the future
outcome, so the population coefficient on $A$ is zero and the coefficient on
$R$ is $\beta$, which proves Proposition~\ref{prop:horse_race}.

\subsection{Modeling Look-Ahead Bias}
Equation~\eqref{eq:cutoffregression} is fully interacted with the binary
indicator $P$. Its population coefficients therefore reproduce the two
within-period regressions: $d_1$ is the slope of $Y$ on $\widetilde S$ when
$P=0$, and $d_1+d_3$ is the corresponding slope when $P=1$.
Under Assumption~\ref{assump:stability}, the same positive scaling can be used
in both periods. After removing the within-period means, normalize the score
and outcome to unit variance and define
\[
 b_S=\Cov(Y,S\mid P=p)=\Corr(Y,S\mid P=p),
 \qquad
 \sigma_\varepsilon^2=\Var(\varepsilon\mid P=p),
\]
for $p\in\{0,1\}$, where $\varepsilon$ is the future innovation of
equation~\eqref{eq:outcome}, common to both scores. The score--outcome slope
and $\sigma_\varepsilon^2$ are stable across the cutoff.

For the raw score, equation~\eqref{eq:outcome} and
Assumption~\ref{assump:stability} give $b_R=\beta$. Because $Y$ and $R$ have
unit variance,
\[
 \sigma_\varepsilon^2
 =1-\beta^2
 =1-b_R^2>0.
\]
For either $S\in\{R,A\}$, Assumption~\ref{assump:stability} gives
$\Cov(S,\varepsilon\mid P=p)=0$, and equation~\eqref{eq:outcome} then implies
\[
 \Cov(Y,\varepsilon\mid P=p)
 =\beta\Cov(R,\varepsilon\mid P=p)+\Var(\varepsilon\mid P=p)
 =\sigma_\varepsilon^2.
\]

After the cutoff, $P=0$ and $\widetilde S=S$, so $d_1=b_S$. Before the cutoff,
$\widetilde S=S+\gamma_S\varepsilon$. Hence
\[
 \Cov(Y,\widetilde S\mid P=1)
 =b_S+\gamma_S\sigma_\varepsilon^2,
 \qquad
 \Var(\widetilde S\mid P=1)
 =1+\gamma_S^2\sigma_\varepsilon^2.
\]
The pre-cutoff slope is therefore
\[
 d_1+d_3
 =\frac{b_S+\gamma_S\sigma_\varepsilon^2}
        {1+\gamma_S^2\sigma_\varepsilon^2},
\]
and
\[
 d_3
 =\frac{\gamma_S\sigma_\varepsilon^2(1-b_S\gamma_S)}
        {1+\gamma_S^2\sigma_\varepsilon^2}.
\]
Because the score is positively oriented but not a perfect predictor,
$0<b_S<1$, and equation~\eqref{eq:lookaheadbias} implies
$0\leq\gamma_S\leq 1$ for either score. Hence, whenever $\gamma_S>0$,
$\gamma_S\leq 1<1/b_S$, so $1-b_S\gamma_S>0$.

There are therefore only two cases. If $\gamma_S=0$, then $d_3=0$. If
$0<\gamma_S\leq 1$, every term in the numerator is positive and the
denominator is positive, so $d_3>0$. Intuitively, the pre-cutoff score becomes
unusually predictive because it contains part of the future innovation itself.
These two cases establish Proposition~\ref{prop:cutoff}.

\subsection{The Anonymization Gap}
Equation~\eqref{eq:gap} defines
\[
 v=\frac{A}{\rho}-R=\frac{u}{\rho},
 \qquad
 \Gap=\lvert v\rvert,
\]
so the anonymized score can be written as
\[
 A=\rho(R+v).
\]
Suppose that $v$ is symmetric and independent of $R$. Conditional on
$\Gap=g$, we have $v^2=g^2$, while symmetry implies
$\E[v\mid\Gap=g]=0$. Independence from the standardized raw score further
implies
\[
 \E[R\mid\Gap=g]=0,\qquad
 \E[R^2\mid\Gap=g]=1,\qquad
 \E[Rv\mid\Gap=g]=0.
\]
These conditions and Assumption~\ref{assump:information_loss} also imply
\[
 \E[A\mid\Gap=g]=0,
 \qquad
 \E[Y\mid\Gap=g]=\mu.
\]
These relations make the conditional variance of $A$ particularly simple.
Because its conditional mean is zero,
\[
\begin{aligned}
 \Var(A\mid\Gap=g)
 &=\rho^2\E[(R+v)^2\mid\Gap=g]\\
 &=\rho^2\left(
     \E[R^2\mid\Gap=g]
     +2\E[Rv\mid\Gap=g]
     +\E[v^2\mid\Gap=g]\right)\\
 &=\rho^2(1+g^2).
\end{aligned}
\]
Next use the outcome equation $Y=\mu+\beta R+\varepsilon$. The condition
$\E[\varepsilon\mid R,A]=0$ removes the covariance between the future
innovation and the anonymized score. Hence
\[
\begin{aligned}
 \Cov(Y,A\mid\Gap=g)
 &=\beta\Cov\!\left(R,\rho(R+v)\mid\Gap=g\right)\\
 &=\rho\beta.
\end{aligned}
\]
The conditional regression slope of $Y$ on $A$ is therefore
\[
 b(g)
 =\frac{\Cov(Y,A\mid\Gap=g)}
        {\Var(A\mid\Gap=g)}
 =\frac{\beta}{\rho(1+g^2)},
\]
which is equation~\eqref{eq:conditionalslope}. For a positively oriented
signal, $\beta>0$, and
\[
 b'(g)=-\frac{2\beta g}{\rho(1+g^2)^2}<0
 \qquad\text{for }g>0.
\]
The covariance with the outcome is unchanged across gap levels, but the
variance of the anonymized score increases with $g$. Under the information-loss
interpretation, the signal therefore becomes less predictive as
the document-specific residual grows.

Finally, equation~\eqref{eq:gapregression} approximates the conditional slope
$b(g)$ by the linear function $c_1+c_3g$. The interaction coefficient is the
slope from a weighted linear fit of $b(\Gap)$ on $\Gap$, with positive weight
$w(g)=\Var(A\mid\Gap=g)$. Equivalently,
\[
 c_3
 =\frac{\Cov_w(\Gap,b(\Gap))}
        {\Var_w(\Gap)}.
\]
Because $b(g)$ is decreasing, the weighted covariance in the numerator is
negative, while the denominator is positive whenever the gap varies across
documents. Thus $c_3<0$, which proves the conditional claim in
Proposition~\ref{prop:gap} and supplies the empirical prediction evaluated in
the paper.

\clearpage

\section{Supplementary Evidence and Results}\label{app:supplementary_results}
\subsection{Additional Empirical Results}
\begin{figure}[htbp]
  \centering
  \begin{subfigure}[b]{0.48\textwidth}
    \centering
    \includegraphics[width=\textwidth]{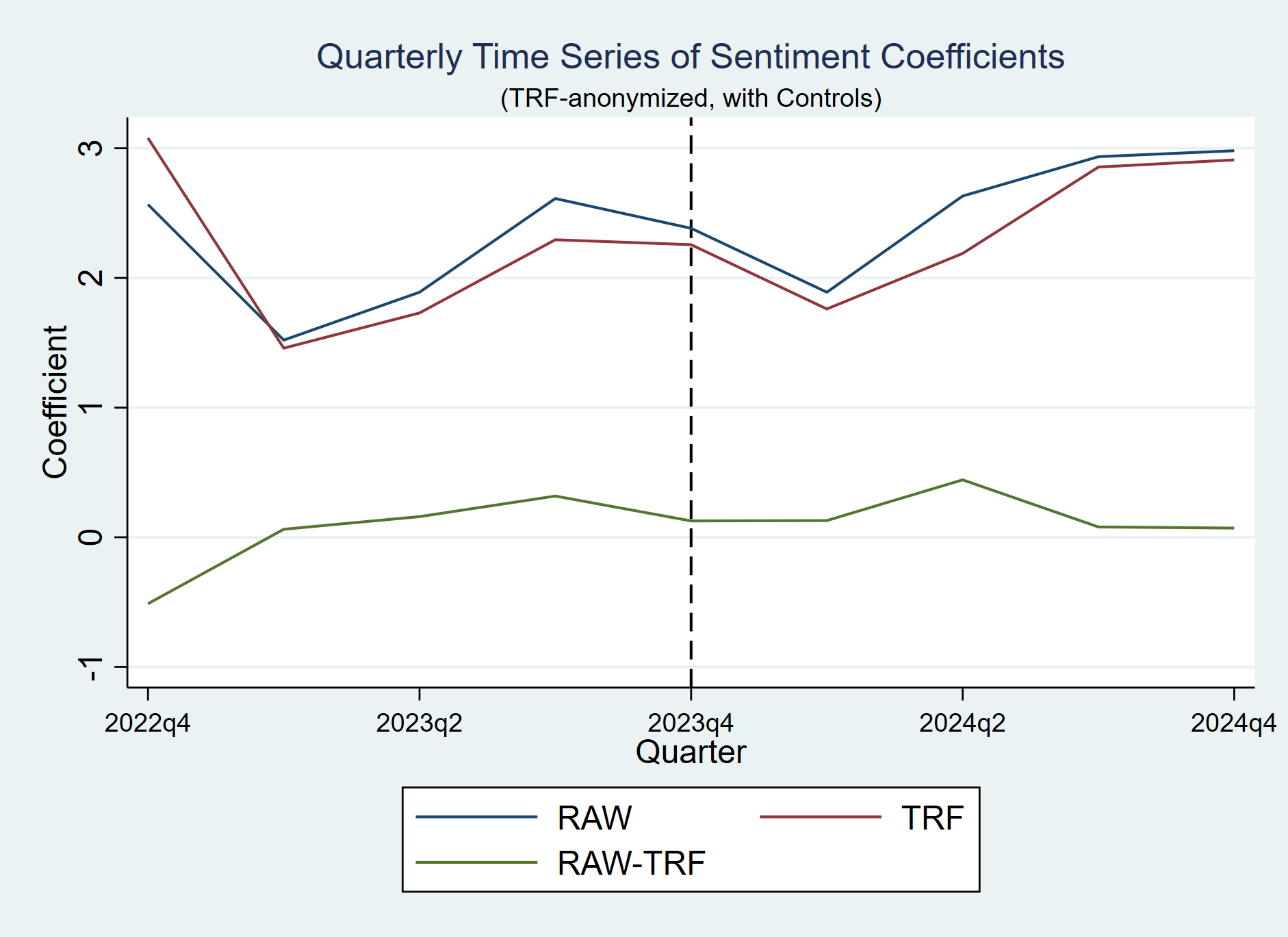}
    \caption{\centering Quarterly Sentiment Coefficients \\ (RAW and TRF-anonymized)}
    \label{appendix-fig:sentiment_trf}
  \end{subfigure}
  \hfill
  \begin{subfigure}[b]{0.48\textwidth}
    \centering
    \includegraphics[width=\textwidth]{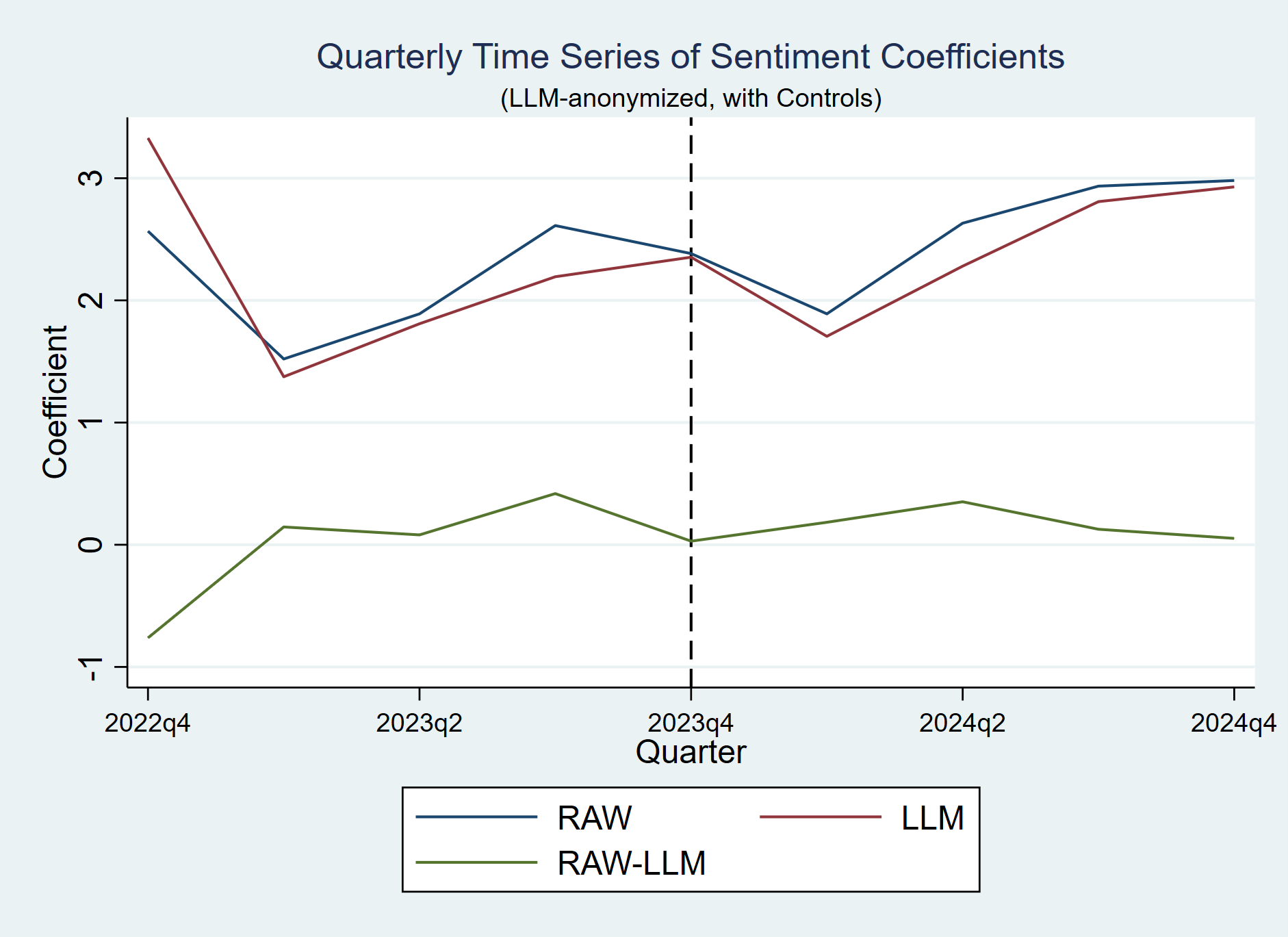}
    \caption{\centering Quarterly Sentiment Coefficients \\ (RAW and LLM-anonymized)}
    \label{appendix-fig:sentiment_llm}
  \end{subfigure}
  \caption{{\bf Quarterly Time Series of Sentiment Coefficients}\\\\
  \footnotesize{Each subfigure plots three quarterly time series: the quarterly regression coefficient of $DGTW_{i,t+1}$ on $Sentiment_{RAW}$ (dark blue), on the anonymized counterpart (red), and the difference between the two coefficients (green). Sentiment scores are extracted using GPT-4o-mini from different versions of earnings call transcripts, and each quarterly regression follows the specification in Panel B of Table~\ref{tab:earnings_call_sentiment}. Subfigure~\ref{appendix-fig:sentiment_trf} compares the coefficients derived from the RAW transcripts against those from the TRF-anonymized samples. Subfigure~\ref{appendix-fig:sentiment_llm} compares the coefficients from the RAW transcripts against those from the LLM-anonymized samples. The sample is the balanced panel of firms with valid observations in both the pre-cutoff (Nov 2022--Oct 2023) and post-cutoff (Nov 2023--Dec 2024) windows. The vertical dashed line marks the LLM knowledge-cutoff (2023 Q4).}}
  \label{appendix-fig:sentiment_quarterly_beta}
\end{figure}

\clearpage
\begingroup
\singlespacing
\footnotesize
\renewcommand{\arraystretch}{1.18}
\setlength{\tabcolsep}{3.2pt}

\begin{longtable}{@{}>{\raggedright\arraybackslash}p{0.25\linewidth}>{\raggedright\arraybackslash}p{0.15\linewidth}>{\raggedright\arraybackslash}p{0.20\linewidth}>{\raggedright\arraybackslash}p{0.22\linewidth}>{\raggedright\arraybackslash}p{0.12\linewidth}@{}}
\caption{{\bf Selected Applications of Anonymization to Address Look-Ahead Bias}\\
\footnotesize{This table summarizes selected finance and accounting studies that use text anonymization to mitigate or assess look-ahead bias. For each study, the table reports the publication outlet or working-paper series, research focus, anonymization approach, and the role of anonymization in the empirical design. The table includes only studies that both implement an anonymization procedure and explicitly relate that procedure to look-ahead bias. We classify each study according to the broadest transformation applied to the text. \textit{Identifier-level anonymization} removes, replaces, or perturbs prespecified explicit identifiers or temporal anchors while largely preserving the remaining wording and sentence structure. Parenthetical labels identify the targeted information: \textit{agents} covers firm, organization, and person identifiers; \textit{products} covers product, brand, and service names; and \textit{time} covers dates, years, and other temporal anchors. \textit{Context-level anonymization} additionally masks indirect contextual clues and rewrites distinctive wording or sentence structure to reduce the recoverability of the source, subject, or time. Studies using both types of transformation are classified as context-level. These categories describe the scope of the transformation rather than the implementation tool and do not imply guaranteed anonymity. The Role column indicates whether anonymization is used in the \textit{main analysis} or \textit{robustness check}. ``Source'' reports the current publication outlet or working-paper series, which may differ from the version supplied for review.}}
\label{appendix-tab:prior_anonymization_applications}\\
\toprule
\textbf{Study and title} & \textbf{Source} & \textbf{Research focus} & \textbf{Anonymization approach} & \textbf{Role} \\
\midrule
\endfirsthead
\multicolumn{5}{l}{\textit{Appendix Table~\thetable\ continued}}\\
\toprule
\textbf{Study and title} & \textbf{Source} & \textbf{Research focus} & \textbf{Anonymization approach} & \textbf{Role} \\
\midrule
\endhead
\midrule
\multicolumn{5}{r}{\textit{Continued on next page}}\\
\endfoot
\bottomrule
\endlastfoot
\citet{GlobalBusinessNet}\newline
\textit{Global Business Networks}
& \textit{Journal of Financial Economics}
& Global business networks, return lead--lag effects, and M\&A prediction
& Identifier-level\newline anonymization\newline (agents, products)
& Robustness check \\
\addlinespace[0.80em]
\citet{glasserman2023assessinglookaheadbiasstock}\newline
\textit{Assessing Look-Ahead Bias in Stock Return Predictions Generated by GPT Sentiment Analysis}
& \textit{The Journal of Financial Data Science}
& Financial-news sentiment and short-horizon stock returns
& Identifier-level\newline anonymization\newline (agents, products)
& Main analysis \\
\addlinespace[0.80em]
\citet{jha2025chatgptcorporatepolicies}\newline
\textit{ChatGPT and Corporate Policies}
& NBER Working Paper No.~32161
& Managerial investment plans and subsequent corporate investment
& Identifier-level\newline anonymization\newline (agents, products, time)
& Robustness check \\
\addlinespace[0.80em]
\citet{jha2025harnessinggenerativeaieconomic}\newline
\textit{Generative AI, Managerial Expectations, and Economic Activity}
& SSRN Working Paper No.~4976759
& Managerial expectations and aggregate, industry, and firm economic activity
& Context-level\newline anonymization
& Robustness check \\
\addlinespace[0.80em]
\citet{KakhbodKermaniMaciel2026}\newline
\textit{In the Fed's Mind}
& NBER Working Paper No.~35016
& FOMC inflation narratives, monetary policy, and asset prices
& Context-level\newline anonymization
& Robustness check \\
\addlinespace[0.80em]
\citet{engelberg2025entity}\newline
\textit{Entity Neutering}
& SSRN Working Paper No.~5182756
& Methodological study of financial-news sentiment and return prediction
& Context-level\newline anonymization
& Main analysis \\
\addlinespace[0.80em]
\citet{GokkayaLiuStulz2025}\newline
\textit{Is There Information in Corporate Acquisition Plans?}
& NBER Working Paper No.~32201
& Acquisition-plan disclosures and subsequent M\&A activity
& Identifier-level\newline anonymization\newline (agents, products, time)
& Robustness check \\
\addlinespace[0.80em]
\citet{Breitung2026Text}\newline
\textit{Text Is All You Need: Asset Pricing Without Returns}
& SSRN Working Paper No.~5616350
& IPO risk disclosures, beta estimation, and initial mispricing
& Identifier-level\newline anonymization\newline (agents)
& Main analysis \\
\addlinespace[0.80em]
\citet{chen2022expected}\newline
\textit{Expected Returns and Large Language Models}
& SSRN Working Paper No.~4416687
& Financial news and next-day cross-sectional stock returns
& Identifier-level\newline anonymization\newline (agents, products, time)
& Robustness check \\
\addlinespace[0.80em]
\citet{GaoXiongYuan2026Beliefs}\newline
\textit{Beliefs That Predict Returns and Beliefs That Attract Flows: Policy Insights and Sentiment Catering in Mutual Funds}
& NBER Working Paper No.~35528
& Mutual-fund beliefs, market timing, and sentiment catering
& Identifier-level\newline anonymization\newline (time)
& Robustness check \\
\end{longtable}
\endgroup

\clearpage

\begin{table}[htbp]
  \centering
  \caption{{\bf Sample Selection Criteria}\\\\
  \footnotesize{This table outlines the filtering procedures used to construct the final samples. Panel A details the selection process for the downgrade-prediction post-cutoff subsample, Panel B for earnings call transcripts, and Panel C for news headlines. Each row sequentially applies a filter, reporting the resulting sample size and the number of observations removed. For the downgrade subsample in Panel A, we additionally exclude utilities firms (SIC 4000--4999) and financial firms (SIC 6000--6299), following standard practice in the credit risk literature \citep{donovan2021measuring}, require non-missing controls, and require an S\&P credit rating observed before the earnings call.}}
  \label{appendix-tab:Sample Selection Criteria}
  \footnotesize
   \begin{tabular}{p{10cm}rr}
    \toprule
    \textbf{Panel A: Downgrade Prediction Subsample (Post-Cutoff)} & & \\
    \midrule
    \textbf{Filter} & \textbf{Sample Size} & \textbf{Obs. Removed} \\
    \midrule
    Earnings call transcripts from Seeking Alpha for iShares Russell 3000 ETF constituents (Nov 2023 - Dec 2024) & 10,812 & \\
    Remove duplicates, match dates and PERMNOs, remove errors & 10,628 & 184 \\
    Remove transcripts with $>$ 15,000 tokens & 10,529 & 99 \\
    Drop utilities and financial firms (SIC 4000--4999, 6000--6299) & 8,566 & 1,963 \\
    Require non-missing controls and an S\&P credit rating observed before the earnings call & 2,488 & 6,078 \\
    \midrule
    \textbf{Panel B: Earnings Call Transcripts} & & \\
    \midrule
    \textbf{Filter} & \textbf{Sample Size} & \textbf{Obs. Removed} \\
    \midrule
    Earnings call transcripts from Seeking Alpha for iShares Russell 3000 ETF constituents (Nov 2023 - Dec 2024) & 10,812 & \\
    Remove duplicates, match dates and PERMNOs, remove errors & 10,628 & 184 \\
    Remove transcripts with $>$ 15,000 tokens & 10,529 & 99 \\
    Match with DGTW returns (Nov 2023 - Dec 2024) & 8,479 & 2,050 \\
    Match with analyst EPS forecast errors & 7,596 & 883 \\
    Removing observations with missing controls & 7,382 & 214 \\
    \midrule
    \textbf{Panel C: News Headlines} & & \\
    \midrule
    \textbf{Filter} & \textbf{Sample Size} & \textbf{Obs. Removed} \\
    \midrule
    Firm news headlines from Finnhub for iShares Russell 3000 ETF constituents (Nov 2023 - Dec 2024) & 1,612,948 & \\
    Remove duplicates, require valid dates and non-missing headlines & 969,077 & 643,871 \\
    Retain headlines from Yahoo, exclude ``stock" or ``stocks" & 271,194 & 697,883 \\
    Retain headlines with a single organization mention & 118,797 & 152,397 \\
    Remove weekend news and match with PERMNO & 112,835 & 5,962 \\
    Aggregate by PERMNO and date & 80,337 & 32,498 \\
    Match with DGTW returns (Nov 2023 - Dec 2024) & 57,603 & 22,734 \\
    Removing observations with missing controls & 30,950 & 26,653 \\
    \bottomrule
  \end{tabular}
\end{table}

\clearpage
\renewcommand{\arraystretch}{0.85}
\setlength{\parskip}{0.8pt}
\setlength{\parindent}{0.8pt}
\linespread{1}\selectfont

\begin{longtable}{@{}p{4.5cm}p{\dimexpr\linewidth-4.5cm-\tabcolsep\relax}@{}}
\caption{\textbf{Variable Definitions}\\\\
{\footnotesize This table defines the variables used in our analysis. Panel A lists measures derived from raw and anonymized earnings call transcripts and news headlines; the extraction prompts are reported in Appendix~\ref{app:prompts}. Panel B lists the dependent variables. Panel C lists the control variables. Unless otherwise specified, stock return data are from CRSP, and firm-level accounting data are from Compustat.}}
\label{appendix-tab:variable_definitions_full}\\
\toprule
\textbf{Variable} & \textbf{Definition} \\
\midrule
\endfirsthead
\multicolumn{2}{c}{\tablename\ \thetable{} -- Continued from previous page} \\[1.5ex]
\toprule
\textbf{Variable} & \textbf{Definition} \\
\midrule
\endhead
\midrule
\multicolumn{2}{r}{Continued on next page} \\
\endfoot
\bottomrule
\addlinespace[2ex]
\endlastfoot
\multicolumn{2}{l}{\textbf{Panel A: LLM Measures}} \\[0.2ex]
\midrule
$\widehat{Downgrade}_{LLM}$ & The predicted probability of an S\&P credit rating downgrade within the 12 months following the earnings call, derived from LLM-anonymized earnings call transcripts.\\[0.2ex]
$\widehat{Downgrade}_{RAW}$ & The predicted probability of an S\&P credit rating downgrade within the 12 months following the earnings call, derived from raw earnings call transcripts.\\[0.2ex]
$\widehat{Downgrade}_{TRF}$ & The predicted probability of an S\&P credit rating downgrade within the 12 months following the earnings call, derived from TRF-anonymized earnings call transcripts.\\[0.2ex]
$\widetilde C_j^v$ & Winsorized and standardized version of the researcher-constructed S\&P-aligned ten-factor composite $C_j^v$ in equation~\eqref{eq:sp10_composite}, for transcript $j$ and text version $v\in\{RAW,TRF,LLM\}$. The composite is the mean score across factors for which the model returns both a score and a supporting quote. It is not an official S\&P assessment or rating.\\[0.2ex]
$\Gap_{v}$ & Anonymization gap for method $v\in\{TRF,LLM\}$, defined as $\Gap_v=\lvert A_v/\rho_v-R\rvert$, where $\rho_v=\Corr(R,A_v)$ is estimated in the pooled balanced sample. The gap is winsorized and standardized before entering the regressions.\\[0.2ex]
$\mathrm{SharedRaw}$ & Mean raw factor score over shared reasons: factors for which the raw transcript and the anonymized transcript both return evidence and cite matched normalized quote anchors.\\[0.2ex]
$\mathrm{SharedAnon}$ & Mean anonymized factor score over the same shared reasons. Comparing this measure with $\mathrm{SharedRaw}$ captures interpretive distortion.\\[0.2ex]
$\mathrm{RawOnly}$ & Mean raw factor score over RAW-only reasons, comprising raw evidence that is absent after anonymization and the raw side of factors whose quote anchor changes.\\[0.2ex]
$\mathrm{AnonOnly}$ & Mean anonymized factor score over anonymized-only reasons, defined symmetrically. Comparing this measure with $\mathrm{RawOnly}$ captures evidence displacement.\\[0.2ex]
$Economy_{RAW}$ & Firm's view on macroeconomic trends of the original earnings call transcript. It is defined as a classification of the firm's anticipated change in optimism about the US economy over the next quarter. The score is categorized into one of five levels (Increase substantially, increase, no change, decrease, decrease substantially) based on the prompt derived from \citet{jha2025harnessinggenerativeaieconomic}. \\[0.2ex]
$Economy_{TRF}$ & Firm's view on macroeconomic trends of the TRF-anonymized earnings call transcript. \\[0.2ex]
$Investment_{RAW}$ & Firm's investment score of the original earnings call transcript. It is defined as a classification of the firm's planned capital spending changes over the next year. The score is categorized into one of five levels (Increase substantially, increase, no change, decrease, decrease substantially) based on the prompt derived from \citet{jha2025chatgptcorporatepolicies}. \\[0.2ex]
$Investment_{TRF}$ & Firm's investment score of the TRF-anonymized earnings call transcript. \\[0.2ex]
$Sentiment_{AGT}$ & Sentiment of the transcript where only AGENTS entity types are preserved while all other entity categories are anonymized by the TRF model. \\[0.2ex]
$Sentiment_{LLM}$ & Sentiment of the earnings call transcript text after anonymization using GPT-4o-mini. \\[0.2ex]
$Sentiment_{NUM}$ & Sentiment of the transcript where only NUMBERS entity types are preserved while all other entity categories are anonymized by the TRF model. \\[0.2ex]
$Sentiment_{OTH}$ & Sentiment of the transcript where only OTHERS entity types are preserved while all other entity categories are anonymized by the TRF model. \\[0.2ex]
$Sentiment_{PLC}$ & Sentiment of the transcript where only PLACES entity types are preserved while all other entity categories are anonymized by the TRF model. \\[0.2ex]
$Sentiment_{RAW}$ & Sentiment of the original earnings call transcript text. It is defined as a measure of its positive or negative implication for stock returns. The sentiment score, which ranges from -1 to 1, is computed with the formula Sentiment = (2 * Direction - 1) * Magnitude, where Direction captures the binary tone and Magnitude captures its intensity. \\[0.2ex]
$Sentiment_{RAW}^{News}$ & Sentiment of original Yahoo news headlines, which is defined as a measure of its positive or negative implication for stock returns. The sentiment score, which ranges from -1 to 1, is computed with the formula Sentiment = (2 * Direction - 1) * Magnitude, where Direction captures the binary tone and Magnitude captures its intensity. \\[0.2ex]
$Sentiment_{TRF}$ & Sentiment of the earnings call transcript text after anonymization using the spaCy \texttt{en\_core\_web\_trf} (TRF) model. \\[0.2ex]
$Sentiment_{TRF}^{News}$ & Sentiment of the TRF-anonymized Yahoo news headlines. \\[0.2ex]
$Uncertainty_{RAW}$ & Uncertainty score of the original earnings call transcript. It is defined as a measure of the level of imprecision and cautious tone conveyed in the text. The score, which ranges from 0 (absolute certainty) to 1 (maximum uncertainty), is extracted based on the prompt derived from \citet{WhenIsaLiability}. \\[0.2ex]
$Uncertainty_{TRF}$ & Uncertainty score of the TRF-anonymized earnings call transcript. \\[0.2ex]
\midrule
\multicolumn{2}{l}{\textbf{Panel B: Dependent Variables}} \\[0.2ex]
\midrule
$Capx_{t+2}$ & Capital expenditure (CAPX) at the end of quarter $t+2$, scaled by total assets. \\[0.2ex]
$DGTW_{i,t+1}$ & DGTW-adjusted stock return that incorporates the information disclosed in the earnings call or news, defined following \citet{DGTW1997}. Under the matching convention in the text, it is measured on the same trading day for events occurring before the U.S. market close (16:00 ET) and on the next trading day for events after the close. \\[0.2ex]
$Downgrade_{i,t+1}$ & An indicator variable equaling 1 if S\&P downgrades the firm's credit rating within the following 12 months from the earnings call release, and 0 otherwise.\\[0.2ex]
$SaleChange_{t+2}$ & Defined as $\log(Sale_{t+2}/Sale_{t})$; represents two-quarter-ahead sales growth. \\[0.2ex]
$ValueAddChange_{t+2}$ & Defined as $\log(ValueAdd_{t+2}/ValueAdd_{t})$; represents two-quarter-ahead value-added growth. \\[0.2ex]
$Vol_{post}$ & Standard deviation of stock returns over the post-announcement window [0, 22], requiring at least 10 daily observations. \\[0.2ex]
\midrule
\multicolumn{2}{l}{\textbf{Panel C: Control Variables}} \\[0.2ex]
$\lvert \text{Abnormal Ret} \rvert$ & Absolute abnormal return, measured as the two-day (day 0 to +1) buy-and-hold return minus the buy-and-hold return of the CRSP value-weighted index. \\[0.2ex]
$Alpha_{pre}$ & Alpha from a market model estimated over the window [-252, -6], requiring at least 60 daily observations. \\[0.2ex]
$BM$ & Book-to-market ratio at the end of June, calculated as book equity (fiscal year-end) divided by market equity (fiscal year-end). \\[0.2ex]
$Capx_t$ & Capital expenditure (CAPX) at the end of quarter $t$, scaled by total assets. \\[0.2ex]
$DGTW_{i,t}$ & DGTW-adjusted stock return on event day 0. \\[0.2ex]
$DGTW_{i,t-1}$ & DGTW-adjusted stock return on event day -1. \\[0.2ex]
$DGTW_{i,t-2}$ & DGTW-adjusted stock return on event day -2. \\[0.2ex]
$DGTW_{i,t-21, t-3}$ & DGTW-adjusted stock return over the event window [-21, -3]. \\[0.2ex]
$DGTW_{i,t-252, t-22}$ & DGTW-adjusted stock return over the event window [-252, -22]. \\[0.2ex]
$FE$ & Analyst forecast error, defined as \citet{HIRSHLEIFER2009}. \\[0.2ex]
$FirmValue$ & Market value of equity plus the book value of debt.
\\[0.2ex]
$Leverage_{Asset}$ & Total debt scaled by total assets ((dlcq + dlttq)/atq).\\[0.2ex]
$Leverage_{Market}$ & Financial leverage, calculated as total debt (dlttq + dlcq) divided by total debt plus market value of equity. \\[0.2ex]
$MTB$ & Market value of equity scaled by book value of equity ((prccq * cshoq) / ceqq).\\[0.2ex]
$Oscore$ & O-Score measure of credit risk following \citet{ohlson1980financial}.\\[0.2ex]
$RevGrowth$ & percentage changes in revenue (revtq) from the same quarter, prior year.\\[0.2ex]
$ROA$ & Income before extraordinary items scaled by total assets (ibq / atq).\\[0.2ex]
$Size$ & Market capitalization at the end of June, calculated as CRSP price times shares outstanding. This equity-size measure is used in the non-downgrade regressions. \\[0.2ex]
$Std(Income)$ & Standard deviation of quarterly income before extraordinary items measured over the previous twenty quarters.\\[0.2ex]
$Std(Revenue)$ & Standard deviation of total revenues measured over the previous twenty quarters.\\[0.2ex]
$Tangibility$ & Plant, Property, and Equipment (PPENTQ) scaled by total assets (atq) at the end of the quarter. \\[0.2ex]
$Total\ Asset$ & Logarithm of quarterly total assets from Compustat (atq). \\[0.2ex]
$Turnover$ & Share turnover at the end of June, calculated as prior year's trading volume divided by shares outstanding. \\[0.2ex]
$Vol_{pre}$ & Standard deviation of stock returns over the pre-announcement window [-252, -1], requiring at least 60 daily observations. \\[0.2ex]
$Zscore$ & Z-Score measure of bankruptcy risk following \citet{altman1968financial}.\\[0.2ex]
\end{longtable}
\renewcommand{\arraystretch}{1.0}
\setlength{\parskip}{\parskip}
\setlength{\parindent}{\parindent}
\linespread{1.0}\selectfont

\clearpage
\begin{landscape}

\begin{table}[htbp]
  \scriptsize
  \centering
  \caption{{\bf Entity Categories}\\\\
  \footnotesize{This table presents the 18 mutually exclusive entity types identified by the \texttt{en\_core\_web\_trf} NER model from Python's spaCy library. The entity types are grouped into four primary categories: NUMBERS, PLACES, AGENTS, and OTHERS. For each entity type, the table reports its description, an illustrative example, and its total frequency in the S\&P-rated post-cutoff downgrade sample of 2,488 earnings call transcripts. Frequency per Transcript reports the mean number of occurrences in each category per transcript. The last two columns report the total frequency of each entity type in the 52,716 news headlines and the mean number of occurrences in each category per headline.}}
  \label{appendix-tab:Entity Category}
  \vspace{0.2cm}
  \setlength{\tabcolsep}{4pt}
  \renewcommand{\arraystretch}{1.15}
  \begin{tabular}{@{}p{2.0cm}p{2.5cm}p{6cm}p{4.0cm}p{1.6cm}p{1.6cm}p{1.6cm}p{1.6cm}@{}}
    \toprule
    Category & Entity Type & Description & Example & \parbox{1.5cm}{\centering Total\\Frequency \\ (Transcript)} & \parbox{1.5cm}{\centering Frequency per\\Transcript} & \parbox{1.5cm}{\centering Total\\Frequency \\ (Headlines)} & \parbox{1.5cm}{\centering Frequency per\\ Headline}\\
    \midrule
    \multirow{7}{*}{NUMBERS}
      & DATE & Absolute or relative dates or periods & Q1 2024 & 463,369 & \multirow{7}{*}{325.11} & 21,862 & \multirow{7}{*}{0.678} \\
      & CARDINAL & Numerals that do not fall under another type & one & 96,713 & & 4,670 &\\
      & MONEY & Monetary values, including unit & \$69.7 billion & 91,265 &  & 5,419 &\\
      & PERCENT & Percentage, including ``\%'' & 6\% & 89,442 &  & 2,322 &\\
      & ORDINAL & ``first,'' ``second,'' etc. & first & 32,485 &  & 1,214 &\\
      & TIME & Times smaller than a day & evening & 29,063 &  & 120 &\\
      & QUANTITY & Measurements, as of weight or distance & 100-foot & 6,532 & & 142 & \\
    \midrule
    \multirow{3}{*}{PLACES}
      & GPE & Countries, cities, states & Korea & 47,021 & \multirow{3}{*}{30.74} & 6,528 & \multirow{3}{*}{0.146}\\
      & LOC & Non-GPE locations, mountain ranges, bodies of water & the Middle East & 24,996 &  & 773 &\\
      & FAC & Buildings, airports, highways, bridges, etc. & San Ciprián & 4,468 & & 384 & \\
    \midrule
    \multirow{3}{*}{AGENTS}
      & PERSON & People, including fictional & Tim Cook & 277,546 & \multirow{3}{*}{197.36} & 5,741 & \multirow{3}{*}{1.125}\\
      & ORG & Firms, agencies, institutions, etc. & Apple Inc. & 207,753 & & 52,714 & \\
      & NORP & Nationalities or religious or political groups & Chinese & 5,731 & & 863 & \\
    \midrule
    \multirow{5}{*}{OTHERS}
      & PRODUCT & Objects, vehicles, foods, etc. (not services) & Watch Ultra & 36,190 & \multirow{5}{*}{17.69} & 3,100 & \multirow{5}{*}{0.084}\\
      & EVENT & Named hurricanes, battles, wars, sports events, etc. & Super Bowl & 4,276 & & 834 & \\
      & LAW & Named documents made into laws & the IRA Inflation Reduction Act & 2,428 & & 76 & \\
      & WORK\_OF\_ART & Titles of books, songs, etc. & America's Greatest Workplaces 2024 & 1,069 & & 395 & \\
      & LANGUAGE & Any named language & English & 40 & & 7 & \\
    \bottomrule
  \end{tabular}
\end{table}
\end{landscape}

\clearpage

\begin{table}[htbp]
  \centering
  \small
  \caption{{\bf Three-Month Forecast Horizon and the Predictive Power of Downgrade Estimates}\\\\
  \footnotesize{This table reproduces the knowledge-cutoff interaction analysis in Table~\ref{tab:downgrade_pre_interaction} using a three-month rather than a twelve-month downgrade horizon. The dependent variable, $Downgrade^{3M}_{i,t+1}$, is an indicator equal to one if S\&P downgrades the firm's credit rating within three months after the earnings call and zero otherwise. $\widehat{Downgrade}^{3M}_{RAW}$, $\widehat{Downgrade}^{3M}_{TRF}$, and $\widehat{Downgrade}^{3M}_{LLM}$ denote the predicted probabilities of a downgrade within the following three months, extracted from raw, TRF-anonymized, and LLM-anonymized transcripts, respectively. The scores are constructed using the same prompts as in the baseline analysis, except that the forecast horizon is changed from twelve months to three months. $Pre$ is an indicator equal to one if the transcript falls in the pre-cutoff window (Nov. 2022--Oct. 2023) and zero if it falls in the post-cutoff window (Nov. 2023--Dec. 2024). The sample is the same balanced panel of firms used in Table~\ref{tab:downgrade_pre_interaction}. All variables are winsorized at the 1st and 99th percentiles, and all independent variables are standardized except for $Pre$. Control variables are identical to those in Panel B of Table~\ref{tab:downgrade_main}. All regressions include SIC3 $\times$ Quarter fixed effects. T-statistics based on SIC3-clustered standard errors are reported in parentheses. * $p<0.10$, ** $p<0.05$, *** $p<0.01$.}}
  \label{appendix-tab:downgrade_pre_interaction_3m}
  \vspace{0.2cm}
  \setlength{\tabcolsep}{8pt}
  \begin{tabular}{@{}lccc@{}}
    \toprule
    & \multicolumn{3}{c}{Dependent Variable: $Downgrade^{3M}_{i,t+1}$} \\
    \cmidrule(lr){2-4}
    VARIABLES & (1) & (2) & (3) \\
    \midrule
    $\widehat{Downgrade}^{3M}_{RAW}$
        & 0.026*** & & \\
        & (4.521) & & \\
    \addlinespace
    $\widehat{Downgrade}^{3M}_{RAW}\times Pre$
        & -0.006 & & \\
        & (-0.684) & & \\
    \addlinespace
    $\widehat{Downgrade}^{3M}_{TRF}$
        & & 0.023*** & \\
        & & (3.857) & \\
    \addlinespace
    $\widehat{Downgrade}^{3M}_{TRF}\times Pre$
        & & -0.010 & \\
        & & (-1.299) & \\
    \addlinespace
    $\widehat{Downgrade}^{3M}_{LLM}$
        & & & 0.020*** \\
        & & & (3.712) \\
    \addlinespace
    $\widehat{Downgrade}^{3M}_{LLM}\times Pre$
        & & & -0.002 \\
        & & & (-0.300) \\
    \midrule
    Controls & Yes & Yes & Yes \\
    SIC3 $\times$ Quarter FE & Yes & Yes & Yes \\
    Observations & 3,943 & 3,943 & 3,943 \\
    Adj. $R^2$ & 0.103 & 0.097 & 0.097 \\
    \bottomrule
  \end{tabular}
\end{table}

\clearpage

\begin{table}[htbp]
  \centering
  \small
    \caption{{\bf Information Loss of Alternative LLMs}\\\\
    \footnotesize{This table compares sentiment scores generated by three distinct models: GPT-4o-mini, GPT-4o, and o3-mini. All regressions include the full set of control variables and use the DGTW-adjusted stock return ($DGTW_{i,t+1}$) as the dependent variable. Following the matching convention in the text, the dependent variable is measured on the same trading day for calls before the U.S. market close (16:00 ET) and on the next trading day for calls after the close. Columns (1) through (3) report the results for sentiment extracted from the raw transcripts by each LLM. Columns (4) through (6) report the results for sentiment extracted from the TRF-anonymized text. Columns (7) through (9) conduct horse-race regressions for each LLM-extracted signal. Control variables are identical to those in Panel B of Table~\ref{tab:earnings_call_sentiment}. All variables are winsorized at the 1st and 99th percentiles. All independent variables are standardized. All regressions include date fixed effects. The sample covers post-cutoff earnings call transcripts (Nov 2023--Dec 2024), and the unit of observation is the earnings call. T-statistics based on time-clustered standard errors are reported in parentheses. * $p < 0.10$, ** $p < 0.05$, *** $p < 0.01$.}}
    \label{appendix-tab:sentiment_3llms}
    \vspace{0.2cm}
    \setlength{\tabcolsep}{2pt}
    \begin{tabular}{@{}lccccccccc@{}}
      \toprule
      & \multicolumn{9}{c}{Dependent Variable: $DGTW_{i,t+1}$} \\
       \cmidrule(lr){2-10}
      VARIABLES  & (1) & (2) & (3) & (4) & (5) & (6) & (7) & (8) & (9) \\
      \midrule

      $Sentiment_{RAW}$ & 2.487*** & 2.528*** & 2.400*** & & & & 1.822*** & 2.105*** & 1.660*** \\
      & (18.832) & (19.972) & (18.968) & & & & (8.475) & (13.157) & (9.504) \\
      \addlinespace

      $Sentiment_{TRF}$  & & & & 2.331*** & 2.005*** & 2.229*** & 0.775*** & 0.611*** & 0.942*** \\
      & & & & (17.287) & (16.879) & (16.412) & (3.662) & (4.337) & (5.069) \\
      \midrule
      LLM & 4o-mini & 4o & o3-mini & 4o-mini & 4o & o3-mini & 4o-mini & 4o & o3-mini \\
      \midrule
    Controls & Yes & Yes & Yes & Yes & Yes & Yes & Yes & Yes & Yes \\
      Date FE & Yes & Yes & Yes & Yes & Yes & Yes & Yes & Yes & Yes \\
      Observations & 7,352 & 7,352 & 7,352 & 7,352 & 7,352 & 7,352 & 7,352 & 7,352 & 7,352 \\
      Adj. $R^2$ & 0.132 & 0.131 & 0.125 & 0.124 & 0.108 & 0.118 & 0.133 & 0.133 & 0.129 \\

      \bottomrule
    \end{tabular}
\end{table}

\clearpage

\begin{table}[htbp]
\centering
\footnotesize
\caption{{\bf Correlations of Sentiment Signals}\\\\
\footnotesize{This table shows correlations among sentiment scores extracted using GPT-4o-mini. Panel A reports correlations among sentiment scores from raw transcripts (RAW) and the two fully anonymized transcripts (TRF and LLM). Panel B reports correlations among sentiment scores from raw transcripts and transcripts where only one category of entities is preserved while the other three categories are anonymized by the TRF model: $Sentiment_{NUM}$, $Sentiment_{PLC}$, $Sentiment_{AGT}$, and $Sentiment_{OTH}$ preserve only the NUMBERS, PLACES, AGENTS, or OTHERS category, respectively. All values are Pearson correlation coefficients, computed on the post-cutoff earnings call sample (Nov 2023--Dec 2024).}}
\label{appendix-tab:sentiment_correlation}
\vspace{0.2cm}
\setlength{\tabcolsep}{6pt}
\begin{tabular}{lccc}
\multicolumn{4}{l}{\textbf{Panel A: Raw and Fully Anonymized Sentiment}}\\
\toprule
& $Sentiment_{RAW}$ &  $Sentiment_{TRF}$ & $Sentiment_{LLM}$ \\
\midrule
$Sentiment_{RAW}$  & 1.000  &  &  \\
$Sentiment_{TRF}$  & 0.864 & 1.000 &  \\
$Sentiment_{LLM}$  & 0.858 & 0.859 & 1.000 \\
\bottomrule
\end{tabular}
\vspace{0.6em}
\setlength{\tabcolsep}{3pt}
\begin{tabular}{lccccc}
\multicolumn{6}{l}{\textbf{Panel B: Raw and Single-Category-Preserved Sentiment}}\\
\toprule
& $Sentiment_{RAW}$ & $Sentiment_{NUM}$ & $Sentiment_{PLC}$ & $Sentiment_{AGT}$ & $Sentiment_{OTH}$ \\
\midrule
$Sentiment_{RAW}$ & 1.000 &   &   &   &   \\
$Sentiment_{NUM}$ & 0.910 & 1.000 &   &   &   \\
$Sentiment_{PLC}$ & 0.872 & 0.878 & 1.000 &   &   \\
$Sentiment_{AGT}$ & 0.880 & 0.871 & 0.907 & 1.000 &   \\
$Sentiment_{OTH}$ & 0.872 & 0.876 & 0.943 & 0.905 & 1.000 \\
\bottomrule
\end{tabular}
\end{table}

\clearpage

\begin{table}[htbp]
  \centering
  \footnotesize
  \caption{{\bf Comparing Explanatory Power When Preserving Different Categories of Entities}\\\\
  \footnotesize{This table reports regressions of DGTW-adjusted stock returns ($DGTW_{i,t+1}$) on sentiment scores extracted from transcripts where only one of the four entity categories is preserved at a time, while the other three categories are anonymized using the TRF model. Following the matching convention in the text, the dependent variable is measured on the same trading day for calls before the U.S. market close (16:00 ET) and on the next trading day for calls after the close. The category definitions are provided in Appendix Table~\ref{appendix-tab:Entity Category}. Specifically, $Sentiment_{NUM}$, $Sentiment_{PLC}$, $Sentiment_{AGT}$, and $Sentiment_{OTH}$ are extracted from transcripts where the TRF model preserves only the NUMBERS, PLACES, AGENTS, or OTHERS category, respectively. Columns (1)--(5) report standalone regressions for the RAW score and the four single-category-preserved variants; Columns (6)--(9) include the RAW score together with each single-category-preserved variant. Control variables are identical to those in Panel B of Table~\ref{tab:earnings_call_sentiment}. All variables are winsorized at the 1st and 99th percentiles, and all independent variables are standardized. All regressions include date fixed effects. The sample covers post-cutoff earnings call transcripts (Nov 2023--Dec 2024), and the unit of observation is the earnings call. T-statistics based on time-clustered standard errors are reported in parentheses. * $p < 0.10$, ** $p < 0.05$, *** $p < 0.01$.}}
  \label{appendix-tab:sentiment_keep_one}
  \vspace{0.2cm}
  \setlength{\tabcolsep}{2pt}
  \begin{tabular}{@{}lccccccccc@{}}
    \toprule
    & \multicolumn{9}{c}{Dependent Variable: $DGTW_{i,t+1}$} \\
    \cmidrule(lr){2-10}
    VARIABLES & (1) & (2) & (3) & (4) & (5) & (6) & (7) & (8) & (9) \\
    \midrule
    $Sentiment_{RAW}$ & 2.487*** & & & & & 1.627*** & 1.898*** & 1.669*** & 1.816*** \\
    & (18.832) & & & & & (6.313) & (9.067) & (7.554) & (8.301) \\
    \addlinespace
    $Sentiment_{NUM}$ & & 2.419*** & & & & 0.950*** & & & \\
    & & (17.743) & & & & (3.628) & & & \\
    \addlinespace
    $Sentiment_{PLC}$ & & & 2.315*** & & & & 0.680*** & & \\
    & & & (16.708) & & & & (3.122) & & \\
    \addlinespace
    $Sentiment_{AGT}$ & & & & 2.393*** & & & & 0.936*** & \\
    & & & & (17.295) & & & & (4.077) & \\
    \addlinespace
    $Sentiment_{OTH}$ & & & & & 2.341*** & & & & 0.775*** \\
    & & & & & (17.423) & & & & (3.540) \\
    \midrule
    Controls & Yes & Yes & Yes & Yes & Yes & Yes & Yes & Yes & Yes \\
    Date FE & Yes & Yes & Yes & Yes & Yes & Yes & Yes & Yes & Yes \\
    Observations & 7,352 & 7,352 & 7,352 & 7,352 & 7,352 & 7,352 & 7,352 & 7,352 & 7,352 \\
    Adj. $R^2$ & 0.132 & 0.128 & 0.123 & 0.127 & 0.124 & 0.133 & 0.133 & 0.134 & 0.133 \\
    \bottomrule
  \end{tabular}
\end{table}

\clearpage

\begin{table}[htbp]
  \centering
  \small
  \caption{{\bf Sample Period and Explanatory Power of Sentiment}\\\\
\footnotesize{This table examines whether sentiment scores extracted from raw and anonymized earnings call transcripts differ in explanatory power between the pre- and post-knowledge-cutoff subsamples. The dependent variable is the DGTW-adjusted stock return ($DGTW_{i,t+1}$). Following the matching convention in the text, it is measured on the same trading day for calls before the U.S. market close (16:00 ET) and on the next trading day for calls after the close. $Sentiment_{TRF}$ and $Sentiment_{LLM}$ denote sentiment scores computed from TRF-anonymized and LLM-anonymized transcripts, respectively. $Pre$ is an indicator variable equal to 1 for observations in the pre-cutoff subsample and 0 for those in the post-cutoff subsample. The three columns use sentiment scores extracted from raw, TRF-anonymized, and LLM-anonymized transcripts, respectively. All variables are winsorized at the 1st and 99th percentiles, and all independent variables are standardized except for $Pre$. All specifications include date fixed effects and the control variables used in the preceding tables. The sample is the balanced panel of firms with valid observations in both the pre-cutoff (Nov 2022--Oct 2023) and post-cutoff (Nov 2023--Dec 2024) windows, and the unit of observation is the earnings call. T-statistics based on time-clustered standard errors are reported in parentheses. * $p < 0.10$, ** $p < 0.05$, *** $p < 0.01$.}}
\label{appendix-tab:sentiment_pre_interaction}
  \vspace{0.2cm}
  \setlength{\tabcolsep}{8pt}
  \begin{tabular}{@{}lccc@{}}
    \toprule
    & \multicolumn{3}{c}{Dependent Variable: $DGTW_{i,t+1}$} \\
    \cmidrule(lr){2-4}
    VARIABLES & (1) & (2) & (3) \\
    \midrule
    $Sentiment_{RAW}$ & 2.506*** & & \\
    & (19.564) & & \\
    \addlinespace
    $Sentiment_{RAW} \times Pre$ & -0.553*** & & \\
    & (-3.130) & & \\
    \addlinespace
    $Sentiment_{TRF}$ & & 2.360*** & \\
    & & (17.715) & \\
    \addlinespace
    $Sentiment_{TRF} \times Pre$ & & -0.488*** & \\
    & & (-2.728) & \\
    \addlinespace
    $Sentiment_{LLM}$ & & & 2.397*** \\
    & & & (17.913) \\
    \addlinespace
    $Sentiment_{LLM} \times Pre$ & & & -0.515*** \\
    & & & (-2.900) \\
    \midrule
    Controls & Yes & Yes & Yes \\
    Date FE & Yes & Yes & Yes \\
    Observations & 12,772 & 12,772 & 12,772 \\
    Adj. $R^2$ & 0.132 & 0.125 & 0.126 \\
    \bottomrule
\end{tabular}
\end{table}

\clearpage

\begin{table}[htbp]
  \centering
  \footnotesize
  \caption{{\bf Anonymization Gap and Information Loss in Sentiment}\\\\
  \footnotesize{This table examines whether the attenuation-adjusted gap tracks where anonymization attenuates the relation between earnings-call sentiment and DGTW-adjusted stock returns. The dependent variable is $DGTW_{i,t+1}$ in percentage points and follows the announcement-timing convention in the text. For each anonymization method, raw sentiment $R$ and anonymized sentiment $A$ are winsorized at the 1st and 99th percentiles and standardized over the pooled balanced sample. We estimate $\rho=\Corr(R,A)$ on that sample and define $\Gap=\lvert A/\rho-R\rvert$. The gap is then winsorized at the 1st and 99th percentiles and standardized. Each specification includes standardized anonymized sentiment, the standardized gap, and their product; the interaction itself is not re-standardized. The estimated values are $\rho_{TRF}=0.8529$ and $\rho_{LLM}=0.8510$, and the same pooled-sample estimate of $\rho$ is used in all columns. Columns (1)--(2) use the full balanced sample, Columns (3)--(4) the pre-cutoff window (Nov. 2022--Oct. 2023), and Columns (5)--(6) the post-cutoff window (Nov. 2023--Dec. 2024). Controls are identical to Panel B of Table~\ref{tab:earnings_call_sentiment}. All regressions include date fixed effects. T-statistics based on date-clustered standard errors are in parentheses. * $p < 0.10$, ** $p < 0.05$, *** $p < 0.01$.}}
  \label{appendix-tab:sentiment_gap_pre_post}
  \vspace{0.2cm}
  \setlength{\tabcolsep}{3pt}
  \begin{tabular}{@{}lcccccc@{}}
    \toprule
    & \multicolumn{6}{c}{Dependent Variable: $DGTW_{i,t+1}$} \\
    \cmidrule(lr){2-7}
    VARIABLES & (1) & (2) & (3) & (4) & (5) & (6) \\
    \midrule
    $Sentiment_{TRF}$ & 2.466*** & & 2.139*** & & 2.725*** & \\
    & (21.623) & & (14.182) & & (17.417) & \\
    \addlinespace
    $Sentiment_{TRF} \times \Gap_{TRF}$ & -0.543*** & & -0.431*** & & -0.635*** & \\
    & (-9.734) & & (-5.860) & & (-7.752) & \\
    \addlinespace
    $\Gap_{TRF}$ & -0.790*** & & -0.687*** & & -0.875*** \\
    & (-6.725) & & (-4.292) & & (-5.095) & \\
    \addlinespace
    $Sentiment_{LLM}$ & & 2.507*** & & 2.171*** & & 2.800*** \\
    & & (19.038) & & (13.401) & & (14.315) \\
    \addlinespace
    $Sentiment_{LLM} \times \Gap_{LLM}$ & & -0.541*** & & -0.401*** & & -0.666*** \\
    & & (-7.863) & & (-4.780) & & (-5.982) \\
    \addlinespace
    $\Gap_{LLM}$ & & -0.865*** & & -0.671*** & & -1.003*** \\
    & & (-7.801) & & (-4.080) & & (-6.781) \\
    \midrule
    Sample & Full & Full & Pre & Pre & Post & Post \\
    Controls & Yes & Yes & Yes & Yes & Yes & Yes \\
    Date FE & Yes & Yes & Yes & Yes & Yes & Yes \\
    Observations & 12,772 & 12,772 & 5,666 & 5,666 & 7,106 & 7,106 \\
    Adj. $R^2$ & 0.132 & 0.133 & 0.126 & 0.128 & 0.138 & 0.139 \\
    \bottomrule
  \end{tabular}
\end{table}

\clearpage

\section{Implementation Details for the S\&P Ten-Factor Credit-Risk Audit}\label{app:sp10_method}
Section~\ref{sec:sp10_audit} explains the audit and reports its results. This
appendix records the supplementary and implementation details needed to reproduce the
results.

\subsection{Factor Scope and Definition}\label{appendixC1}
\begin{figure}[!htbp]
  \centering
  \includegraphics[width=1\textwidth]{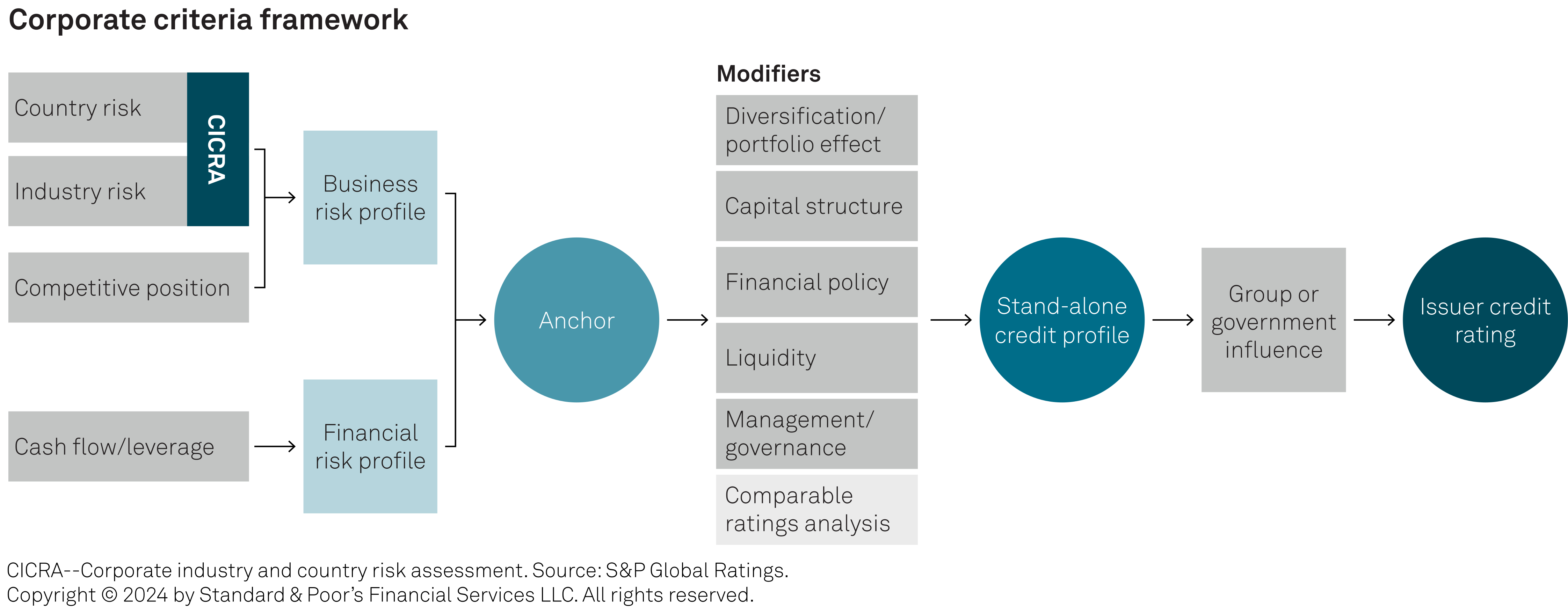}
  \caption{\textbf{S\&P Corporate Criteria Framework and Audit Coverage}\\
  \footnotesize{The figure summarizes the sequence from the business and financial risk profiles to the anchor, post-anchor modifiers, stand-alone credit profile, and issuer credit rating. The ten dimensions used in our transcript audit are IR, CR, CP, CFL, DIV, CS, FP, LIQ, MG, and CRA. Group or government influence is part of the broader rating process but is not included in the ten-factor transcript audit. Source: S\&P Global Ratings, \textit{Corporate Methodology}.}}
  \label{fig:sp_corporate_criteria_framework}
\end{figure}

The ten factors are drawn from S\&P Global Ratings'
\href{https://www.spglobal.com/ratings/en/regulatory/article/-/view/sourceId/12913251}{\textit{Corporate Methodology}}. Figure~\ref{fig:sp_corporate_criteria_framework} locates them in
the broader criteria sequence, and Table~\ref{appendix-tab:sp10_factor_scope}
gives the exact factor scopes supplied to the model. Group and government
influence lie outside the audit because they are separate from the ten
corporate-methodology dimensions and generally cannot be assessed from an
earnings call.

\subsection{Quote Matching and Evidence Decomposition}\label{appendixC2}
Using the notation introduced in Section~\ref{sec:sp10_audit}, we compare the
raw transcript separately with each anonymized version. Let
$s_{jk}^v\in[0,1]$ denote the downgrade-risk score assigned to factor $k$ for
transcript $j$ when the model reads text version
$v\in\{RAW,TRF,LLM\}$. The score $s_{jk}^v$ is included in the analysis only
when the model returns both a non-null score and a non-null supporting quote
for that factor.

When both the raw and anonymized versions contain valid evidence for a factor,
we normalize their supporting quotes before matching. We remove named-entity
spans identified by spaCy \texttt{en\_core\_web\_trf} from the raw quote and
remove explicit placeholders and tokens containing numbers from the anonymized
quote. We then lowercase both quotes, remove punctuation and extra whitespace,
and calculate a fuzzy token-sort ratio. Two quotes are treated as matched when
this ratio is at least 75.

For method $a\in\{TRF,LLM\}$, define three sets. $S_j^a$ contains factors with
matched raw and anonymized quote anchors, $R_j^a$ contains RAW-only reasons, and
$A_j^a$ contains anonymized-only reasons. Equivalently,
\[
R_j^a=\mathcal K_j^{RAW}\setminus S_j^a,
\qquad
A_j^a=\mathcal K_j^a\setminus S_j^a.
\]
If both versions populate a factor but cite different anchors, that factor
belongs to both $R_j^a$ and $A_j^a$ because each version supplies a different
reason.

Throughout, vertical bars denote set cardinality rather than absolute value.

Thus, $|S_j^a|$ is the number of factors with valid score--quote pairs in both
versions whose quote anchors match, while $|R_j^a|$ and $|A_j^a|$ count the
valid RAW-only and anonymized-only reasons, respectively. For example, if both
versions populate a factor but cite different quote anchors, that factor
contributes one to $|R_j^a|$ and one to $|A_j^a|$, but contributes nothing to
$|S_j^a|$.
The two means over shared reasons are

\begin{equation}
\begin{aligned}
 \mathrm{SharedRaw}_{j,a}
   &=\frac{1}{|S_j^a|}\sum_{k\in S_j^a}s_{jk}^{RAW},\\
 \mathrm{SharedAnon}_{j,a}
   &=\frac{1}{|S_j^a|}\sum_{k\in S_j^a}s_{jk}^{a}.
\end{aligned}
\label{eq:sp10_matched}
\end{equation}
The two means over version-specific reasons are

\begin{equation}
\begin{aligned}
 \mathrm{RawOnly}_{j,a}
   &=\frac{1}{|R_j^a|}\sum_{k\in R_j^a}s_{jk}^{RAW},\\
 \mathrm{AnonOnly}_{j,a}
   &=\frac{1}{|A_j^a|}\sum_{k\in A_j^a}s_{jk}^{a}.
\end{aligned}
\label{eq:sp10_displaced}
\end{equation}
Each conditional mean is defined only when the corresponding set is nonempty.

If a set is empty, its conditional mean is coded as missing rather than zero.

Because $S_j^a$ and $R_j^a$ partition the valid raw reasons, while $S_j^a$ and
$A_j^a$ partition the valid anonymized reasons, the full raw and anonymized
composites can be recovered as

\begin{equation}
\begin{aligned}
 C_j^{RAW}
 &=\frac{\sum_{k\in S_j^a}s_{jk}^{RAW}
          +\sum_{k\in R_j^a}s_{jk}^{RAW}}
         {|S_j^a|+|R_j^a|},\\
 C_j^{a}
 &=\frac{\sum_{k\in S_j^a}s_{jk}^{a}
          +\sum_{k\in A_j^a}s_{jk}^{a}}
         {|S_j^a|+|A_j^a|}.
\end{aligned}
\label{eq:sp10_identity}
\end{equation}
Equation~\eqref{eq:sp10_identity} also provides a computational check on the
decomposition. In each composite, the relevant factor scores are pooled before
division by the total number of valid reasons; the separate conditional means
are never added directly.

\subsection{Shapley Decomposition}\label{appendixC3}
Equation~\eqref{eq:sp10_identity} provides the two endpoint states used in the
Shapley decomposition. Specifically,
$C_{j,a}^{\mathrm{Raw}}\equiv C_j^{RAW}$ is the original raw composite, while
$C_{j,a}^{\mathrm{Anon}}\equiv C_j^a$ is the fully anonymized composite. The
two intermediate states are constructed directly from the underlying score
sums and reason counts:

\begin{equation}
\begin{aligned}
 C_{j,a}^{\mathrm{Replace}}
   &=\frac{\sum_{k\in S_j^a}s_{jk}^{RAW}
            +\sum_{k\in A_j^a}s_{jk}^{a}}
           {|S_j^a|+|A_j^a|},\\
 C_{j,a}^{\mathrm{Reread}}
   &=\frac{\sum_{k\in S_j^a}s_{jk}^{a}
            +\sum_{k\in R_j^a}s_{jk}^{RAW}}
           {|S_j^a|+|R_j^a|}.
\end{aligned}
\label{eq:sp10_hybrid_composites}
\end{equation}
The \textit{Replace} state introduces evidence displacement alone by replacing
the RAW-only reasons with the anonymized-only reasons while retaining the raw
scores on shared reasons. The \textit{Reread} state introduces interpretive
distortion alone by retaining the raw evidence set while replacing the raw
scores on shared reasons with their anonymized counterparts. Each numerator
pools the relevant factor scores before division; the conditional group means
are never added directly.

Let
\[
\mathcal Q=\{\mathrm{Raw},\mathrm{Replace},
             \mathrm{Reread},\mathrm{Anon}\}
\]
denote the four states. For each anonymization method $a$ and each state
$q\in\mathcal Q$, we estimate equation~\eqref{eq:sp10_composite_regression}
separately by setting the composite score equal to $C_{j,a}^{q}$. Equivalently,
each state is entered into the following single-score regression:

\begin{equation}
 D_j
 =\alpha+\beta_q C_{j,a}^{q}
 +\gamma X_j+\delta_t+\varepsilon_j,
 \qquad q\in\mathcal Q.
\label{eq:sp10_state_regression}
\end{equation}
Thus, the four regressions differ only in which state composite is used as the
primary explanatory variable. The outcome, controls, SIC3 $\times$ Quarter
fixed effects, and estimation sample are held fixed across the four
regressions.

Before estimation, the four state composites are winsorized and standardized
separately. For each anonymization method, we restrict the analysis to the
common sample for which all four composites, the outcome, and the controls are
observed. This common-sample restriction ensures that differences in adjusted
$R^2$ reflect differences in the predictive content of the four composites
rather than differences in sample composition.

Suppressing the method subscript $a$ for notational simplicity, let
$R^2_{\mathrm{Raw}}$, $R^2_{\mathrm{Replace}}$,
$R^2_{\mathrm{Reread}}$, and $R^2_{\mathrm{Anon}}$ denote the adjusted $R^2$
values obtained from the four regressions:
\[
\begin{aligned}
 C_{j,a}^{\mathrm{Raw}}
   &\longrightarrow R^2_{\mathrm{Raw}},&
 C_{j,a}^{\mathrm{Replace}}
   &\longrightarrow R^2_{\mathrm{Replace}},\\
 C_{j,a}^{\mathrm{Reread}}
   &\longrightarrow R^2_{\mathrm{Reread}},&
 C_{j,a}^{\mathrm{Anon}}
   &\longrightarrow R^2_{\mathrm{Anon}}.
\end{aligned}
\]
Total information loss is therefore
\[
L=R^2_{\mathrm{Raw}}-R^2_{\mathrm{Anon}}.
\]
The Shapley decomposition averages each channel's marginal contribution across
the two possible orderings:
\[
\mathrm{Raw}\rightarrow\mathrm{Replace}\rightarrow\mathrm{Anon}
\quad\text{and}\quad
\mathrm{Raw}\rightarrow\mathrm{Reread}\rightarrow\mathrm{Anon}.
\]

\FloatBarrier
\begingroup
\footnotesize
\setlength{\tabcolsep}{3pt}
\renewcommand{\arraystretch}{1.08}

\begin{longtable}{@{}
  >{\raggedright\arraybackslash}p{0.8cm}
  >{\raggedright\arraybackslash}p{3.3cm}
  >{\raggedright\arraybackslash}p{11.6cm}@{}}
\caption{\textbf{S\&P-Aligned Factor Scope Used in the Credit-Risk Audit}}
\label{appendix-tab:sp10_factor_scope}\\
\toprule
\textbf{ID} & \textbf{Factor} & \textbf{Definition} \\
\midrule
\endfirsthead
\multicolumn{3}{c}{\tablename\ \thetable{} -- Continued from previous page}\\[1ex]
\toprule
\textbf{ID} & \textbf{Factor} & \textbf{Definition} \\
\midrule
\endhead
\midrule
\multicolumn{3}{r}{Continued on next page}\\
\endfoot
\bottomrule
\endlastfoot
IR & Industry Risk & Assesses the relative health and stability of the issuer's industry markets through revenue and profit cyclicality, competitive risk, and the growth environment, including barriers to entry, industry margin levels and trends, secular change and substitution risk, and industry growth trends. \\
\addlinespace
CR & Country Risk & Assesses country-related economic risk, institutional and governance effectiveness risk, financial-system risk, and payment culture or rule-of-law risk across the countries to which the issuer is exposed, taking account of the scale and concentration or diversification of those exposures. \\
\addlinespace
CP & Competitive Position & Assesses four components: competitive advantage; scale, scope, and diversity; operating efficiency; and profitability, including the level and volatility of profitability, in the context of the issuer's industry characteristics. \\
\addlinespace
CFL & Cash Flow/Leverage & Assesses current and prospective cash flow relative to financial obligations, centered on FFO to debt and debt to EBITDA and supplemented, where relevant, by CFO to debt, FOCF to debt, DCF to debt, and interest coverage, together with cash-flow needs and volatility. \\
\addlinespace
DIV & Diversification/ Portfolio Effect & For entities that qualify for the diversification or portfolio-effect assessment, assesses whether diversity across material business lines and the correlation of their cash flows reduce consolidated cash-flow volatility and strengthen debt-service capacity across cycles. Under the methodology, this modifier is positive or neutral. \\
\addlinespace
CS & Capital Structure & Assesses capital-structure risks or benefits not fully captured in cash flow and leverage, including debt-currency risk, debt-maturity profile, debt interest-rate risk, and the potential credit benefit of investments. \\
\addlinespace
FP & Financial Policy & Assesses how owners' and management's financial policy can affect credit quality beyond the base-case forecast, including financial-sponsor influence, financial discipline, the financial-policy framework, leverage tolerance, shareholder returns, acquisitions and growth, and event risk. \\
\addlinespace
LIQ & Liquidity & Assesses the sufficiency and reliability of liquidity sources relative to uses, covenant headroom, capacity to absorb high-impact low-probability events, banking relationships, standing in credit markets, and financial risk management. \\
\addlinespace
MG & Management and Governance & Assesses ownership; board structure, composition, and effectiveness; risk management, internal controls, and audit; transparency and reporting; and management, including capabilities not already reflected in other rating factors. \\
\addlinespace
CRA & Comparable Ratings Analysis & Provides the final holistic review of the issuer's relative positioning after the other factors, considering where aggregate strengths and weaknesses place it within assessment categories, proximity to financial-risk boundaries, and material residual factors not fully captured elsewhere. It may adjust the SACP by one notch. \\
\end{longtable}
\endgroup

\clearpage

\section{Prompts}\label{app:prompts}
\subsection{LLM Anonymization Sequence and Prompts}\label{subsec:anonymization_prompt}
We adopt the entity-neutering methodology proposed by \citet{engelberg2025entity} and implement it in the following operational order: (1) masking, (2) an initial re-identification test, (3) conditional paraphrasing if the masked output fails that test, and (4) a second re-identification test. A passing masked output is accepted immediately and is not paraphrased. If the paraphrased output also fails, the next round uses more aggressive versions of the masking and paraphrasing prompts. The procedure stops when an output passes or after three rounds.

\subsubsection{Step 1: Masking}
The masking strategy replaces specific identifiers with generic placeholders. The initial masking prompt is defined as follows, where \{NAME\}, \{TICKER\}, and \{INDUSTRY\} are placeholders for the specific firm's details:
\\
\\
\textit{Your role is to ANONYMIZE all text that is provided by the user. The text you need to anonymize will be about the firm: \{NAME\} whose ticker is: \{TICKER\} and is in the industry of: \{INDUSTRY\}. Any instance of the firm, their products, the industry or industry related areas they are involved in should be anonymized. Absolutely no mention of key words related to that industry can be left in the text. After you have anonymized a text, NOBODY, not even an expert financial analyst should be able to use the text to identify the company or the industry the company operates in. For example, if the text is: The country's largest phone producer Apple had great phone-related earnings but Google did not in 2024, likely because of Apple's slogan Think Different, then you should ANONYMIZE it to: The country's largest product\_type\_1 producer Company\_1 had great product\_type\_1 related earnings but Company\_2 did not in time\_1 likely because of Company\_1's slogan slogan\_1. You should also ANONYMIZE any other information which one could use to identify the company or to make an educated guess about its identity. Stock tickers are identifiers, are usually four or fewer capitalized letters (use TIKR as a stand-in for an arbitrary ticker), and can be referenced in the text in the following formats; SYMBOL:TIKR, \$TIKR, $>$TIKR, \$ TIKR, SYMBOL TIKR, SYMBOL: TIKR, $>$ TIKR. Make sure you ANONYMIZE TIKR to ticker\_x. Also ANONYMIZE any other identifiers related to companies, including the names of individuals, locations, industries, sectors, product names and types and generic product lines, services, times, years, dates and all numbers and percentages in the text, including units. These should be replaced with name\_x, location\_x, industry\_x, sector\_x, product\_x, product\_type\_x, product\_line\_x, service\_x, time\_x, year\_x, date\_x and number a, b, c, respectively. Also replace any website or internet links with link\_x. You should never just delete an identifier; instead, always replace it with an anonymized analog. After you read and ANONYMIZE the text, you should output the anonymized text and nothing else.}
\\
\\
In the iterative masking phase, the prompt is modified to be more aggressive to ensure anonymity. The beginning of the prompt is replaced with: \textit{``Your role is to ANONYMIZE all text that is provided by the user. You have already tried to anonymize the following text and failed, leaving details which were used to identify the text by an expert. Be more aggressive in your efforts to anonymize and pay more attention, so that you do not leave even the most minute details in the anonymized text.''}
\\
Additionally, the instructions at the end are reinforced with: \textit{``You have already failed me before, take no chances this time and be aggressive in censoring the text. After you read and ANONYMIZE the text, you should output the anonymized text and nothing else.''}

\subsubsection{Step 2: Initial Re-identification Test}\label{subsec:anonymization_test_prompt}
Immediately after masking, we test whether an LLM can re-identify the firm, its industry, or the date from the masked text, following \citet{engelberg2025entity}. The test prompt is:
\\
\\
\textit{You will receive a body of text which has been anonymized. You are omniscient. Use all of your knowledge and any clues in the text to identify which company the text is about, its industry, and the date it was written. Provide your best guess if you are unsure. Your options for industry guesses are the following: Agriculture, Food Products, Candy \& Soda, Beer \& Liquor, Tobacco Products, Recreation, Entertainment, Printing and Publishing, Consumer Goods, Apparel, Healthcare, Medical Equipment, Pharmaceutical Products, Chemicals, Rubber and Plastic Products, Textiles, Construction Materials, Construction, Steel Works Etc, Fabricated Products, Machinery, Electrical Equipment, Automobiles and Trucks, Aircraft, Shipbuilding \& Railroad Equipment, Defense, Precious Metals, Non-Metallic and Industrial Metal Mining, Coal, Petroleum and Natural Gas, Utilities, Communication, Personal Services, Business Services, Computers, Electronic Equipment, Measuring and Control Equipment, Business Supplies, Shipping Containers, Transportation, Wholesale, Retail, Restaurants Hotels \& Motels, Banking, Insurance, Real Estate, Trading, Almost Nothing. Provide your estimate EXACTLY in the following format, with NO other text at all (TIKR is your estimate of the ticker, NAME is your estimate of the firm name, IND is your estimate of the firm's industry, YYYY-MM-DD is your estimate of the date): **Ticker Estimate: TIKR**,**Name Estimate: NAME**,**Industry Estimate: IND**,**Date Estimate: YYYY-MM-DD**}
If neither the firm nor the date is identified under the matching criteria in Section~\ref{subsec:anonymization_algorithms}, the masked output is accepted. If either is identified, the output proceeds to Step 3.

\subsubsection{Step 3: Conditional Paraphrasing}
The paraphrasing strategy rewrites the text to remove identifying writing styles and structures while preserving core information. The initial paraphrasing prompt is defined as follows:
\\
\\
\textit{You are to paraphrase text ensuring that its core information content is unchanged, such that someone who has memorized the text could not recognize it. Make sure to replace all unique or identifying words and phrases. Also change the sentence structure so that the resulting text is completely different from the original. This includes changing quotes and other highly recognizable structures in the text. The text provided will be about the firm: \{NAME\} whose ticker is: \{TICKER\} and is in the industry: \{INDUSTRY\}. Any information which could ever be used to identify the firm the text is about or when the text was written MUST be anonymized. Absolutely no mention of key words related to that industry or any other industries can be left in the text. If you observe any names, proper nouns, aliases, acronyms, dates, industry references (related and unrelated to the industry of the firm in question), government agency names, product names or types, etc., replace them with anonymous strings. Anonymity is the prime objective, leave zero hints for someone who has memorized the text! Ensure that the paraphrased text has approximately the same number of words as the original text. After you read and ANONYMIZE the text, you should output the anonymized text and nothing else.}
\\
\\
Similar to the masking strategy, the iterative paraphrasing prompt is modified to strictly enforce anonymity. The beginning is modified to: \textit{``You are to paraphrase text ensuring that its core information content is unchanged, such that someone who has memorized the text could not recognize it. You have already failed to paraphrase the text to make it anonymous. Be more aggressive in your paraphrasing and anonymizing of the text. Make sure to replace all unique or identifying words and phrases. Also change the sentence structure so that the resulting text is completely different from the original. This includes changing quotes and other highly recognizable structures in the text.''}
\\
The prompt concludes with the reinforced instruction: \textit{``You have already failed once before, take no chances this time, be aggressive in your effort to anonymize the text. After you read and ANONYMIZE the text, you should output the anonymized text and nothing else.''}

\subsubsection{Step 4: Final Re-identification Test}
The same re-identification prompt and the same passing criteria from Step 2 are applied to the paraphrased output. If neither the firm nor the date is identified, the paraphrased output is accepted as the final LLM-anonymized text. If either is identified, the output fails and a new round begins with the more aggressive masking prompt, followed by the same test--paraphrase--test sequence. The procedure is capped at three rounds; if the third-round output still fails, the last paraphrased output is retained because the round limit has been reached. Thus, the implemented order is always masking $\rightarrow$ test $\rightarrow$ conditional paraphrasing $\rightarrow$ test; masking and paraphrasing are never performed consecutively without an intervening re-identification test.

\subsection{Downgrade Prediction Prompt}\label{subsec:downgrade_prompt}
We adapt the approach of \citet{donovan2021measuring} to design a prompt that extracts the predicted probability of an S\&P credit rating downgrade within one year of the earnings call:
\textit{You are an expert Credit Risk Analyst. Your task is to read quarterly company earnings call transcripts and estimate the probability that the company will experience an S\&P credit rating downgrade within the next twelve months from the date of this call. Based on your analysis, output only a single probability score between 0 and 1, where a score near 1 indicates a very high likelihood of a downgrade (severe credit risk) and a score near 0 indicates a very low likelihood (stable or improving credit profile). Provide your output exactly in the following format, with no other text at all: **Downgrade Estimate: [score]**.}

\subsection{Uncertainty Score Prompt}\label{subsec:uncertainty_prompt}
We refer to the definition in \citet{WhenIsaLiability} to extract the uncertainty score from the transcript, with the following prompt:\\
\textit{Analyze the following conference call transcript. As a financial analyst specializing in textual analysis, your task is to evaluate the level of Uncertainty. Your evaluation must be on a continuous scale from 0 (representing absolute certainty) to 1 (representing maximum uncertainty) ONLY. This assessment should not be based on a simple count of keywords. Instead, you must analyze the contextual meaning, usage, and significance of words and phrases that convey a cautious or uncertain tone, emphasizing the general notion of imprecision rather than exclusively focusing on risk. Your goal is to determine the genuine, unmitigated uncertainty conveyed by the transcript. Provide your output only in the following format, with no other text: **Uncertainty Score: UNCERTAINTY**.\\
}

\subsection{Investment Score Prompt}\label{subsec:investment_prompt}
We refer to the prompt in \citet{jha2025chatgptcorporatepolicies} to extract the Investment score from the transcript, with the following prompt:\\
\textit{The following text is a company's earnings call transcript. You are a finance expert. Based on this text only, please answer the following question. How does the firm plan to change its capital spending over the next year? There are five choices: Increase substantially, increase, no change, decrease, and decrease substantially. Please select one of the above five choices for each question and provide a one-sentence explanation of your choice for each question. The format for the answer to each question should be **choice - explanation**. If no relevant information is provided related to the question, answer **no information is provided**.}

\subsection{Economy Score Prompt}\label{subsec:economy_prompt}
We refer to the prompt in \citet{jha2025harnessinggenerativeaieconomic} to extract the Economy score from the transcript, with the following prompt:\\
\textit{The following text is a company's earnings call transcript. You are a finance expert. Based on this text only, please answer the following question. Over the next quarter, how does the firm anticipate a change in optimism about the US economy? There are five choices: Increase substantially, increase, no change, decrease, and decrease substantially. Please select one of the above five choices for each question and provide a one-sentence explanation of your choice for each question. The format for the answer to each question should be **choice - explanation**. If no relevant information is provided related to the question, answer **no information is provided**.}

\subsection{Sentiment Prompt for Transcripts}\label{subsec:sentiment_prompt}
In designing the prompt for our analysis, we reference the methodology for controlling LLM output tokenization from \citet{engelberg2025entity}. They demonstrate that by forcing the model to generate a number in a structured format—such as 0.X for a sentiment score—one can isolate the magnitude of the estimate into the log probability of a single, unambiguous token (the digit X). We adopt this principle to ensure that the numerical values generated by our prompt are free from tokenization artifacts, such as leading or trailing spaces. This structured approach allows us to reliably use the model's log probabilities as a direct measure of its assessment. We use the following prompt:\\
\\
\textit{You are an expert Financial Analyst. Your task is to read quarterly company earnings call transcripts and infer if the information in the transcripts is a good or bad signal for the future returns of the security of the company whose earnings call is being transcribed. Based on your answer, I will choose to buy or sell that security. After reading the transcripts, provide a bull or bear signal as two numbers: the direction and the magnitude. For direction, output only 0,1, or NA, where 0 is bearish, 1 is bullish, and NA means there is no information relevant to security prices. For magnitude, output a number ranging from 0 to 1 ONLY, where numbers near 0 indicate slightly bearish (if the direction is 0) or slightly bullish (if the direction is 1), and numbers near 1 indicate highly bearish (if the direction is 0) or highly bullish (if the direction is 1). Provide your output exactly in the following format, with no other text at all: **Direction Estimate: DIRECTION**,**Magnitude Estimate: MAGNITUDE**.\\}

\subsection{Sentiment Prompt for News Headlines}\label{subsec:sentiment_news_prompt}
We used a similar LLM prompt to extract sentiment from news:\\
\textit{You are an expert Financial Analyst. Your task is to read news headlines and infer if the reported headline is a good or bad signal for the future returns of the security being written about. Based on your answer, I will choose to buy or sell that security. After reading the headline, provide a bull or bear signal as two numbers: the direction and the magnitude. For direction, output only 0,1, or NA, where 0 is bearish, 1 is bullish, and NA means there is no information relevant to security prices. For magnitude, output a number ranging from 0 to 1 ONLY, where numbers near 0 indicate slightly bearish (if the direction is 0) or slightly bullish (if the direction is 1), and numbers near 1 indicate highly bearish (if the direction is 0) or highly bullish (if the direction is 1). Provide your output exactly in the following format, with no other text at all: **Direction Estimate: DIRECTION**,**Magnitude Estimate: MAGNITUDE**.}

\subsection{S\&P Ten-Factor Score-and-Quote Prompt}\label{subsec:sp10_prompt}

For the audit in Section~\ref{sec:sp10_audit}, the user message contains the complete earnings call transcript and the system message below is held fixed across raw, TRF-anonymized, and LLM-anonymized versions. We use GPT-4o-mini-2024-07-18 with temperature zero and a strict structured-output schema whose only top-level keys are IR, CR, CP, CFL, DIV, CS, FP, LIQ, MG, and CRA. Each key must contain exactly two fields, \texttt{score} and \texttt{quote}; each field is nullable, and additional properties are prohibited. We use the following prompt:

\begingroup
\itshape
\setlength{\parindent}{0pt}

You are an expert Credit Risk Analyst. The user message is the transcript of a quarterly company earnings call. Using only this transcript, score the focal company on each of the ten S\&P Corporate Methodology factors below: the likelihood that this factor will contribute to an S\&P credit rating downgrade within the next twelve months from the date of this call.

\medskip

Each score is a single number between 0 and 1: near 1 indicates a very high likelihood of a downgrade driven by this factor (severe credit risk on this dimension); near 0 indicates a very low likelihood (stable or improving credit profile).

\medskip

Use only concrete issuer-specific evidence stated in the transcript. Do not use outside knowledge or infer unstated facts. Analyst questions without company confirmation, generic boilerplate, and vague possibilities are not qualifying evidence.

\medskip

IR Industry Risk: Assesses the relative health and stability of the issuer's industry markets through revenue and profit cyclicality, competitive risk, and the growth environment, including barriers to entry, industry margin levels and trends, secular change and substitution risk, and industry growth trends.\\
CR Country Risk: Assesses country-related economic risk, institutional and governance effectiveness risk, financial-system risk, and payment culture or rule-of-law risk across the countries to which the issuer is exposed, taking account of the scale and concentration or diversification of those exposures.\\
CP Competitive Position: Assesses four components: competitive advantage; scale, scope, and diversity; operating efficiency; and profitability, including the level and volatility of profitability, in the context of the issuer's industry characteristics.\\
CFL Cash Flow/Leverage: Assesses current and prospective cash flow relative to financial obligations, centered on FFO to debt and debt to EBITDA and supplemented, where relevant, by CFO to debt, FOCF to debt, DCF to debt, and interest coverage, together with cash-flow needs and volatility.\\
DIV Diversification/Portfolio Effect: For entities that qualify for the diversification or portfolio effect assessment under the methodology, assesses whether diversity across material business lines and the correlation of their cash flows reduce consolidated cash-flow volatility and strengthen debt-service capacity across cycles. Under the methodology, this modifier is positive or neutral.\\
CS Capital Structure: Assesses capital-structure risks or benefits not fully captured in cash flow and leverage, including debt currency risk, debt maturity profile, debt interest-rate risk, and the potential credit benefit of investments.\\
FP Financial Policy: Assesses how owners' and management's financial policy can affect credit quality beyond the base-case forecast, including financial-sponsor influence, financial discipline, the financial-policy framework, leverage tolerance, shareholder returns, acquisitions and growth, and event risk.\\
LIQ Liquidity: Assesses the sufficiency and reliability of liquidity sources relative to uses, covenant headroom, capacity to absorb high-impact low-probability events, banking relationships, standing in credit markets, and financial risk management.\\
MG Management and Governance: Assesses ownership; board structure, composition, and effectiveness; risk management, internal controls, and audit; transparency and reporting; and management, including capabilities not already reflected in other rating factors.\\
CRA Comparable Ratings Analysis: Provides the final holistic review of the issuer's relative positioning after the other factors, considering where aggregate strengths and weaknesses place it within assessment categories, proximity to financial-risk boundaries, and material residual factors not fully captured elsewhere. It may adjust the SACP by one notch.

\medskip

For every factor return exactly two fields:\\
1. score: a number between 0 and 1, or null when the transcript contains no qualifying evidence within this factor's S\&P-defined scope.\\
2. quote: one exact contiguous verbatim quote copied from the transcript that is most decision-relevant for the score, or null when score is null.

\medskip

Mandatory rules:\\
- score and quote must either both be non-null or both be null.\\
- If the transcript contains no qualifying issuer-specific evidence within a factor's defined scope, return score=null and quote=null. Do not force a score merely to fill every factor.\\
- The quote must directly support the score under that factor's definition. Do not paraphrase it.\\
- Classify evidence by the factor scopes above and avoid double counting. Do not assign evidence to a factor outside its defined scope merely to avoid null.\\
- Evaluate each transcript independently and do not assume that all ten factors will be discussed in an earnings call.

\medskip

Output only the ten factor keys as one JSON object, with no text outside JSON.

\par
\endgroup
\end{document}